%% file: paper.tex
\documentclass[fleqn]{llncs}
\title{Benchmarks for Parity Games\\(extended version)}

\author{Jeroen J.A. Keiren\inst{1,2} \quad \email{j.j.a.keiren@vu.nl}
\institute{Theoretical Computer Science,
    VU University Amsterdam, The Netherlands
    \and
    Faculty of Management, Science \& Technology,
    Open Universiteit, Heerlen, The Netherlands
    }
}

\usepackage{amsmath}
\usepackage{amssymb}
\usepackage[UKenglish]{babel}
\usepackage{booktabs}
\usepackage{graphicx}

\usepackage{pgfplots}
\usepackage{stmaryrd}
\usepackage{subfig}
\usepackage{tikz}
\usetikzlibrary{plotmarks}
\usepgfplotslibrary{statistics}
\usepackage{xspace}

\input{macros}

\begin{document}
\maketitle

\begin{abstract} 
We propose a benchmark suite for parity games that includes the benchmarks that
have been used in the literature, and make it available online. We give an 
overview of the parity games, including a description of how they have been 
generated. We also describe structural properties of parity games, and using 
these properties we show that our benchmarks are representative. With this work 
we provide a starting point for further experimentation with parity games.
\end{abstract}

\section{Introduction}

Parity games (see, \eg, \cite{EJ:91,McN:93,Zie:98})
play an important role in model checking research. The
$\mu$-calculus model checking problem is polynomial time reducible to the
problem of deciding the winner in parity games \cite{Sti:99}. Other problems 
that are expressible in parity games are equivalence checking of labelled 
transition systems \cite{Sti:99}, as well as synthesis, satisfiability and 
validity of temporal logics \cite{Saf:88}.

Besides their practical interest for verification, solving (deciding the winner of)
parity games is known to be in the complexity class $\mathsf{NP} \cap \mathsf{co-NP}$, and more 
specifically in $\mathsf{UP} \cap \mathsf{co-UP}$ \cite{Jur:98}. Parity game
solving is one of 
the few problems in this complexity class that is not known to be in 
$\mathsf{P}$, yet there is hope that a polynomial time algorithm exists. In 
recent years this has led to the development of (1) a large number of 
algorithms for solving parity games, such as \cite{JPZ:06,Sch:07,Sche:08csl}, 
all of which were recently shown to be exponential, and (2) the study of 
(polynomial time) reduction techniques for 
parity games \cite{FL:09atva,KW:11,CKW:11,CKW:12}.

So far, practical evaluation of parity game algorithms has been based on ad-hoc 
benchmarks, mainly consisting of random games or synthetic benchmarks.
Friedmann and Lange observed in 2009 \cite{FL:09atva} that no standard 
benchmark set for parity games was available. They introduced a small benchmark 
set in the context of their comprehensive comparison of parity game solving 
algorithms and their related heuristics \cite{FL:09atva}. The set of benchmarks 
was extended in \cite{KW:11,CKW:11,CKW:12} using model checking and equivalence 
checking cases. To the best of our knowledge, the situation has not improved 
since then, and the benchmarks in these papers still are the most comprehensive 
benchmarks included in a single paper.
The number of games and the diversity of parity games in each set in 
isolation are however limited. The lack of standard benchmarks makes 
it hard to compare the different tools and algorithms presented in the 
literature.
\medskip

To improve the current situation, in this paper we propose a set of parity 
games for benchmarking purposes that (1) is diverse, (2) contains games that 
originate from different verification problems, and (3) includes those games 
that have been used to experimentally evaluate algorithms in the literature.

In general, parity game examples in the literature can be classified as follows 
(we indicate their origins):
\begin{enumerate}
  \item Encodings of problems such as model checking, equivalence checking and
  complementation of B\"uchi automata to parity games
  \cite{Mad:97,Mat:03,PW:08,KW:11,CKW:11,CKW:12,FL:09atva}.
  \item Synthetic parity games for which a certain solving algorithm requires 
  exponential
  time
  \cite{Mad:97,Jur:00,Obdr:06,Frie:09lics,FL:10pgsolver,Frie:11rairo,GW:13}. 
  \item Random games 
  \cite{BV:01,Lan:05,Sche:08csl,Sche:08thesis,FL:09atva,FL:10pgsolver}.
\end{enumerate}
Our benchmarks include games from each of these categories.

Additionally, inspired by the properties for explicit state spaces in 
\cite{Pel:04} we introduce a set of structural properties for parity games, and
in the spirit of \cite{Pel:06,Pel:07} we analyse our benchmarks. Among others, 
we introduce a novel notion of alternation depth for parity games.

The structure of the paper is as follows. We first introduce parity games and
their structural properties in Section~\ref{sec:structural_properties}. Next 
we describe the benchmarks (Section~\ref{sec:benchmark_games}) and the way in
which they have been generated (Section~\ref{sec:implementation}). Finally we
illustrate diversity of our benchmarks with respect to the structural 
properties in Section~\ref{sec:games_properties}. This paper is based on
the PhD thesis of the author \cite[Chapter 5]{Kei:13}. The current paper is
an extended version of the work in \cite{Kei:15}.
We plan to update this version when new benchmarks are added, and we invite
the community to contribute benchmarks.

\section{Parity Games and Their Structural Properties} 
\label{sec:structural_properties}

A parity game is a two-player game played on a finite, directed graph by two
players, \emph{even} and \emph{odd}, denoted $\even$ and $\odd$, respectively.
We use $\player \in \{ \even, \odd \}$ to denote an arbitrary player.
Formally, a parity game is a structure $(V_{\even}, V_{\odd}, \to,
\priosym)$, where $V_{\even}$ and $V_{\odd}$ are disjoint sets
of vertices. We say that $\player$ owns $v$ if $v \in V_{\player}$, we
write $V$ for $V_{\even} \cup V_{\odd}$;
$\to \subseteq V \times V$ provides the total edge relation---hence each vertex
has a successor---and $\priosym \colon V \to \mathbb{N}$ assigns a non-negative 
integer priority to every vertex. The parity game is played
by placing a token on some initial vertex, and then the players take turns
moving the token: if the token is on a vertex $v \in V_{\player}$ then
$\player$ plays the token to one of the successors of $v$. This way, an infinite
play through the game is constructed. If the largest priority that occurs
infinitely often on this play is \emph{even} (resp. \emph{odd}) then $\even$ 
(resp. $\odd$) wins the play.

The time required for parity game solving and reduction algorithms depends on
the structure of the game. Typically the algorithmic complexity of parity
game algorithms are expressed in terms of the size of the game graph, \ie the 
number of vertices and edges, and the number of priorities in the game. Although
other structural properties may not affect the asymptotic running times of the
algorithms, in general they do affect the actual running time. We therefore
describe a number structural properties that could be used for the further study
of parity games.

\paragraph{Sizes.}
As basic parity game properties, we consider the numbers of vertices 
$\cardinality{V}$, $\cardinality{V_{\even}}$ and $\cardinality{V_{\odd}}$, and 
the number of edges $\cardinality{\to}$. We write $\prio{V}$ for the 
set of 
priorities $\{ \prio{v} \mid v \in V \}$, and denote the number of priorities 
in the game by $\cardinality{\prio{V}}$. The number of vertices with priority 
$k$ is represented by $\cardinality{\inverseprio{k}}$. The complexity of most
parity game algorithms is expressed in these quantities. For parity games in 
which either $\cardinality{V_{\odd}} = 0$ or $\cardinality{V_{\even}} = 0$,
special polynomial time solving algorithms are available, see \cite{FL:09atva}.

\paragraph{Degrees.}
Typical structural properties in the graph are the in- and out-degrees of 
vertices, \ie, the number of incoming and outgoing edges of vertices. Formally, 
for vertex $v \in V$, $\indegree{v} = \cardinality{\{ u \in V \mid u \to v \}}$,
$\outdegree{v} = \cardinality{\{ w \in V \mid v \to w \}}$,
and $\degree{v} = \cardinality{\{ w \in V \mid v \to w \lor w \to 
v \}}$ are the in-degree, out-degree and degree of $v$. We consider the 
minimum, maximum and average of these values.

The degrees of vertices might have an effect on, \eg, algorithms that use
lifting strategies to propagate information between vertices. Examples of
such algorithms are small progress measures \cite{Jur:00} and the strategy
improvement algorithm \cite{Sche:08csl}.

\paragraph{Strongly Connected Components.}
The strongly connected components (SCCs) of a graph are the maximal strongly 
connected subgraphs. More formally, a strongly connected component is a maximal 
set $\mathcal{C} \subseteq V$ for which, for all $u, v \in \mathcal{C}$, $u 
\to^* v$, \ie, each vertex in $\mathcal{C}$ can reach every other vertex in 
$\mathcal{C}$.

The strongly connected components in a graph induce a quotient graph. Let 
$\sccs{G}$ denote the strongly connected components of the graph. The quotient 
graph is the graph $(\sccs{G}, \to')$ and for $\mathcal{C}_1, \mathcal{C}_2 \in
\sccs{G}$, there is an edge $\mathcal{C}_1 \to' \mathcal{C}_2$ if and only if
$\mathcal{C}_1 \neq \mathcal{C}_2$ and there exist $u \in \mathcal{C}_1$ and $v
\in \mathcal{C}_2$ such that $u \to v$. Observe that the quotient graph is a
directed acyclic graph.

We say that an SCC $\mathcal{C}$ is \emph{trivial} if $\cardinality{\mathcal{C}}
= 1$ and $\mathcal{C} \not \to \mathcal{C}$, \ie, it only contains one vertex
and no edges, and we say that $\mathcal{C}$ is \emph{terminal} if $\mathcal{C} 
\not \to'$, \ie, its outdegree in the quotient graph is $0$.
The \emph{SCC quotient height} of a graph is the length of the longest path in
the quotient graph.

Parity game algorithms and heuristics can benefit from a decomposition into 
strongly connected components (SCCs). One prominent example of this is the 
global parity game solving algorithm presented by Friedmann and Lange 
\cite{FL:09atva}, for which it was shown that SCC decomposition generally works 
well in practice.

\paragraph{Properties of Search Strategies.}
Given some initial vertex $v_0 \in V$, breadth-first search (BFS) and 
depth-first search (DFS) are search strategies that can be used to 
systematically explore all vertices in the graph. The fundamental difference 
between BFS and DFS is that the BFS maintains a queue of vertices that still 
need to be processed, whereas the DFS maintains a stack of vertices. We record 
the queue and stack sizes during the search.

Breadth-first search induces a natural notion of levels, where a vertex is at
level $k$ if it has least distance $k$ to $v_0$. The \emph{BFS height} of a 
graph is $k$ if $k$ is the maximal non-empty level of the BFS. For each level 
the number of vertices at that level is recorded. During a BFS, three kinds of 
edges can be detected, \viz edges that go to a vertex that was not yet seen, 
edges that go to a vertex that was seen, but has not yet been processed (\ie, 
vertices in the queue) and edges that go back to a vertex on a previous level. 
This last type of edges is also referred to as a \emph{back-level edge}.
Formally it is an edge $u \to v$ where the level of $u$, say $k_u$ is larger 
than the level of $v$, say $k_v$. The length of a back-level edge $u \to v$ is 
$k_u - k_v$.

Graph algorithms are typically based on a search strategy like BFS or DFS,
given some initial vertex $v_0 \in V$. The characteristics of these search
strategies are therefore likely to affect the performance of such graph 
algorithms.

\paragraph{Distances.}
The \emph{diameter} of a graph is the maximal length of a 
shortest path
between any pair of vertices. The \emph{girth} is the length of 
the shortest cycle in
the graph. Both measures require solving the all-sources-shortest
path problem with unit edge-weights, which is quadratic in the size of
the graph.

For undirected graphs the diameter can be computed more efficiently using the
techniques from Takes and Koster \cite{TK:11}. For directed graphs, however,
no more efficient algorithm is known.

The diameter and the girth characterise global properties of graphs. 
Intuitively,
they describe how hard it is to get from one vertex in the graph to another,
or back to itself.
A girth of $1$ denotes that the graph contains a self-loop. We expect
to see this value quite often when analysing parity games due to the occurrence
of vertices that are trivially won by one of the two players.

\paragraph{Local Structure.}
P\'{e}lanek also studied some local graph properties.
A \emph{diamond}\index{diamond} rooted at a vertex $u$ is a quadruple $(u, v, 
v', w)$ such
that $v \neq v'$, $u \to v$, $u \to v'$, $v \to w$, and $v' \to w$. 
For parity games, we characterise two more specific classes of diamonds.
A diamond $(u, v, v', w)$ is defined to
be \emph{even}\index{diamond!even diamond} if $\getplayer{u} = \getplayer{v} =
\getplayer{v'} = \even$, and \emph{odd}\index{diamond!odd diamond} if 
$\getplayer{u} = \getplayer{v} =
\getplayer{v'} = \odd$. These structures might prove to be interesting in the 
sense
that from vertex $u$, $\getplayer{u}$ has at least two strategies to play to $w$
in two steps. The question is open whether these
kinds of structures can be used to improve parity game solving.

The \emph{$k$-neighbourhood of $v$}\index{neighbourhood} is the set of vertices 
that can be reached
from $v$ in at most $k$ steps (not counting $v$). The \emph{k-clustering
  coefficient of $v$}\index{clustering coefficient} is the ratio of the number 
  of edges and the number of
vertices in the $k$-neighbourhood of $v$. The $k$-neighbourhood can be thought
of as a generalisation of the out-degree, except that we exclude a vertex from
its own neighbourhood.

\paragraph{Width-measures on Graphs.} \label{sec:graph_width}
Width-measures of graphs are based on cops-and-robbers games 
\cite{NW:83,Qui:78}, where different measures are obtained by varying 
the rules of the game. For various measures, specialised algorithms are known 
that can solve games polynomially if their width is bounded. Most of the 
measures have an alternative characterisation using graph decompositions.

The classical width notion for \emph{undirected graphs} is \emph{treewidth} 
\cite{RS:86,Bodl:97}. Intuitively, the treewidth of a graph expresses how
tree-like the graph is---the treewidth of a tree is $1$. This corresponds to 
the idea that some problems are easier to solve for trees, or graphs that are 
almost trees, than for arbitrary graphs. 
For directed graphs, the treewidth is defined as the treewidth of the graph
obtained by forgetting the direction of the edges. The complexity for solving 
parity games is bounded in the treewidth \cite{Obdr:03}; this means that, for 
parity games with a small, constant treewidth, parity game solving is 
polynomial.

Treewidth has been lifted to directed graphs in a number of different ways. For
instance, \emph{Directed treewidth} \cite{JRST:01} is bounded by the
treewidth \cite{Adl:07}. \emph{DAG-width} \cite{BDHKO:12} describes how much a 
graph is like a directed acyclic graph. DAG-width bounds the directed tree width of a 
graph from above, and is at most the treewidth. The \emph{Kelly-width} \cite{HK:08} is yet
another generalitation of treewidth to directed graphs. If the Kelly-width 
of a graph is bounded, then also a bound on its directed 
treewidth can be given, however, classes of directed graphs with bounded 
directed treewidth and unbounded Kelly-width exist.
\emph{Entanglement} \cite{BG:05,BGKR:12} is a graph measure that aims to
express how much the cycles in a graph are intertwined. If an undirected graph 
has bounded treewidth or bounded DAG-with, then it also has bounded 
entanglement. Finally, \emph{clique-width} \cite{CO:00} measures how close a 
graph is to a complete bipartite graph. For every directed graph with bounded 
treewidth an exponential upper bound on its clique-width can be given.
Unlike the other width measures that we discussed clique-width does not have
a characterisation in terms of cops-and-robbers games.

If a parity game is bounded to a constant in any of the measures
introduced above, it can be solved in polynomial time.

\paragraph{Alternation Depth.}
Typically, the complexity of parity game algorithms is expressed in the number
of vertices, the number of edges, and the number of priorities in the game.
If we look at other verification problems, such as $\mu$-calculus model 
checking, or solving Boolean equation systems, the complexity is typically 
expressed in terms of the \emph{alternation depth}. Different versions of 
alternation depth (with varying precision) have been coined, see \cite{BS:00}.
Intuitively, the alternation depth of a formula captures the number of 
alternations between different fixed point symbols.

We aim to develop a characterisation of the `important' alternations between 
priorities in parity games, inspired by the notion of alternation depth. We 
base our definition on the notion of alternation depth for modal equation 
systems as it was defined by Cleaveland \etal \cite{CKS:93}.

Analogous to this definition, the alternation depth that we define comes in two 
stages. First we define the nesting depth of a strongly connected component 
within a parity game, next we define the alternation depth of the parity game 
as the maximum of the nesting depths of its strongly connected components.
\begin{definition}
  Let $G = (V_{\even}, V_{\odd}, \to, \priosym)$ be a parity game, and let
  $\sccs{G}$ be the set of strongly connected components of $G$. Let
  $\mathcal{C} \in \sccs{G}$ be a strongly connected component.
  The nesting depth of $v_i$ in $\mathcal{C}$ is given by
  \begin{align*}
  \nd{v_i, \mathcal{C}} \isdef \max\{ & 1, \\
  & \max\{\nd{v_j, \mathcal{C}} \mid v_j \to^*_{\mathcal{C},\priority(v_i)} 
  v_i, 
  v_j \neq v_i 
  \text{ and }\priority(v_i) \equiv_2 \priority(v_j) \}, \\
  & \max\{\nd{v_j, \mathcal{C}} + 1 \mid v_j \to^*_{\mathcal{C},\priority(v_i)} 
  v_i \text{ and 
  }\priority(v_i) \not \equiv_2 \priority(v_j) \} \\
  \} & 
  \end{align*}
  where $v_j \to_{\mathcal{C}, k} v_i$ if $v_j \to v_i$ is an edge in the SCC 
  $\mathcal{C}$ with
  $\priority(v_j) \leq k$ and $\priority(v_i) \leq k$. Intuitively, the nesting
  depth of a vertex $v$ counts the number of alternations between even and odd
  priorities on paths of descending priorities in the SCC of $v$. Note that this
  is well-defined since we forbid paths between identical nodes.
\end{definition}
The nesting depth of an SCC $\mathcal{C} \in \sccs{G}$ is defined as the
maximum nesting depth of any vertices in $\mathcal{C}$, \ie, 
$\nd{\mathcal{C}} \isdef \max\{ \nd{v, \mathcal{C}} \mid v \in \mathcal{C} \}$. 
The \emph{alternation depth} of a parity game is defined as the maximal
nesting depth of its SCCs.
\begin{definition}
  Let $G = (V_{\even}, V_{\odd}, \to, \priosym)$ be a parity game, and let 
  $\sccs{G}$
  be the set of strongly connected components of $G$. Then the \emph{alternation
    depth}
  of $G$ is defined as $\ad{}{G} \isdef \max\{ \nd{\mathcal{C}} \mid 
  \mathcal{C} 
  \in \sccs{G} 
  \}$.
\end{definition}
There are reasonable translations of the $\mu$-calculus model checking problem
into parity games, such that the alternation depth of the resulting parity
game is at most the fixed point alternation depth of the
$\mu$-calculus formula as described by Emerson and Lei \cite{EL:86},
see \cite[Proposition 5.4]{Kei:13}. Note that the 
alternation depth of a game can be smaller than the number of priorities in the 
game, and could provide an interesting alternative to the number of priorities 
in computing the complexity of parity game algorithms.

\section{Benchmarks} \label{sec:benchmark_games}
For benchmarking parity game algorithms, it makes sense to distinguish three
classes of parity games, (1) the games that are the result of encoding a
problem into parity games, (2) games that represent hard cases for certain
algorithms, and (3) random games. All three classes of games occur in the
literature, and our benchmark set contains games from each of these classes.
In the rest of this section we discuss our benchmarks. In the next section we 
briefly discuss these games with respect to the properties described in 
Section~\ref{sec:structural_properties}.

\subsection{Encodings}
A broad range of verification problems can be encoded as a parity game. The
most prominent examples of these are the $\mu$-calculus model checking
problem---does a model satisfy a given property?---, equivalence checking 
problems---are two models equivalent?---, decision procedures---is a formula 
valid or satis\-fiable?--- and synthesis---given a property, give a model that 
satisfies the property. 

\paragraph{Model Checking.}
The model checking problems we consider are mainly selected from the
literature. All of the systems are encodings that, given a model $L$ of a
system, and a property $\varphi$, encode the model checking problem
$L \models \varphi$, \ie, does $L$ satisfy property $\varphi$. Most sensible
encodings of model checking problems typically lead to a low number of 
priorities, corresponding to the low alternation depths of these properties. We 
verify fairness, liveness and safety properties. This set includes, but is not
limited to, the model checking problems described in 
\cite{Mat:03,PW:08,FL:09atva,CKW:11,CKW:12}.

We take a number of communication protocols from the literature, see, 
\eg, \cite{BSW:69,CK:74,KM:90,GP:96}: two variations of the
\emph{Alternating Bit Protocol} (ABP), the \emph{Concurrent Alternating Bit
  Protocol} (CABP), the \emph{Positive Acknowledgement with Retransmission
  Protocol} (PAR), the \emph{Bounded Retransmission Protocol} (BRP),
the \emph{Onebit} sliding window protocol, and the
\emph{Sliding Window Protocol} (SWP). All protocols are parameterised
with the number of messages that can be sent, and the sliding window
protocol is parameterised by the window size. For these protocols a number of
properties of varying complexity was considered, ranging from alternation
free properties, \eg deadlock freedom, to fairness properties.

A \emph{Cache Coherence Protocol} (CCP) \cite{VHBJB:01} and a \emph{wait-free 
handshake register} (Hesselink) \cite{Hes:98} are considered. For the cache 
coherence protocol we consider a number of properties from \cite{PFHV:07} and 
for the register we consider properties from \cite{Hes:98}. Additionally we 
consider a \emph{leader election protocol} for which we verify whether it 
eventually stabilises.

To obtain parity games with a high degree of alternation between vertices owned
by different players we also consider a number of two-player board games,
\viz \emph{Clobber} \cite{AGNW:05}, \emph{Domineering} \cite{Gar:74},
\emph{Hex}, see \eg \cite{BBC:00,Maa:05}, \emph{Othello}, also known
as reversi, see \eg \cite{Ros:05}, and \emph{Snake}. For these games we check 
for each of the players whether the player has a winning strategy starting from 
the initial configuration of the game. The games are parameterised by their 
board size.

Additionally, we consider a number of industrial model checking problems.
The first is a system for lifting trucks (Lift) \cite{GPW:03}, of which we
consider both a correct and an incorrect version. We verify the liveness and
safety properties described in \cite{GPW:03}. For the \emph{IEEE 1394 Link
Layer Protocol} (1394) we verify the properties from \cite{Lut:97fmsd}. We
translated the ACTL properties from \cite{SM:98} to the $\mu$-calculus.

Finally, we check the \emph{Elevator} described by Friedmann and Lange,
in a version in which requests are treated on a first-in-first-out basis (FIFO),
and on a last-in-first-out basis (LIFO). We then check whether, globally, if
the lift is requested on the top floor, then it is eventually served. This
holds for the FIFO version, but does not hold for the LIFO version of the model.
The elevator model is parameterised by the strategy and the number of floors.
Furthermore we consider the parity games generated using an encoding of an LTS
with a $\mu$-calculus formula, as well as the direct encoding presented in
\cite{FL:09atva}. In a similar way we consider the Hanoi towers from
\cite{FL:09atva} as well as our own version of this problem.

\paragraph{Equivalence Checking.}
Given two processes $L_1, L_2$, the problem whether $L_1 \equiv L_2$, for 
relations
$\equiv$, denoting that $L_1$ and $L_2$ are equivalent under some process
equivalence, can be encoded as a parity game \cite{Lar:93,VL:94}.
We consider strong bisimulation,
weak bisimulation, branching bisimulation and branching simulation equivalence
in our benchmarks, using the approach described in \cite{CPPW:07}. The number
of different priorities in these parity games is limited to 2, but they do 
include alternations between vertices owned by different players.

Here we again use the specifications of the communication protocols that we 
also used
for model checking, \ie, two ABP versions, CABP, PAR, Onebit and SWP. In 
addition
we include a model of a buffer. We vary the capacity of the buffer, the
number of messages that can be transmitted, and the window size in the
sliding window protocol. We compare each pair of protocols using all four 
equivalences, resulting in both positive and negative cases. These cases are a 
superset
of the ones described in \cite{CKW:11,CKW:12}.

In addition, we include a comparison of the implementation of the wait-free
handshake register with a possible specification. The implementation is trace
equivalent to the specification, but it is not equivalent with respect to the
equivalences that we consider here.

\paragraph{Decision Procedures.}
Parity games can also be obtained from decision procedures for temporal logics
such as $\LTL$, $\CTL$, $\CTLs$, $\PDL$ and the \mucalc. %Classical approaches 
%rely on testing non-emptiness of a tree automaton, see, \eg, \cite{EJ:00}.
Friedmann \etal presented a decision procedure that is based on a
combination of infinite tableaux in which the existence of a tableau is coded
as a parity game \cite{FLL:10}. %The priorities in the parity game originate 
%from
%a deterministic parity automaton that is the result of complementing a
%non-deterministic B\"{u}chi automaton. In \ibid, the authors also present a 
%tool that, 
For a given formula, it is checked whether it is (1) \emph{valid}, \ie, 
whether the formula holds in all models, or (2) \emph{satisfiable}, \ie, 
whether the formula is satisfiable in some model.

Our benchmark set includes a number of scalable satisfiability and validity
problems that are provided as examples for the MLSolver tool \cite{FL:10entcs}.
In particular, we include the benchmarks used in \cite{FL:10entcs}: encoding 
that a deterministic parity condition is expressible as a nondeterministic 
B\"uchi condition, and nesting Kleene stars in different logics. Additionally we
consider formulas that involve encodings of a binary counter in various logics.

\paragraph{Synthesis.}
Another problem that involves solving parity games is the LTL synthesis problem.
Traditional synthesis approaches convert a formula into a non-deterministic
B\"uchi automaton, which is, in turn, transformed into a deterministic parity
automaton using Safra's construction \cite{Saf:88}. Emptiness of this
deterministic parity automaton can then be checked using parity games with 
three priorities. Synthesis tools have been implemented that employ parity 
games internally, most notably GOAL \cite{TCTCL:08} and Gist \cite{CHJR:10}.
All synthesis tools that we are aware of, however, are research quality tools, 
of which we have not been able to obtain working versions on current computing 
platforms. As a result, our benchmark set currently does not include parity 
games obtained from the synthesis problem. We plan to extend our benchmarks
with such games, and update this paper accordingly.

\subsection{Hard Games}
The interesting complexity of solving parity games, and its link to the model
checking problem, have led to the conception of a large number of parity game
solving algorithms. For most of these algorithms it has long been an open 
problem whether they have exponential lower bounds.

We consider the games described by Jurdzi\'{n}ski that shows
the exponential lower bound for small progress measures \cite{Jur:00},
the \emph{ladder games} described by Friedmann \cite{Frie:11lmcs} defeating 
strategy guessing heuristics, \emph{recursive ladder games} that give a lower 
bound for the recursive algorithms, and \emph{model checker ladder games} 
\cite{Frie:10} for which the algorithm by Stevens and Stirling \cite{SS:98} 
behaves exponentially.

\subsection{Random Games}
The final class of games that is typically used in publications that empirically
evaluate the performance of algorithms on parity games are random parity games
\cite{BV:01,Sche:08thesis,Sche:08csl,Lan:05,FL:09atva}. We study three classes 
of random games.
We expect that the structural properties of random games are, typically, 
different from parity games obtained in the previous classes. This class is, 
therefore, unlikely to give insights in the performance of parity game 
algorithms on practical problems.

\section{Implementation} \label{sec:implementation}
All games were generated on a 1TB main memory,
56-core Linux machine, where each core was running at 2.27GHz. Executions of
tools generating and solving parity games, and tools collecting statistics 
about parity games, were limited to running times of 1 hour and their memory 
usage was limited to 32GB.

To systematically generate the benchmarks, we have implemented tooling that 
allows the parallel execution of individual cases. Here a case is either 
generating or solving a game, or collecting a single measure. Each individual 
case only uses a single core.
The tools are implemented in an extensible way, \ie, additional parity games,
additional encodings, as well as additional measures can be added
straightforwardly. The tools are available for download from
\url{https://github.com/jkeiren/paritygame-generator}.

\subsection{Generating Parity Games}
For the generation of our benchmarks we rely on a number of external tools:
version 3.3 of PGSolver \cite{FL:10pgsolver} for generating random
games, and games that prove to be hard for certain algorithms; version 1.2 of
MLSolver to generate the games for satisfiability and validity problems
\cite{FL:10entcs}; and revision 11703 of the mCRL2 toolset \cite{CGK+:13} for 
the model checking and equivalence checking problems.
For all games we have collected the information described in
Section~\ref{sec:structural_properties} to the extent in which this is
feasible.

\subsection{Collecting Statistics}
We developed the tool \texttt{pginfo} for collecting structural information
from parity games. The tool is available from 
\url{https://github.com/jkeiren/pginfo} and accepts parity games in the file 
format used by PGSolver. The tool reads a parity game, and writes statistics to a file
in a structured way.

The implementation is built on top of the Boost Graph library \cite{SLL:02},
which provides data structures and basic algorithms for manipulating graphs.
Computing the exact value for the width-measures is problematic: it is known to 
be NP-complete \cite{ACP:87}. Approximation algorithms are known that compute 
upper- and lower bound for these measures; especially for treewidth
these have been thoroughly studied \cite{BK:10,BK:11}. 
To determine feasibility of computing width-measures for our benchmarks we have 
implemented three approximation algorithms. For computing upper and lower 
bounds on treewidth we implemented the greedy degree algorithm \cite{BK:10} and 
the minor min-width algorithm \cite{GD:04}, respectively. For computing an 
upper bound of the Kelly-width we implemented the elimination ordering 
described in \cite{HK:08}. Even these approximation algorithms have proven to be
impractical due to their complexity. Computing (bounds) on the other width
measures is equally complex.

\subsection{Availability of Parity Games}
All parity games that are described in this paper are available for download
from \url{http://www.github.com/jkeiren/paritygame-generator} in bzip2
compressed PGSolver format \cite{FL:10pgsolver}. The dataset
is approximately 10GB in size, and includes the structural information that was 
collected from these games.

\section{Analysis of Benchmarks} \label{sec:games_properties}
We have presented benchmarks originating from different problems. Next we 
analyse them with respect to the measures described in 
Section~\ref{sec:structural_properties}. This analysis illustrates that our 
benchmarks exhibit a wide variety of properties. Furthermore, this gives us 
some insights in the characteristics of typical parity games.
For each of the statistics, we only consider games for which that specific 
statistic could be computed within an hour, and we only include those 
statistics that can feasibly be computed for the majority of games, as a consequence
the width measure are excluded from the analysis we present here.
We used this selection to avoid timeouts for computing the measures that are
expensive to compute, such as the diameter and the girth.
All graphs in this section are labelled by their class.
Note that the satisfiability and validity problems are labelled by ``mlsolver''
and the games that are hard for some solving algorithms are labelled by
``specialcases''. The full data presented in this chapter is also available
from \url{http://www.github.com/jkeiren/paritygame-generator}.

\begin{figure}[ht]
  \centering
  \subfloat[\label{fig:vertices_vs_edges}]{
  \input{vertices_vs_edges_orig}
  }
  %\quad
  \subfloat[\label{fig:size_vs_girth_orig}]{
  \input{size_vs_girth_orig}
  }
  \caption{Relation between number of vertices and (a)
  number of edges, and (b) girth.}
  %\label{fig:vertices_vs_edges}
\end{figure}
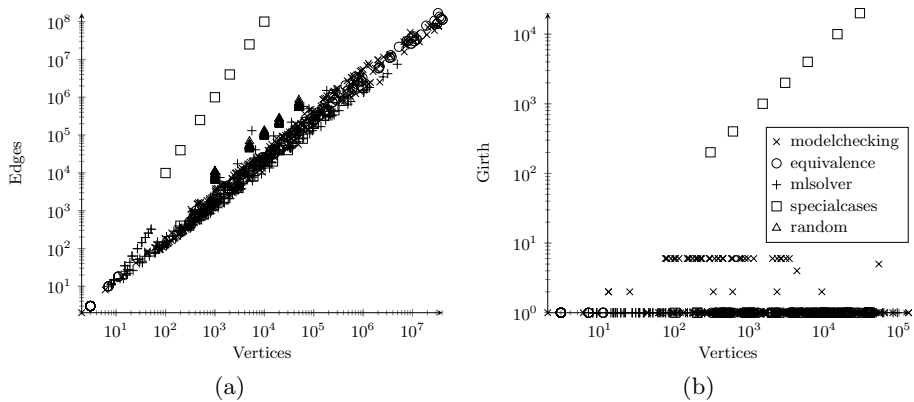

Our data set contains 1037 parity games
%SELECT count(*) from query_gamesizes where reduction = "orig";
that range from 2 vertices to 40 million vertices,
and on average they have about 95,000 vertices. The number of edges ranges from
2 to 167 million, with an average of about 3.1 million.
%SELECT min(vertices), max(vertices), avg(vertices), min(edges), max(edges), 
%avg(edges) from query_gamesizes where reduction = "orig";
The 59 parity games are games in which all vertices are owned by a single player,
the so-called solitaire games \cite{BG:04}, the rest are parity
games in which both players own non-empty sets of vertices.
% select count(*)
% from gamesizes, games
% where gamesizes.id = games.id
%    and games.reduction = "orig"
%    and (gamesizes.even_vertices = 0
%      or gamesizes.odd_vertices = 0)
The parity games that we consider have differing degrees. There are instances
in which the average degree is 1,
the average degree is maximally 9999, but it is typically below 10.
The ratio between the number of vertices
and the number of edges is, therefore, relatively small in general.
This can also be observed from
Figure~\ref{fig:vertices_vs_edges}, which displays the correlation between
the two. The games in which these numbers coincide are on the line $x = y$,
the other games lie around this line due to the log scale that we use.
Our parity games generally contain a vertex with in-degree 0, which is the 
starting
vertex. Most of the games contain
vertices with a high in-degree---typically representing vertices that are 
trivially
won by either of the players---, and vertices with a high out-degree.

% SELECT min(gamesizes.vertices) 'min_vertices', max(gamesizes.vertices) 
%'max_vertices', avg(gamesizes.vertices) 'avg_vertices',
%        min(gamesizes.edges) 'min_edges', max(gamesizes.edges) 'max_edges', 
%avg(gamesizes.edges) 'avg_edges',
%        min((gamesizes.edges)/(gamesizes.vertices)) 'min_ratio', 
%max((gamesizes.edges)/(gamesizes.vertices)) 'max_ratio', 
%avg((gamesizes.edges)/(gamesizes.vertices)) 'avg_ratio'
% FROM gamesizes, games
% WHERE gamesizes.id = games.id
%   AND reduction = "orig"
%   AND gamesizes.vertices is not NULL

In general, the
SCC quotient height ranges up to 513 for the parity games that we consider with
an average of around 14. The
number of non-trivial SCCs can grow large, up to 1.4 million for our games.
% SELECT min(scc_quotient_height), max(scc_quotient_height), 
%avg(scc_quotient_height), min(nontrivial_sccs), max(nontrivial_sccs), 
%avg(nontrivial_sccs) from query_gamesizes

The diameter and girth have been computed only for smaller parity games, and
the data we present for them, therefore, considers a subset of the parity games
only.
We expect that typical parity games contain self-loops, which leads to a
small girth---the girth is 1 if the game contains a self-loop.
This is confirmed by the data in Figure~\ref{fig:size_vs_girth_orig}. Note that 
the
girth is large for some of the hard cases that we consider. A closer
investigation shows that this is solely due to the model checker ladder games
\cite{Frie:10}. %\todo{uitleg?}

\begin{figure}[ht]
  \centering
  \subfloat[\label{fig:size_vs_diameter}]{
    \input{size_vs_diameter_orig}
  }
  \subfloat[\label{fig:bfs_levels_vs_diameter}]
  {
    \input{bfs_levels_vs_diameter_orig}
  }
  \caption{Relation between (a) number of vertices and diameter, and (b) BFS height and diameter.}
\end{figure}
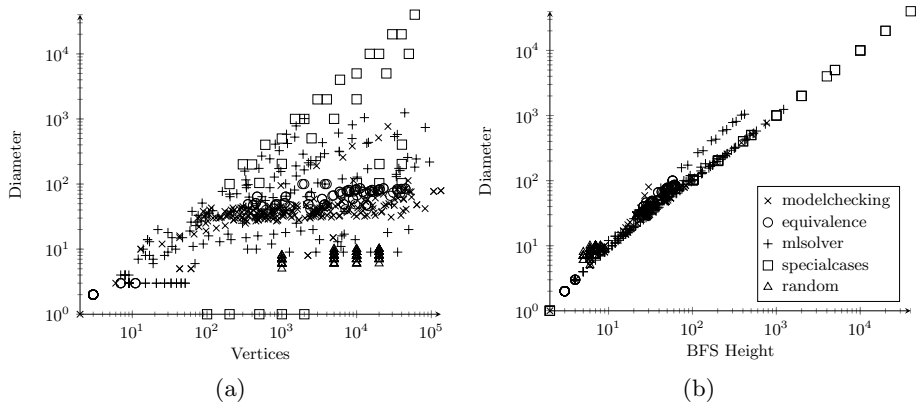

The diameters of the parity games, \ie, the maximal length of any shortest path
in the game, are nicely distributed over the sizes of the game.
Figure~\ref{fig:size_vs_diameter} shows that for every size of game we have
parity games of a large range of different diameters. For the hard
cases the diameter is, generally, large due to variations of 
ladder
games. Generally, the diameter for satisfiability and validity problems is
larger than the diameter for model checking problems.  For random games the
diameter is typically small.

Figure~\ref{fig:bfs_levels_vs_diameter} indicates that the diameter and the
number of BFS levels are correlated, the number of BFS levels is therefore
likely to be a good approximation of the diameter, also for larger instances.
Note that this corresponds to a similar observation made by P\'{e}lanek, who
stated that typically the diameter is smaller than 1.5 times the number of
BFS levels for state spaces \cite{Pel:04}.  

Of the parity games that we consider, 882 contain diamonds. Of these,
494 contain even diamonds, and 656 contain odd diamonds, 382 contain both.
This indicates that
it is worth investigating techniques, such as confluence reduction,
that use these diamonds to either simplify parity games or speed up solving.
In general, the number of diamonds is independent of the number of vertices in
the game.

\begin{figure}[ht]
  \centering
  \subfloat[\label{fig:neighbourhoods_orig}]{
    \input{vertices_vs_avg_3_neighbourhood_orig}
  }
  \subfloat[\label{fig:size_vs_ad}]
  {
    \input{size_vs_ad_orig}
  }
  \caption{Number of vertices in relation to (a) size of 3-neighbourhoods, and (b) alternation depth.}
\end{figure}
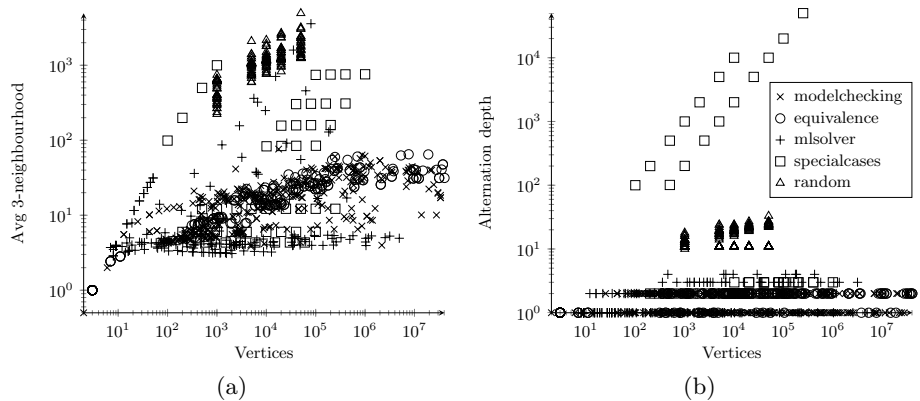

The average $3$-neighbourhoods range from
3 to 5000 across the sizes of the games, as can be seen from
Figure~\ref{fig:neighbourhoods_orig}. Note that the average size of the
3-neighbourhoods is typically high for random games, and limited to 100 for
most other classes of games.

% select count(*)
% from gamesizes, games
% where gamesizes.id = games.id
%    and games.reduction = "orig"
%    and gamesizes.odd_diamonds > 0

We have included parity games with alternation
depths up to 50,000 as shown in Figure~\ref{fig:size_vs_ad}. Observe that the
games for model checking and equivalence checking included in our benchmarks 
all have alternation depth
at most 2. Model checking problems could be formulated that have a
higher alternation depth---up to arbitrary numbers---however, in practice
properties have limited alternation depth because they become too hard to
understand otherwise. The satisfiability and validity properties
have alternation depths between 1 and 4. The alternation
depths of the random games are between 10 and 15. All parity games with more
than 50 priorities represent special cases. Closer investigation shows that
these special cases are the clique games and recursive ladder games.

To summarise, we have presented a large set of parity games. For a selection
of the structural properties introduced in Section~\ref{sec:structural_properties}
we have shown that the games cover a large range of values. 
Also observe that, for parity game specific
properties such as alternation depth, higher values are only available for
smaller games due to generation times. Unsurprisingly, the random games 
considered in this paper are not structurally similar to
parity games that represent encodings of verification problems.

\section{Closing Remarks}\label{sec:benchmarks:conclusions}
No standard benchmarks for parity game algorithms existed. As a consequence, it
was virtually impossible to make a good comparison between algorithms and 
applications described in the literature.
In this paper we have addressed this issue by presenting a comprehensive set of 
parity game benchmarks.
Our benchmarks include the games that appear in the literature, and provides 
a first step towards standardising experimental evaluation of parity 
game algorithms. All games have been generated in an extensible way, and are 
available on-line.

We also presented a set of structural properties for parity games, and analysed 
our benchmarks with respect to these properties. Of particular interest is a 
new notion of alternation depth for parity games, that 
is always at most the number of priorities in a parity game, and that is bounded
also by the alternation depth of $\mu$-calculus formulae given
a reasonable translation of the model checking problem.

\paragraph{Future work.}
Some of the structural properties, such as treewidth, cannot be computed for 
all games in the benchmark suite due to their complexity. An interesting 
algorithmic question is, therefore, whether algorithms or heuristics can be 
devised that can compute or approximate these measures for large graphs.

Additionally, we have presented a selection of structural properties in this
paper. One can wonder whether there are other structural properties of parity 
games that are relevant  to the practical performance of parity game 
algorithms. The question whether the 
theoretical complexity of existing parity game algorithms can be made tighter 
using structural properties, such as our notion of alternation depth.

We believe our work also paves the way for a full-scale comparison of parity 
game algorithms and the effect of heuristics in the spirit of \cite{FL:09atva},
including the comparison of alternative implementations of algorithms 
\cite{CGK+:13,di_stasio_solving_2014}. Here also the impact of the structural
properties on the performance of implementations should be studied, since we
have only scratched the surface of this aspect in this paper.

Finally, we welcome the addition of problems and properties to our benchmark 
suite to establish and maintain a corpus for experimentation
with parity game algorithms. In particular parity games with a large number of
priorities and a high alternation depth stemming from encodings of, \eg,
verification and synthesis problems form a welcome addition.

\subsubsection*{Acknowledgements}
For generating the parity games described in the paper, a large number
of tools have been used. The author would like to thank the developers
of, in particular, Gist, GOAL, mCRL2, MLSolver and PGSolver. Thanks also
go to Wan Fokkink and Tim Willemse for helpful feedback on earlier versions
of this paper, and remarks by anonymous reviewers that led to usability
improvements of the benchmarks and tools presented.

\bibliographystyle{plain}
\bibliography{references}

\end{document}

%% file: macros.tex
\newcommand{\ie}{\textit{i.e.}\xspace }
\newcommand{\eg}{\textit{e.g.}\xspace }
\newcommand{\etal}{\textit{et al.}\xspace }

\newcommand{\viz}{\textit{viz.}\xspace }

\makeatletter
\def\slashedarrowfill@#1#2#3#4#5{%
  $\m@th\thickmuskip0mu\medmuskip\thickmuskip\thinmuskip\thickmuskip
   \relax#5#1\mkern-7mu%
   \cleaders\hbox{$#5\mkern-2mu#2\mkern-2mu$}\hfill
   \mathclap{#3}\mathclap{#2}%
   \cleaders\hbox{$#5\mkern-2mu#2\mkern-2mu$}\hfill
   \mkern-7mu#4$%
} 
\def\rightslashedarrowfill@{%
  \slashedarrowfill@\relbar\relbar{\raisebox{.12em}{\tiny/}}\rightarrow}
\newcommand\xslashedrightarrow[2][]{%
  \ext@arrow 0055{\rightslashedarrowfill@}{#1}{#2}}
  
\makeatletter
\renewcommand{\bigsqcup}{%
  \mathop{%
    \mathpalette\@updown\bigsqcap
  }%
}
\newcommand*{\@updown}[2]{%
  \rotatebox[origin=c]{180}{$\m@th#1#2$}%
}
\makeatother

\newcommand{\isdef}{\buildrel\mbox{\tiny $\Delta$}\over=}

\newcommand{\cardinality}[1]{\left\vert{#1}\right\vert} % Cardinality of a set
\renewcommand{\lor}{\mathrel{\vee}}

\newcommand{\LTL}{\mathsf{LTL}\xspace}
\newcommand{\mucalc}{$\mu$-calculus\xspace}
\newcommand{\CTL}{\mathsf{CTL}}
\newcommand{\CTLs}{\CTL^{*}}
\newcommand{\PDL}{\mathsf{PDL}}

\newcommand{\ndname}{\mathsf{nd}}
\newcommand{\nd}[1]{\ndname(#1)}
\newcommand{\adname}{\mathsf{ad}}
\newcommand{\ad}[2]{\adname^{\mathsf{#1}}(#2)}

\usepackage{centernot}

\newlength{\simw}
\newcommand{\indegreename}{\mathsf{indeg}}
\newcommand{\indegree}[1]{\indegreename(#1)}
\newcommand{\outdegreename}{\mathsf{outdeg}}
\newcommand{\outdegree}[1]{\outdegreename(#1)}
\newcommand{\degreename}{\mathsf{deg}}
\newcommand{\degree}[1]{\degreename(#1)}
\newcommand{\sccsname}{\mathsf{sccs}}
\newcommand{\sccs}[1]{\sccsname(#1)}

\tikzstyle{even}=[draw,shape=diamond,inner sep=1pt, minimum size=8pt]
\tikzstyle{odd}=[draw,shape=rectangle,inner sep=1pt, minimum size=8pt]  

\newcommand{\priosym}{\Omega}
\newcommand{\prio}[1]{\priosym(#1)}
\newcommand{\inverseprio}[1]{\priosym^{-1}(#1)}
\newcommand{\player}{\bigcirc\xspace}
\newcommand{\priority}{\ensuremath{\priosym}\xspace}
\newcommand{\getplayername}{\ensuremath{\mathcal{P}}}
\newcommand{\getplayer}[1]{\getplayername(#1)}

\newcommand{\even}{\scalebox{0.7071}{\rotatebox{45}{$\Box$}}}
\newcommand{\odd}{\Box}
 
\newcounter{soscounter}

%% file: vertices_vs_edges_orig.tex
  \begin{tikzpicture}[mark size=2.5pt,remember picture,scale=0.7]
    \begin{axis}[axis x line=bottom,
                       axis y line=left,
                       xlabel={Vertices},
                       ylabel={Edges},
                       xmode={log},
                       ymode={log}
                       ,
                       xmin=0,
                       ymin=0,
                       scatter/classes={
                         modelchecking={mark=x},
                         equivalence={mark=o},
                         mlsolver={mark=+},
                         specialcases={mark=square},
                         random={mark=triangle}
                       },
                       legend pos=outer north east,
                       legend image post style={xshift=0.5cm},
                       legend cell align=left,
                       legend style={
                         nodes={right}
                       }]
      \addplot+[black,only marks,mark=x,scatter,scatter src=explicit symbolic] 
      table[col sep=comma,x=x, y=y, meta=cluster] { 
        x, y, cluster
        2, 2, modelchecking
        972, 1698, modelchecking
        2688, 4544, modelchecking
        15684, 26354, modelchecking
        588, 1096, modelchecking
        111456, 287964, modelchecking
        656, 1349, modelchecking
        308, 511, modelchecking
        227, 360, modelchecking
        328, 549, modelchecking
        649, 1134, modelchecking
        656, 1349, modelchecking
        649, 1134, modelchecking
        227, 360, modelchecking
        8626, 21275, modelchecking
        2126, 4462, modelchecking
        4313, 9919, modelchecking
        4021, 9267, modelchecking
        6359, 13357, modelchecking
        6359, 13357, modelchecking
        8626, 21275, modelchecking
        2126, 4462, modelchecking
        2832, 5924, modelchecking
        16356, 38194, modelchecking
        2916, 5100, modelchecking
        8748, 15306, modelchecking
        78732, 137778, modelchecking
        564, 950, modelchecking
        876780, 2484252, modelchecking
        2338, 3253, modelchecking
        147, 185, modelchecking
        260, 322, modelchecking
        747, 945, modelchecking
        442, 557, modelchecking
        731, 921, modelchecking
        439, 661, modelchecking
        598, 753, modelchecking
        173, 211, modelchecking
        150, 188, modelchecking
        236, 298, modelchecking
        9282, 12969, modelchecking
        294, 372, modelchecking
        341, 419, modelchecking
        2643, 3361, modelchecking
        291, 369, modelchecking
        516, 642, modelchecking
        2034, 2585, modelchecking
        871, 1317, modelchecking
        2346, 2977, modelchecking
        2611, 3313, modelchecking
        468, 594, modelchecking
        594, 819, modelchecking
        78, 96, modelchecking
        156, 193, modelchecking
        132, 162, modelchecking
        75, 93, modelchecking
        231, 289, modelchecking
        89, 107, modelchecking
        223, 277, modelchecking
        78, 95, modelchecking
        223, 333, modelchecking
        120, 150, modelchecking
        63907, 240612, modelchecking
        31954, 121625, modelchecking
        63907, 240612, modelchecking
        13222, 32797, modelchecking
        13222, 32797, modelchecking
        11977, 39577, modelchecking
        11977, 39577, modelchecking
        26996, 99348, modelchecking
        236196, 413340, modelchecking
        108336, 180898, modelchecking
        26244, 45924, modelchecking
        2946, 10597, modelchecking
        5649, 19809, modelchecking
        1041, 3809, modelchecking
        5201, 19041, modelchecking
        3121, 9697, modelchecking
        4486, 15561, modelchecking
        1122, 3890, modelchecking
        1619, 4387, modelchecking
        16642, 44933, modelchecking
        708588, 1240026, modelchecking
        107276, 298842, modelchecking
        18625, 47653, modelchecking
        18625, 47653, modelchecking
        44773, 107216, modelchecking
        53638, 154793, modelchecking
        51220, 148411, modelchecking
        44773, 107216, modelchecking
        107276, 298842, modelchecking
        1028, 3365, modelchecking
        514, 1682, modelchecking
        691, 1859, modelchecking
        3714, 9699, modelchecking
        465, 1633, modelchecking
        1393, 4897, modelchecking
        1553, 5153, modelchecking
        450, 1523, modelchecking
        1393, 4193, modelchecking
        12130, 17081, modelchecking
        2658, 3425, modelchecking
        3452, 4457, modelchecking
        3412, 4393, modelchecking
        380, 489, modelchecking
        3066, 3945, modelchecking
        438, 547, modelchecking
        1138, 1735, modelchecking
        384, 493, modelchecking
        564914, 2185853, modelchecking
        564914, 2185853, modelchecking
        4860, 6922, modelchecking
        4860, 6922, modelchecking
        2229, 7060, modelchecking
        2229, 7060, modelchecking
        5861, 14892, modelchecking
        2245634, 6978309, modelchecking
        512018, 2504114, modelchecking
        653574, 2444297, modelchecking
        163394, 611074, modelchecking
        858114, 3433490, modelchecking
        388418, 1125698, modelchecking
        140353, 588033, modelchecking
        211955, 876954, modelchecking
        869569, 3200129, modelchecking
        353089, 800769, modelchecking
        421057, 1456769, modelchecking
        147458, 406665, modelchecking
        4898, 17378, modelchecking
        41762, 154498, modelchecking
        39178, 139025, modelchecking
        12818, 49890, modelchecking
        13825, 43393, modelchecking
        28964, 106938, modelchecking
        10369, 22849, modelchecking
        45937, 160561, modelchecking
        4609, 17089, modelchecking
        9986, 26994, modelchecking
        21506, 58405, modelchecking
        5958, 20041, modelchecking
        6866, 24674, modelchecking
        1490, 5010, modelchecking
        3344, 11922, modelchecking
        3530, 13266, modelchecking
        4033, 12417, modelchecking
        2977, 6497, modelchecking
        2690, 7322, modelchecking
        1345, 4865, modelchecking
        7897, 26073, modelchecking
        196, 249, modelchecking
        222, 275, modelchecking
        3058, 4285, modelchecking
        192, 245, modelchecking
        782, 997, modelchecking
        578, 737, modelchecking
        956, 1221, modelchecking
        976, 1253, modelchecking
        574, 871, modelchecking
        112514, 345315, modelchecking
        3949, 15256, modelchecking
        17810, 60770, modelchecking
        35620, 121541, modelchecking
        54322, 203914, modelchecking
        14065, 57025, modelchecking
        32402, 95314, modelchecking
        42618, 196506, modelchecking
        42193, 142177, modelchecking
        34673, 77633, modelchecking
        56241, 192609, modelchecking
        3458, 9195, modelchecking
        1046, 3714, modelchecking
        1621, 5005, modelchecking
        433, 1513, modelchecking
        326, 1086, modelchecking
        1370, 4714, modelchecking
        770, 2118, modelchecking
        1012, 3173, modelchecking
        506, 1586, modelchecking
        937, 2017, modelchecking
        1297, 3889, modelchecking
        11362, 16249, modelchecking
        3188, 4193, modelchecking
        3228, 4249, modelchecking
        356, 473, modelchecking
        2490, 3313, modelchecking
        1066, 1655, modelchecking
        2874, 3817, modelchecking
        360, 477, modelchecking
        414, 531, modelchecking
        6563, 19681, modelchecking
        19685, 59047, modelchecking
        730, 1027, modelchecking
        282, 367, modelchecking
        274, 419, modelchecking
        96, 121, modelchecking
        92, 119, modelchecking
        192, 247, modelchecking
        96, 123, modelchecking
        108, 135, modelchecking
        272, 353, modelchecking
        43, 85, modelchecking
        13, 18, modelchecking
        175, 342, modelchecking
        97, 186, modelchecking
        133, 258, modelchecking
        67, 126, modelchecking
        82434, 231945, modelchecking
        2577, 9889, modelchecking
        24593, 91553, modelchecking
        17154, 65225, modelchecking
        21770, 80273, modelchecking
        23185, 88993, modelchecking
        7729, 24929, modelchecking
        2722, 10034, modelchecking
        4243, 11555, modelchecking
        43, 78, modelchecking
        2866, 4077, modelchecking
        734, 965, modelchecking
        542, 713, modelchecking
        184, 241, modelchecking
        538, 831, modelchecking
        210, 267, modelchecking
        180, 237, modelchecking
        892, 1169, modelchecking
        912, 1197, modelchecking
        17773, 81348, modelchecking
        59051, 177145, modelchecking
        2395, 8767, modelchecking
        343, 957, modelchecking
        177149, 531439, modelchecking
        926213, 5127415, modelchecking
        3471554, 13612756, modelchecking
        607753, 2008153, modelchecking
        867889, 4933009, modelchecking
        308738, 1709138, modelchecking
        638065, 3680785, modelchecking
        579745, 3354841, modelchecking
        289297, 1689697, modelchecking
        1350433, 7068673, modelchecking
        1191962, 6907934, modelchecking
        1278433, 7843609, modelchecking
        474914, 2739604, modelchecking
        579745, 3354841, modelchecking
        25, 42, modelchecking
        13, 18, modelchecking
        655362, 2541187, modelchecking
        170753, 556993, modelchecking
        297089, 1483457, modelchecking
        177668, 950149, modelchecking
        81921, 468161, modelchecking
        88834, 475074, modelchecking
        245761, 1376833, modelchecking
        153985, 870977, modelchecking
        258050, 1456802, modelchecking
        164353, 928289, modelchecking
        153985, 870977, modelchecking
        65506, 365971, modelchecking
        185089, 1094241, modelchecking
        540737, 1115713, modelchecking
        1081474, 2231426, modelchecking
        1093761, 2246401, modelchecking
        1787138, 5852835, modelchecking
        22873, 94230, modelchecking
        742970, 3925130, modelchecking
        223393, 964513, modelchecking
        294434, 1035554, modelchecking
        588868, 2071109, modelchecking
        944090, 3685946, modelchecking
        586658, 1782434, modelchecking
        576385, 1317505, modelchecking
        670177, 2375809, modelchecking
        917713, 3283153, modelchecking
        1559, 2180, mlsolver
        636, 1156, mlsolver
        298, 556, mlsolver
        693, 940, mlsolver
        2858, 4040, mlsolver
        1143, 2050, mlsolver
        1126, 1560, mlsolver
        467, 858, mlsolver
        12, 16, mlsolver
        22, 26, mlsolver
        974, 1752, mlsolver
        2425, 3420, mlsolver
        805, 1454, mlsolver
        1992, 2800, mlsolver
        1312, 2348, mlsolver
        3291, 4660, mlsolver
        6817, 15208, mlsolver
        14053, 18499, mlsolver
        1562, 3401, mlsolver
        3029, 4035, mlsolver
        3724, 5280, mlsolver
        1481, 2646, mlsolver
        336, 394, mlsolver
        15, 35, mlsolver
        191, 392, mlsolver
        365, 479, mlsolver
        76, 95, mlsolver
        9, 15, mlsolver
        912, 1029, mlsolver
        21, 63, mlsolver
        45, 255, mlsolver
        8776, 9329, mlsolver
        5736, 6150, mlsolver
        39, 195, mlsolver
        27, 99, mlsolver
        1924, 2120, mlsolver
        3492, 3787, mlsolver
        33, 143, mlsolver
        9, 15, mlsolver
        319, 344, mlsolver
        160, 179, mlsolver
        9, 15, mlsolver
        1156, 1199, mlsolver
        9, 15, mlsolver
        817, 854, mlsolver
        9, 15, mlsolver
        61, 74, mlsolver
        9, 15, mlsolver
        538, 569, mlsolver
        9, 15, mlsolver
        29335, 66905, mlsolver
        62707, 82920, mlsolver
        30, 42, mlsolver
        9, 15, mlsolver
        51, 323, mlsolver
        12732, 13444, mlsolver
        2014, 2069, mlsolver
        9, 15, mlsolver
        140, 210, mlsolver
        21, 63, mlsolver
        9, 15, mlsolver
        1555, 1604, mlsolver
        15, 35, mlsolver
        75, 110, mlsolver
        33, 143, mlsolver
        330, 506, mlsolver
        200, 224, mlsolver
        47, 63, mlsolver
        455, 702, mlsolver
        39, 195, mlsolver
        45, 255, mlsolver
        600, 930, mlsolver
        765, 1190, mlsolver
        51, 323, mlsolver
        105, 118, mlsolver
        34, 46, mlsolver
        63, 80, mlsolver
        19, 27, mlsolver
        1490, 1664, mlsolver
        119, 159, mlsolver
        1190, 1329, mlsolver
        107, 143, mlsolver
        924, 1032, mlsolver
        95, 127, mlsolver
        83, 111, mlsolver
        692, 773, mlsolver
        33969, 44709, mlsolver
        6182, 24833, mlsolver
        456, 1027, mlsolver
        2497, 3341, mlsolver
        6431, 8516, mlsolver
        730, 1709, mlsolver
        8, 11, mlsolver
        16, 20, mlsolver
        543, 717, mlsolver
        69, 149, mlsolver
        348, 933, mlsolver
        3772, 5112, mlsolver
        6346, 15349, mlsolver
        60011, 79752, mlsolver
        38075, 51593, mlsolver
        3195, 9044, mlsolver
        110328, 251339, mlsolver
        259455, 339108, mlsolver
        1264, 7531, mlsolver
        36735, 51267, mlsolver
        25971, 75480, mlsolver
        325637, 441635, mlsolver
        49830, 122701, mlsolver
        492107, 655428, mlsolver
        2861, 17366, mlsolver
        94631, 130441, mlsolver
        19, 43, mlsolver
        354, 430, mlsolver
        8, 11, mlsolver
        16, 20, mlsolver
        6831, 19792, mlsolver
        33977, 48764, mlsolver
        9585, 101651, mlsolver
        174667, 248994, mlsolver
        6650, 40753, mlsolver
        234577, 319754, mlsolver
        970, 1371, mlsolver
        65, 122, mlsolver
        334, 458, mlsolver
        30, 53, mlsolver
        7, 10, mlsolver
        14, 16, mlsolver
        2443, 3482, mlsolver
        148, 287, mlsolver
        332, 654, mlsolver
        5837, 8343, mlsolver
        778, 1545, mlsolver
        13687, 19598, mlsolver
        89, 147, mlsolver
        2265, 2921, mlsolver
        8, 12, mlsolver
        34, 42, mlsolver
        17, 28, mlsolver
        250, 316, mlsolver
        203, 295, mlsolver
        27, 51, mlsolver
        32233, 46339, mlsolver
        1794, 3576, mlsolver
        17, 22, mlsolver
        9, 15, mlsolver
        38, 62, mlsolver
        795, 1016, mlsolver
        35037, 45534, mlsolver
        1123, 1891, mlsolver
        478, 801, mlsolver
        14399, 18701, mlsolver
        5914, 7665, mlsolver
        209, 349, mlsolver
        2611, 4394, mlsolver
        82968, 107716, mlsolver
        105, 207, mlsolver
        1546, 2302, mlsolver
        15427, 95008, mlsolver
        567753, 766649, mlsolver
        5596, 130825, mlsolver
        140309, 167007, mlsolver
        44028, 66368, mlsolver
        2163, 4323, mlsolver
        48211, 96419, mlsolver
        1031612, 1559132, mlsolver
        10183, 20363, mlsolver
        214775, 324327, mlsolver
        228787, 457571, mlsolver
        4924413, 7445509, mlsolver
        467, 931, mlsolver
        8691, 13051, mlsolver
        100, 9900, specialcases
        200, 39800, specialcases
        200, 400, specialcases
        400, 800, specialcases
        1000, 2000, specialcases
        2000, 4000, specialcases
        4000, 8000, specialcases
        10000, 20000, specialcases
        2000, 3998000, specialcases
        40000, 80000, specialcases
        20000, 40000, specialcases
        500, 249500, specialcases
        301, 401, specialcases
        601, 801, specialcases
        1501, 2001, specialcases
        100000, 200000, specialcases
        3001, 4001, specialcases
        6001, 8001, specialcases
        1000, 999000, specialcases
        15001, 20001, specialcases
        5000, 24995000, specialcases
        30001, 40001, specialcases
        10000, 99990000, specialcases
        1000, 10390, random
        5000, 51817, random
        10000, 104720, random
        1000, 10685, random
        5000, 52230, random
        500, 1097, specialcases
        2500, 5497, specialcases
        1000, 2197, specialcases
        5000, 10997, specialcases
        10000, 105013, random
        10000, 21997, specialcases
        1000, 10499, random
        60001, 80001, specialcases
        5000, 52450, random
        20000, 208373, random
        25000, 54997, specialcases
        1000, 10826, random
        10000, 104191, random
        5000, 51966, random
        50000, 109997, specialcases
        1000, 10809, random
        5000, 52862, random
        10000, 104188, random
        20000, 209336, random
        1000, 10723, random
        10000, 105286, random
        50000, 525759, random
        5000, 52235, random
        20000, 209484, random
        1000, 10343, random
        10000, 106096, random
        50000, 526588, random
        20000, 210196, random
        5000, 52862, random
        1000, 10251, random
        10000, 104527, random
        20000, 210006, random
        5000, 52521, random
        50000, 524507, random
        10000, 104392, random
        1000, 10682, random
        20000, 207502, random
        5000, 52823, random
        50000, 525193, random
        1000, 10351, random
        50000, 525345, random
        5000, 52067, random
        20000, 210077, random
        10000, 105449, random
        10000, 105071, random
        1000, 10460, random
        5000, 52199, random
        50000, 525412, random
        20000, 209265, random
        1000, 10424, random
        10000, 105089, random
        5000, 52094, random
        20000, 208604, random
        50000, 525810, random
        10000, 105602, random
        1000, 10378, random
        5000, 52910, random
        50000, 526107, random
        20000, 209557, random
        1000, 10461, random
        10000, 104653, random
        5000, 52120, random
        50000, 525548, random
        20000, 210993, random
        1000, 10496, random
        10000, 106357, random
        5000, 52546, random
        50000, 525378, random
        20000, 211061, random
        1000, 10733, random
        5000, 52512, random
        10000, 105236, random
        50000, 526563, random
        10000, 104944, random
        20000, 210481, random
        1000, 10641, random
        50000, 522281, random
        5000, 52856, random
        20000, 209796, random
        1000, 10755, random
        50000, 527772, random
        10000, 104207, random
        5000, 52630, random
        20000, 211569, random
        1000, 10595, random
        5000, 52787, random
        10000, 103997, random
        50000, 525003, random
        20000, 210626, random
        1000, 10548, random
        50000, 526968, random
        10000, 105120, random
        20000, 209878, random
        5000, 52742, random
        1000, 10704, random
        5000, 52605, random
        10000, 105547, random
        1000, 10657, random
        50000, 525568, random
        20000, 210250, random
        10000, 105305, random
        5000, 52507, random
        50000, 524594, random
        1000, 10750, random
        20000, 209474, random
        10000, 105345, random
        5000, 52289, random
        50000, 527400, random
        20000, 208566, random
        1000, 10499, random
        10000, 104788, random
        20000, 210190, random
        50000, 526864, random
        5000, 52399, random
        1000, 10368, random
        10000, 105795, random
        5000, 52699, random
        50000, 525539, random
        20000, 209585, random
        10000, 106168, random
        50000, 523291, random
        1000, 9357, random
        20000, 210186, random
        1000, 6860, random
        1000, 7589, random
        1000, 7185, random
        1000, 6034, random
        1000, 6970, random
        1000, 6926, random
        1000, 7449, random
        1000, 7835, random
        1000, 7554, random
        50000, 526826, random
        1000, 6492, random
        1000, 6176, random
        1000, 6504, random
        1000, 7851, random
        20000, 210544, random
        1000, 7808, random
        1000, 7854, random
        1000, 8630, random
        1000, 9011, random
        1000, 6762, random
        1000, 7750, random
        1000, 8658, random
        1000, 11189, random
        1000, 8750, random
        1000, 7400, random
        1000, 7452, random
        5000, 41501, random
        5000, 57805, random
        50000, 526732, random
        20000, 209780, random
        5000, 50443, random
        5000, 45931, random
        5000, 43425, random
        5000, 48918, random
        5000, 44623, random
        5000, 44612, random
        5000, 47513, random
        5000, 45116, random
        5000, 52754, random
        5000, 56081, random
        5000, 49837, random
        5000, 44687, random
        5000, 46583, random
        5000, 57789, random
        5000, 50813, random
        50000, 525611, random
        5000, 45532, random
        5000, 45563, random
        5000, 51959, random
        5000, 68826, random
        5000, 52016, random
        5000, 54754, random
        5000, 41380, random
        5000, 49564, random
        50000, 524843, random
        20000, 230193, random
        20000, 234858, random
        20000, 212965, random
        20000, 193426, random
        20000, 216324, random
        20000, 225152, random
        20000, 229972, random
        20000, 275660, random
        20000, 224917, random
        20000, 278603, random
        20000, 221533, random
        20000, 196836, random
        20000, 265528, random
        20000, 235034, random
        20000, 181963, random
        20000, 236466, random
        20000, 211497, random
        20000, 212209, random
        20000, 201195, random
        20000, 264246, random
        20000, 272503, random
        20000, 191616, random
        20000, 208638, random
        20000, 201159, random
        20000, 210762, random
        50000, 616050, random
        50000, 544367, random
        50000, 719280, random
        50000, 632296, random
        50000, 570049, random
        50000, 649493, random
        50000, 552353, random
        50000, 592216, random
        50000, 567435, random
        50000, 651438, random
        50000, 600297, random
        50000, 574223, random
        50000, 623393, random
        50000, 591893, random
        50000, 661142, random
        50000, 715668, random
        50000, 571970, random
        50000, 649382, random
        10000, 108037, random
        10000, 101768, random
        50000, 661654, random
        50000, 727957, random
        50000, 593284, random
        10000, 97147, random
        10000, 95995, random
        10000, 93391, random
        50000, 542708, random
        10000, 105372, random
        10000, 111298, random
        10000, 114637, random
        10000, 122675, random
        10000, 123116, random
        10000, 101999, random
        10000, 91370, random
        10000, 104537, random
        50000, 537773, random
        10000, 126174, random
        10000, 109528, random
        10000, 111173, random
        10000, 132175, random
        10000, 90141, random
        10000, 101342, random
        10000, 89214, random
        50000, 849930, random
        10000, 97015, random
        10000, 101640, random
        10000, 107295, random
        50000, 574556, random
        10000, 99239, random
        778, 1079, modelchecking
        114, 139, modelchecking
        102, 127, modelchecking
        292, 439, modelchecking
        204, 255, modelchecking
        292, 367, modelchecking
        102, 125, modelchecking
        302, 383, modelchecking
        98, 123, modelchecking
        10000, 98217, random
        1691, 4825, modelchecking
        1691, 4825, modelchecking
        882, 1939, modelchecking
        882, 1939, modelchecking
        846, 2172, modelchecking
        763, 1903, modelchecking
        486, 1126, modelchecking
        486, 1126, modelchecking
        27, 99, mlsolver
        225, 342, mlsolver
        59, 79, mlsolver
        330, 369, mlsolver
        71, 95, mlsolver
        494, 552, mlsolver
        86, 111, mlsolver
        15, 25, mlsolver
        7, 10, equivalence
        447, 788, equivalence
        447, 788, equivalence
        532, 859, equivalence
        310, 516, equivalence
        5925, 11971, equivalence
        4921, 11463, equivalence
        5925, 11971, equivalence
        3, 3, equivalence
        6722, 16466, equivalence
        6722, 16466, equivalence
        7106, 17154, equivalence
        816, 1377, equivalence
        16119, 32732, equivalence
        16119, 32732, equivalence
        13425, 31402, equivalence
        614, 1028, equivalence
        12133, 24579, equivalence
        12133, 24579, equivalence
        10121, 23559, equivalence
        3, 3, equivalence
        8961, 25073, equivalence
        9553, 26273, equivalence
        8961, 25073, equivalence
        3, 3, equivalence
        966897, 3278913, equivalence
        912641, 2762529, equivalence
        912641, 2762529, equivalence
        412, 691, equivalence
        7855, 15904, equivalence
        7855, 15904, equivalence
        6511, 15242, equivalence
        3, 3, equivalence
        3457, 8945, equivalence
        3761, 9457, equivalence
        3457, 8945, equivalence
        3, 3, equivalence
        2076302, 5857418, equivalence
        2076302, 5857418, equivalence
        2162594, 6445442, equivalence
        3, 3, equivalence
        3697, 9561, equivalence
        3985, 10665, equivalence
        3697, 9561, equivalence
        11, 18, equivalence
        1095, 2052, equivalence
        1095, 2052, equivalence
        1264, 2193, equivalence
        3, 3, equivalence
        13409, 36625, equivalence
        14177, 40049, equivalence
        13409, 36625, equivalence
        11, 18, equivalence
        1127, 2060, equivalence
        1127, 2060, equivalence
        1306, 2207, equivalence
        3, 3, equivalence
        3644173, 12968893, equivalence
        3471553, 11083645, equivalence
        3471553, 11083645, equivalence
        34954, 85706, equivalence
        34954, 85706, equivalence
        30922, 95682, equivalence
        3, 3, equivalence
        3, 3, equivalence
        97762, 240530, equivalence
        86802, 268594, equivalence
        97762, 240530, equivalence
        1883, 3288, equivalence
        6124, 14585, equivalence
        7383, 15080, equivalence
        7383, 15080, equivalence
        3759, 6572, equivalence
        15111, 30956, equivalence
        12590, 29963, equivalence
        15111, 30956, equivalence
        7, 10, equivalence
        467, 808, equivalence
        467, 808, equivalence
        558, 883, equivalence
        3, 3, equivalence
        2266, 5418, equivalence
        2410, 5666, equivalence
        2266, 5418, equivalence
        7, 10, equivalence
        353, 607, equivalence
        423, 665, equivalence
        353, 607, equivalence
        3, 3, equivalence
        57905, 147937, equivalence
        57905, 147937, equivalence
        50865, 168209, equivalence
        134097, 345057, equivalence
        118225, 391905, equivalence
        3, 3, equivalence
        134097, 345057, equivalence
        3977, 6973, equivalence
        20317, 41753, equivalence
        16910, 40450, equivalence
        20317, 41753, equivalence
        1993, 3489, equivalence
        8206, 19646, equivalence
        9907, 20295, equivalence
        9907, 20295, equivalence
        3, 3, equivalence
        604354, 1658402, equivalence
        604354, 1658402, equivalence
        631474, 1832290, equivalence
        3, 3, equivalence
        77529, 198673, equivalence
        77529, 198673, equivalence
        68025, 226281, equivalence
        3, 3, equivalence
        178233, 459921, equivalence
        178233, 459921, equivalence
        156729, 523185, equivalence
        816, 1377, equivalence
        19045, 44820, equivalence
        22915, 46698, equivalence
        22915, 46698, equivalence
        3, 3, equivalence
        167247, 434239, equivalence
        147323, 500573, equivalence
        167247, 434239, equivalence
        3, 3, equivalence
        74039, 191547, equivalence
        74039, 191547, equivalence
        65131, 221717, equivalence
        3, 3, equivalence
        39614, 123530, equivalence
        44926, 110170, equivalence
        44926, 110170, equivalence
        11, 18, equivalence
        845, 1531, equivalence
        983, 1645, equivalence
        845, 1531, equivalence
        3, 3, equivalence
        109890, 343274, equivalence
        124458, 306210, equivalence
        124458, 306210, equivalence
        3, 3, equivalence
        107824, 265669, equivalence
        107824, 265669, equivalence
        95288, 299105, equivalence
        4743, 8732, equivalence
        16162, 39527, equivalence
        19423, 40348, equivalence
        19423, 40348, equivalence
        3, 3, equivalence
        35234, 110927, equivalence
        39822, 98011, equivalence
        39822, 98011, equivalence
        9205, 21682, equivalence
        412, 691, equivalence
        11135, 22616, equivalence
        11135, 22616, equivalence
        2375, 4368, equivalence
        9507, 19696, equivalence
        7878, 19287, equivalence
        9507, 19696, equivalence
        104074, 315882, equivalence
        700289, 3449345, equivalence
        788225, 2505729, equivalence
        788225, 2505729, equivalence
        28226, 86306, equivalence
        214897, 1007185, equivalence
        240097, 733857, equivalence
        240097, 733857, equivalence
        38502, 115446, equivalence
        245697, 1211969, equivalence
        276225, 869985, equivalence
        276225, 869985, equivalence
        54758, 168878, equivalence
        440569, 2057941, equivalence
        488629, 1500793, equivalence
        488629, 1500793, equivalence
        12239050, 33790242, equivalence
        12239050, 33790242, equivalence
        3, 3, equivalence
        10685466, 40545186, equivalence
        40556396, 112380248, equivalence
        40556396, 112380248, equivalence
        3, 3, equivalence
        35431922, 134886692, equivalence
        9488018, 26181506, equivalence
        9488018, 26181506, equivalence
        3, 3, equivalence
        8326786, 31295586, equivalence
        31611530, 87556070, equivalence
        31611530, 87556070, equivalence
        3, 3, equivalence
        27799634, 104694734, equivalence
        7626354, 30467442, equivalence
        33702306, 76550466, modelchecking
        33702306, 76550466, modelchecking
        43, 73, modelchecking
        29868274, 78747250, modelchecking
        61, 105, modelchecking
        33702306, 76550466, modelchecking
        33702306, 76550466, modelchecking
        43, 73, modelchecking
        29868274, 78747250, modelchecking
        61, 105, modelchecking
        5322498, 21604226, equivalence
        19026506, 78220622, equivalence
        10927074, 30319218, equivalence
        10927074, 30319218, equivalence
        3, 3, equivalence
        9558194, 36746354, equivalence
        3, 3, equivalence
        37636481, 120755185, equivalence
        37636481, 120755185, equivalence
        3782172, 11533061, equivalence
        32926785, 167527601, equivalence
        14808231, 45377590, equivalence
        2544209, 3452747, mlsolver
        196527, 580276, mlsolver
        1167042, 1638060, mlsolver
        62953, 572688, mlsolver
        1348959, 1807450, mlsolver
        35388, 218527, mlsolver
        3155971, 4201801, mlsolver
        80133, 495630, mlsolver
        1021740, 1337584, mlsolver
        426185, 986070, mlsolver
        1042752, 1380194, mlsolver
        183781, 1506551, mlsolver
        2177202, 2532590, modelchecking
        9101, 10445, modelchecking
        48861, 56113, modelchecking
        6, 8, modelchecking
        531443, 1594321, modelchecking
        1594325, 4782967, modelchecking
        13834801, 29028241, modelchecking
        27876961, 58309201, modelchecking
        188570, 328544, modelchecking
        377113, 819008, modelchecking
        417016, 787653, modelchecking
        1285576, 2268975, modelchecking
        571378, 996970, modelchecking
        1411274, 2454775, modelchecking
        2341446, 7935608, modelchecking
        1997579, 9752561, modelchecking
        1997579, 9752561, modelchecking
        134162, 372852, modelchecking
        134162, 372852, modelchecking
        998790, 5412890, modelchecking
        788879, 4146139, modelchecking
        267378, 1257302, modelchecking
        267378, 1257302, modelchecking
        20712450, 70290691, modelchecking
        8964338, 51952370, modelchecking
        6823297, 15799809, modelchecking
        3487362, 12463874, modelchecking
        6974724, 24927749, modelchecking
        7767169, 28309249, modelchecking
        11488274, 45840722, modelchecking
        94710, 404736, modelchecking
        2589057, 11565569, modelchecking
        10782593, 39508737, modelchecking
        7310722, 22667778, modelchecking
        19550209, 45106177, modelchecking
        8835074, 34391042, modelchecking
        22288897, 80830465, modelchecking
        6690605, 29413398, modelchecking
        7429633, 32985601, modelchecking
        24565250, 73675010, modelchecking
        861780, 1431610, modelchecking
        40200, 80199, specialcases
        80400, 160599, specialcases
        20100, 39999, specialcases
        201000, 401799, specialcases
        80200, 159999, specialcases
        160400, 320399, specialcases
        40100, 79799, specialcases
        401000, 801599, specialcases
        20200, 40299, specialcases
        40400, 80699, specialcases
        10100, 20099, specialcases
        101000, 201899, specialcases
        200200, 399399, specialcases
        400400, 799799, specialcases
        100100, 199199, specialcases
        1001000, 2000999, specialcases
        100000, 219997, specialcases
        250000, 549997, specialcases 
      };
      \legend{}
    \end{axis}
  \end{tikzpicture}

%% file: size_vs_girth_orig.tex
  \begin{tikzpicture}[mark size=2.5pt,remember picture,scale=0.7]
    \begin{axis}[axis x line=bottom,
                       axis y line=left,
                       xlabel={Vertices},
                       ylabel={Girth},
                       xmode={log},
                       ymode={log}
                       ,
                       xmin=0,
                       ymin=0,
                       scatter/classes={
                         modelchecking={mark=x},
                         equivalence={mark=o},
                         mlsolver={mark=+},
                         specialcases={mark=square},
                         random={mark=triangle}
                       },
                       legend image post style={xshift=0.5cm},
                       legend cell align=left,
                       legend style={
                         at={(1,0.62)},
                         nodes={right}
                       }]
      \addplot+[black,only marks,mark=x,scatter,scatter src=explicit symbolic] 
      table[col sep=comma,x=x, y=y, meta=cluster] { 
        x, y, cluster
        2, 1, modelchecking
        972, 1, modelchecking
        2688, 1, modelchecking
        15684, 1, modelchecking
        588, 1, modelchecking
        656, 1, modelchecking
        308, 1, modelchecking
        227, 1, modelchecking
        328, 2, modelchecking
        649, 1, modelchecking
        656, 1, modelchecking
        649, 1, modelchecking
        227, 1, modelchecking
        8626, 1, modelchecking
        2126, 1, modelchecking
        4313, 4, modelchecking
        4021, 1, modelchecking
        6359, 1, modelchecking
        6359, 1, modelchecking
        8626, 1, modelchecking
        2126, 1, modelchecking
        2832, 1, modelchecking
        16356, 1, modelchecking
        2916, 1, modelchecking
        8748, 1, modelchecking
        564, 1, modelchecking
        2338, 2, modelchecking
        147, 6, modelchecking
        260, 1, modelchecking
        747, 6, modelchecking
        442, 6, modelchecking
        731, 6, modelchecking
        439, 6, modelchecking
        598, 6, modelchecking
        173, 6, modelchecking
        150, 6, modelchecking
        236, 1, modelchecking
        9282, 2, modelchecking
        294, 6, modelchecking
        341, 6, modelchecking
        2643, 6, modelchecking
        291, 6, modelchecking
        516, 1, modelchecking
        2034, 6, modelchecking
        871, 6, modelchecking
        2346, 6, modelchecking
        2611, 6, modelchecking
        468, 1, modelchecking
        594, 2, modelchecking
        78, 6, modelchecking
        156, 6, modelchecking
        132, 1, modelchecking
        75, 6, modelchecking
        231, 6, modelchecking
        89, 6, modelchecking
        223, 6, modelchecking
        78, 6, modelchecking
        223, 6, modelchecking
        120, 1, modelchecking
        63907, 1, modelchecking
        31954, 1, modelchecking
        63907, 1, modelchecking
        13222, 1, modelchecking
        13222, 1, modelchecking
        11977, 1, modelchecking
        11977, 1, modelchecking
        26996, 1, modelchecking
        26244, 1, modelchecking
        2946, 1, modelchecking
        5649, 1, modelchecking
        1041, 1, modelchecking
        5201, 1, modelchecking
        3121, 1, modelchecking
        4486, 1, modelchecking
        1122, 1, modelchecking
        1619, 1, modelchecking
        16642, 1, modelchecking
        18625, 1, modelchecking
        18625, 1, modelchecking
        44773, 1, modelchecking
        53638, 5, modelchecking
        51220, 1, modelchecking
        44773, 1, modelchecking
        1028, 1, modelchecking
        514, 1, modelchecking
        691, 1, modelchecking
        3714, 1, modelchecking
        465, 1, modelchecking
        1393, 1, modelchecking
        1553, 1, modelchecking
        450, 1, modelchecking
        1393, 1, modelchecking
        12130, 1, modelchecking
        2658, 6, modelchecking
        3452, 1, modelchecking
        3412, 6, modelchecking
        380, 6, modelchecking
        3066, 6, modelchecking
        438, 6, modelchecking
        1138, 6, modelchecking
        384, 6, modelchecking
        4860, 1, modelchecking
        4860, 1, modelchecking
        2229, 1, modelchecking
        2229, 1, modelchecking
        5861, 1, modelchecking
        4898, 1, modelchecking
        41762, 1, modelchecking
        39178, 1, modelchecking
        12818, 1, modelchecking
        13825, 1, modelchecking
        28964, 1, modelchecking
        10369, 1, modelchecking
        45937, 1, modelchecking
        4609, 1, modelchecking
        9986, 1, modelchecking
        21506, 1, modelchecking
        5958, 1, modelchecking
        6866, 1, modelchecking
        1490, 1, modelchecking
        3344, 1, modelchecking
        3530, 1, modelchecking
        4033, 1, modelchecking
        2977, 1, modelchecking
        2690, 1, modelchecking
        1345, 1, modelchecking
        7897, 1, modelchecking
        196, 6, modelchecking
        222, 6, modelchecking
        3058, 1, modelchecking
        192, 6, modelchecking
        782, 6, modelchecking
        578, 6, modelchecking
        956, 6, modelchecking
        976, 1, modelchecking
        574, 6, modelchecking
        3949, 1, modelchecking
        17810, 1, modelchecking
        35620, 1, modelchecking
        54322, 1, modelchecking
        14065, 1, modelchecking
        32402, 1, modelchecking
        42618, 1, modelchecking
        42193, 1, modelchecking
        34673, 1, modelchecking
        56241, 1, modelchecking
        3458, 1, modelchecking
        1046, 1, modelchecking
        1621, 1, modelchecking
        433, 1, modelchecking
        326, 1, modelchecking
        1370, 1, modelchecking
        770, 1, modelchecking
        1012, 1, modelchecking
        506, 1, modelchecking
        937, 1, modelchecking
        1297, 1, modelchecking
        11362, 1, modelchecking
        3188, 1, modelchecking
        3228, 1, modelchecking
        356, 1, modelchecking
        2490, 1, modelchecking
        1066, 1, modelchecking
        2874, 1, modelchecking
        360, 1, modelchecking
        414, 1, modelchecking
        6563, 1, modelchecking
        19685, 1, modelchecking
        730, 1, modelchecking
        282, 1, modelchecking
        274, 1, modelchecking
        96, 1, modelchecking
        92, 1, modelchecking
        192, 1, modelchecking
        96, 1, modelchecking
        108, 1, modelchecking
        272, 1, modelchecking
        43, 1, modelchecking
        13, 2, modelchecking
        175, 1, modelchecking
        97, 1, modelchecking
        133, 1, modelchecking
        67, 1, modelchecking
        82434, 1, modelchecking
        2577, 1, modelchecking
        24593, 1, modelchecking
        17154, 1, modelchecking
        21770, 1, modelchecking
        23185, 1, modelchecking
        7729, 1, modelchecking
        2722, 1, modelchecking
        4243, 1, modelchecking
        43, 1, modelchecking
        2866, 1, modelchecking
        734, 1, modelchecking
        542, 1, modelchecking
        184, 1, modelchecking
        538, 1, modelchecking
        210, 1, modelchecking
        180, 1, modelchecking
        892, 1, modelchecking
        912, 1, modelchecking
        17773, 1, modelchecking
        2395, 1, modelchecking
        343, 1, modelchecking
        25, 2, modelchecking
        13, 2, modelchecking
        22873, 1, modelchecking
        1559, 1, mlsolver
        636, 1, mlsolver
        298, 1, mlsolver
        693, 1, mlsolver
        2858, 1, mlsolver
        1143, 1, mlsolver
        1126, 1, mlsolver
        467, 1, mlsolver
        12, 1, mlsolver
        22, 1, mlsolver
        974, 1, mlsolver
        2425, 1, mlsolver
        805, 1, mlsolver
        1992, 1, mlsolver
        1312, 1, mlsolver
        3291, 1, mlsolver
        6817, 1, mlsolver
        14053, 1, mlsolver
        1562, 1, mlsolver
        3029, 1, mlsolver
        3724, 1, mlsolver
        1481, 1, mlsolver
        336, 1, mlsolver
        15, 1, mlsolver
        191, 1, mlsolver
        365, 1, mlsolver
        76, 1, mlsolver
        9, 1, mlsolver
        912, 1, mlsolver
        21, 1, mlsolver
        45, 1, mlsolver
        8776, 1, mlsolver
        5736, 1, mlsolver
        39, 1, mlsolver
        27, 1, mlsolver
        1924, 1, mlsolver
        3492, 1, mlsolver
        33, 1, mlsolver
        9, 1, mlsolver
        319, 1, mlsolver
        160, 1, mlsolver
        9, 1, mlsolver
        1156, 1, mlsolver
        9, 1, mlsolver
        817, 1, mlsolver
        9, 1, mlsolver
        61, 1, mlsolver
        9, 1, mlsolver
        538, 1, mlsolver
        9, 1, mlsolver
        29335, 1, mlsolver
        62707, 1, mlsolver
        30, 1, mlsolver
        9, 1, mlsolver
        51, 1, mlsolver
        12732, 1, mlsolver
        2014, 1, mlsolver
        9, 1, mlsolver
        140, 1, mlsolver
        21, 1, mlsolver
        9, 1, mlsolver
        1555, 1, mlsolver
        15, 1, mlsolver
        75, 1, mlsolver
        33, 1, mlsolver
        330, 1, mlsolver
        200, 1, mlsolver
        47, 1, mlsolver
        455, 1, mlsolver
        39, 1, mlsolver
        45, 1, mlsolver
        600, 1, mlsolver
        765, 1, mlsolver
        51, 1, mlsolver
        105, 1, mlsolver
        34, 1, mlsolver
        63, 1, mlsolver
        19, 1, mlsolver
        1490, 1, mlsolver
        119, 1, mlsolver
        1190, 1, mlsolver
        107, 1, mlsolver
        924, 1, mlsolver
        95, 1, mlsolver
        83, 1, mlsolver
        692, 1, mlsolver
        33969, 1, mlsolver
        6182, 1, mlsolver
        456, 1, mlsolver
        2497, 1, mlsolver
        6431, 1, mlsolver
        730, 1, mlsolver
        8, 1, mlsolver
        16, 1, mlsolver
        543, 1, mlsolver
        69, 1, mlsolver
        348, 1, mlsolver
        3772, 1, mlsolver
        6346, 1, mlsolver
        60011, 1, mlsolver
        38075, 1, mlsolver
        3195, 1, mlsolver
        1264, 1, mlsolver
        36735, 1, mlsolver
        25971, 1, mlsolver
        49830, 1, mlsolver
        2861, 1, mlsolver
        94631, 1, mlsolver
        19, 1, mlsolver
        354, 1, mlsolver
        8, 1, mlsolver
        16, 1, mlsolver
        6831, 1, mlsolver
        33977, 1, mlsolver
        9585, 1, mlsolver
        6650, 1, mlsolver
        970, 1, mlsolver
        65, 1, mlsolver
        334, 1, mlsolver
        30, 1, mlsolver
        7, 1, mlsolver
        14, 1, mlsolver
        2443, 1, mlsolver
        148, 1, mlsolver
        332, 1, mlsolver
        5837, 1, mlsolver
        778, 1, mlsolver
        13687, 1, mlsolver
        89, 1, mlsolver
        2265, 1, mlsolver
        8, 1, mlsolver
        34, 1, mlsolver
        17, 1, mlsolver
        250, 1, mlsolver
        203, 1, mlsolver
        27, 1, mlsolver
        32233, 1, mlsolver
        1794, 1, mlsolver
        17, 1, mlsolver
        9, 1, mlsolver
        38, 1, mlsolver
        795, 1, mlsolver
        35037, 1, mlsolver
        1123, 1, mlsolver
        478, 1, mlsolver
        14399, 1, mlsolver
        5914, 1, mlsolver
        209, 1, mlsolver
        2611, 1, mlsolver
        82968, 1, mlsolver
        105, 1, mlsolver
        1546, 1, mlsolver
        15427, 1, mlsolver
        5596, 1, mlsolver
        44028, 1, mlsolver
        2163, 1, mlsolver
        48211, 1, mlsolver
        10183, 1, mlsolver
        467, 1, mlsolver
        8691, 1, mlsolver
        100, 1, specialcases
        200, 1, specialcases
        200, 1, specialcases
        400, 1, specialcases
        1000, 1, specialcases
        2000, 1, specialcases
        4000, 1, specialcases
        10000, 1, specialcases
        40000, 1, specialcases
        20000, 1, specialcases
        500, 1, specialcases
        301, 201, specialcases
        601, 401, specialcases
        1501, 1001, specialcases
        3001, 2001, specialcases
        6001, 4001, specialcases
        1000, 1, specialcases
        15001, 10001, specialcases
        30001, 20001, specialcases
        1000, 1, random
        5000, 1, random
        10000, 1, random
        1000, 1, random
        5000, 1, random
        500, 1, specialcases
        2500, 1, specialcases
        1000, 1, specialcases
        5000, 1, specialcases
        10000, 1, random
        10000, 1, specialcases
        1000, 1, random
        5000, 1, random
        20000, 1, random
        25000, 1, specialcases
        1000, 1, random
        10000, 1, random
        5000, 1, random
        1000, 1, random
        5000, 1, random
        10000, 1, random
        20000, 1, random
        1000, 1, random
        10000, 1, random
        5000, 1, random
        20000, 1, random
        1000, 1, random
        10000, 1, random
        20000, 1, random
        5000, 1, random
        1000, 1, random
        10000, 1, random
        20000, 1, random
        5000, 1, random
        10000, 1, random
        1000, 1, random
        20000, 1, random
        5000, 1, random
        1000, 1, random
        5000, 1, random
        20000, 1, random
        10000, 1, random
        10000, 1, random
        1000, 1, random
        5000, 1, random
        20000, 1, random
        1000, 1, random
        10000, 1, random
        5000, 1, random
        20000, 1, random
        10000, 1, random
        1000, 1, random
        5000, 1, random
        20000, 1, random
        1000, 1, random
        10000, 1, random
        5000, 1, random
        20000, 1, random
        1000, 1, random
        10000, 1, random
        5000, 1, random
        20000, 1, random
        1000, 1, random
        5000, 1, random
        10000, 1, random
        10000, 1, random
        20000, 1, random
        1000, 1, random
        5000, 1, random
        20000, 1, random
        1000, 1, random
        10000, 1, random
        5000, 1, random
        20000, 1, random
        1000, 1, random
        5000, 1, random
        10000, 1, random
        20000, 1, random
        1000, 1, random
        10000, 1, random
        20000, 1, random
        5000, 1, random
        1000, 1, random
        5000, 1, random
        10000, 1, random
        1000, 1, random
        20000, 1, random
        10000, 1, random
        5000, 1, random
        1000, 1, random
        20000, 1, random
        10000, 1, random
        5000, 1, random
        20000, 1, random
        1000, 1, random
        10000, 1, random
        20000, 1, random
        5000, 1, random
        1000, 1, random
        10000, 1, random
        5000, 1, random
        20000, 1, random
        10000, 1, random
        1000, 1, random
        20000, 1, random
        1000, 1, random
        1000, 1, random
        1000, 1, random
        1000, 1, random
        1000, 1, random
        1000, 1, random
        1000, 1, random
        1000, 1, random
        1000, 1, random
        1000, 1, random
        1000, 1, random
        1000, 1, random
        1000, 1, random
        20000, 1, random
        1000, 1, random
        1000, 1, random
        1000, 1, random
        1000, 1, random
        1000, 1, random
        1000, 1, random
        1000, 1, random
        1000, 1, random
        1000, 1, random
        1000, 1, random
        1000, 1, random
        5000, 1, random
        5000, 1, random
        20000, 1, random
        5000, 1, random
        5000, 1, random
        5000, 1, random
        5000, 1, random
        5000, 1, random
        5000, 1, random
        5000, 1, random
        5000, 1, random
        5000, 1, random
        5000, 1, random
        5000, 1, random
        5000, 1, random
        5000, 1, random
        5000, 1, random
        5000, 1, random
        5000, 1, random
        5000, 1, random
        5000, 1, random
        5000, 1, random
        5000, 1, random
        5000, 1, random
        5000, 1, random
        5000, 1, random
        20000, 1, random
        20000, 1, random
        20000, 1, random
        20000, 1, random
        20000, 1, random
        20000, 1, random
        20000, 1, random
        20000, 1, random
        20000, 1, random
        20000, 1, random
        20000, 1, random
        20000, 1, random
        20000, 1, random
        20000, 1, random
        20000, 1, random
        20000, 1, random
        20000, 1, random
        20000, 1, random
        20000, 1, random
        20000, 1, random
        20000, 1, random
        20000, 1, random
        20000, 1, random
        10000, 1, random
        10000, 1, random
        10000, 1, random
        10000, 1, random
        10000, 1, random
        10000, 1, random
        10000, 1, random
        10000, 1, random
        10000, 1, random
        10000, 1, random
        10000, 1, random
        10000, 1, random
        10000, 1, random
        10000, 1, random
        10000, 1, random
        10000, 1, random
        10000, 1, random
        10000, 1, random
        10000, 1, random
        10000, 1, random
        10000, 1, random
        10000, 1, random
        10000, 1, random
        10000, 1, random
        778, 1, modelchecking
        114, 6, modelchecking
        102, 6, modelchecking
        292, 6, modelchecking
        204, 6, modelchecking
        292, 6, modelchecking
        102, 6, modelchecking
        302, 1, modelchecking
        98, 6, modelchecking
        10000, 1, random
        1691, 1, modelchecking
        1691, 1, modelchecking
        882, 1, modelchecking
        882, 1, modelchecking
        846, 1, modelchecking
        763, 1, modelchecking
        486, 1, modelchecking
        486, 1, modelchecking
        27, 1, mlsolver
        225, 1, mlsolver
        59, 1, mlsolver
        330, 1, mlsolver
        71, 1, mlsolver
        494, 1, mlsolver
        86, 1, mlsolver
        15, 1, mlsolver
        7, 1, equivalence
        447, 1, equivalence
        447, 1, equivalence
        532, 1, equivalence
        310, 1, equivalence
        5925, 1, equivalence
        4921, 1, equivalence
        5925, 1, equivalence
        3, 1, equivalence
        6722, 1, equivalence
        6722, 1, equivalence
        7106, 1, equivalence
        816, 1, equivalence
        16119, 1, equivalence
        16119, 1, equivalence
        13425, 1, equivalence
        614, 1, equivalence
        12133, 1, equivalence
        12133, 1, equivalence
        10121, 1, equivalence
        3, 1, equivalence
        8961, 1, equivalence
        9553, 1, equivalence
        8961, 1, equivalence
        3, 1, equivalence
        412, 1, equivalence
        7855, 1, equivalence
        7855, 1, equivalence
        6511, 1, equivalence
        3, 1, equivalence
        3457, 1, equivalence
        3761, 1, equivalence
        3457, 1, equivalence
        3, 1, equivalence
        3, 1, equivalence
        3697, 1, equivalence
        3985, 1, equivalence
        3697, 1, equivalence
        11, 1, equivalence
        1095, 1, equivalence
        1095, 1, equivalence
        1264, 1, equivalence
        3, 1, equivalence
        13409, 1, equivalence
        14177, 1, equivalence
        13409, 1, equivalence
        11, 1, equivalence
        1127, 1, equivalence
        1127, 1, equivalence
        1306, 1, equivalence
        3, 1, equivalence
        34954, 1, equivalence
        34954, 1, equivalence
        30922, 1, equivalence
        3, 1, equivalence
        3, 1, equivalence
        1883, 1, equivalence
        6124, 1, equivalence
        7383, 1, equivalence
        7383, 1, equivalence
        3759, 1, equivalence
        15111, 1, equivalence
        12590, 1, equivalence
        15111, 1, equivalence
        7, 1, equivalence
        467, 1, equivalence
        467, 1, equivalence
        558, 1, equivalence
        3, 1, equivalence
        2266, 1, equivalence
        2410, 1, equivalence
        2266, 1, equivalence
        7, 1, equivalence
        353, 1, equivalence
        423, 1, equivalence
        353, 1, equivalence
        3, 1, equivalence
        3, 1, equivalence
        3977, 1, equivalence
        20317, 1, equivalence
        16910, 1, equivalence
        20317, 1, equivalence
        1993, 1, equivalence
        8206, 1, equivalence
        9907, 1, equivalence
        9907, 1, equivalence
        3, 1, equivalence
        3, 1, equivalence
        3, 1, equivalence
        816, 1, equivalence
        19045, 1, equivalence
        22915, 1, equivalence
        22915, 1, equivalence
        3, 1, equivalence
        3, 1, equivalence
        3, 1, equivalence
        39614, 1, equivalence
        11, 1, equivalence
        845, 1, equivalence
        983, 1, equivalence
        845, 1, equivalence
        3, 1, equivalence
        3, 1, equivalence
        4743, 1, equivalence
        16162, 1, equivalence
        19423, 1, equivalence
        19423, 1, equivalence
        3, 1, equivalence
        35234, 1, equivalence
        39822, 1, equivalence
        39822, 1, equivalence
        9205, 1, equivalence
        412, 1, equivalence
        11135, 1, equivalence
        11135, 1, equivalence
        2375, 1, equivalence
        9507, 1, equivalence
        7878, 1, equivalence
        9507, 1, equivalence
        28226, 1, equivalence
        38502, 1, equivalence
        3, 1, equivalence
        3, 1, equivalence
        3, 1, equivalence
        3, 1, equivalence
        43, 1, modelchecking
        61, 1, modelchecking
        43, 1, modelchecking
        61, 1, modelchecking
        3, 1, equivalence
        3, 1, equivalence
        35388, 1, mlsolver
        9101, 1, modelchecking
        48861, 1, modelchecking
        6, 1, modelchecking
        134162, 1, modelchecking
        134162, 1, modelchecking
        40200, 1, specialcases
        20100, 1, specialcases
        40100, 1, specialcases
        20200, 1, specialcases
        40400, 1, specialcases
        10100, 1, specialcases 
      };
      \legend{~modelchecking, ~equivalence, ~mlsolver, ~specialcases, ~random}
    \end{axis}
  \end{tikzpicture}

%% file: size_vs_diameter_orig.tex
  \begin{tikzpicture}[mark size=2.5pt,remember picture,scale=0.7]
    \begin{axis}[axis x line=bottom,
                       axis y line=left,
                       xlabel={Vertices},
                       ylabel={Diameter},
                       xmode={log},
                       ymode={log}
                       ,
                       xmin=0,
                       ymin=0,
                       scatter/classes={
                         modelchecking={mark=x},
                         equivalence={mark=o},
                         mlsolver={mark=+},
                         specialcases={mark=square},
                         random={mark=triangle}
                       },
                       legend pos=outer north east,
                       legend image post style={xshift=0.5cm},
                       legend cell align=left,
                       legend style={
                         nodes={right}
                       }]
      \addplot+[black,only marks,mark=x,scatter,scatter src=explicit symbolic] 
      table[col sep=comma,x=x, y=y, meta=cluster] { 
        x, y, cluster
        2, 1, modelchecking
        972, 95, modelchecking
        2688, 59, modelchecking
        15684, 80, modelchecking
        588, 29, modelchecking
        656, 29, modelchecking
        308, 29, modelchecking
        227, 28, modelchecking
        328, 28, modelchecking
        649, 33, modelchecking
        656, 29, modelchecking
        649, 33, modelchecking
        227, 28, modelchecking
        8626, 49, modelchecking
        2126, 48, modelchecking
        4313, 48, modelchecking
        4021, 48, modelchecking
        6359, 54, modelchecking
        6359, 54, modelchecking
        8626, 49, modelchecking
        2126, 48, modelchecking
        2832, 38, modelchecking
        16356, 47, modelchecking
        2916, 191, modelchecking
        8748, 383, modelchecking
        564, 38, modelchecking
        2338, 34, modelchecking
        147, 30, modelchecking
        260, 30, modelchecking
        747, 40, modelchecking
        442, 30, modelchecking
        731, 38, modelchecking
        439, 32, modelchecking
        598, 31, modelchecking
        173, 38, modelchecking
        150, 31, modelchecking
        236, 30, modelchecking
        9282, 34, modelchecking
        294, 31, modelchecking
        341, 38, modelchecking
        2643, 40, modelchecking
        291, 30, modelchecking
        516, 30, modelchecking
        2034, 30, modelchecking
        871, 32, modelchecking
        2346, 31, modelchecking
        2611, 38, modelchecking
        468, 30, modelchecking
        594, 34, modelchecking
        78, 31, modelchecking
        156, 31, modelchecking
        132, 30, modelchecking
        75, 30, modelchecking
        231, 40, modelchecking
        89, 38, modelchecking
        223, 38, modelchecking
        78, 22, modelchecking
        223, 32, modelchecking
        120, 30, modelchecking
        63907, 48, modelchecking
        31954, 47, modelchecking
        63907, 48, modelchecking
        13222, 54, modelchecking
        13222, 54, modelchecking
        11977, 47, modelchecking
        11977, 47, modelchecking
        26996, 50, modelchecking
        26244, 767, modelchecking
        2946, 36, modelchecking
        5649, 48, modelchecking
        1041, 36, modelchecking
        5201, 46, modelchecking
        3121, 38, modelchecking
        4486, 38, modelchecking
        1122, 38, modelchecking
        1619, 46, modelchecking
        16642, 42, modelchecking
        107276, 77, modelchecking
        18625, 76, modelchecking
        18625, 76, modelchecking
        44773, 84, modelchecking
        53638, 76, modelchecking
        51220, 76, modelchecking
        44773, 84, modelchecking
        107276, 77, modelchecking
        1028, 38, modelchecking
        514, 38, modelchecking
        691, 46, modelchecking
        3714, 42, modelchecking
        465, 36, modelchecking
        1393, 46, modelchecking
        1553, 48, modelchecking
        450, 27, modelchecking
        1393, 38, modelchecking
        12130, 39, modelchecking
        2658, 35, modelchecking
        3452, 45, modelchecking
        3412, 43, modelchecking
        380, 35, modelchecking
        3066, 36, modelchecking
        438, 43, modelchecking
        1138, 37, modelchecking
        384, 36, modelchecking
        4860, 15, modelchecking
        4860, 15, modelchecking
        2229, 8, modelchecking
        2229, 8, modelchecking
        5861, 32, modelchecking
        4898, 33, modelchecking
        41762, 39, modelchecking
        39178, 33, modelchecking
        12818, 32, modelchecking
        13825, 33, modelchecking
        28964, 31, modelchecking
        10369, 43, modelchecking
        45937, 41, modelchecking
        4609, 31, modelchecking
        9986, 32, modelchecking
        21506, 37, modelchecking
        5958, 33, modelchecking
        6866, 39, modelchecking
        1490, 33, modelchecking
        3344, 31, modelchecking
        3530, 32, modelchecking
        4033, 33, modelchecking
        2977, 43, modelchecking
        2690, 32, modelchecking
        1345, 31, modelchecking
        7897, 41, modelchecking
        196, 36, modelchecking
        222, 43, modelchecking
        3058, 39, modelchecking
        192, 35, modelchecking
        782, 36, modelchecking
        578, 35, modelchecking
        956, 43, modelchecking
        976, 45, modelchecking
        574, 37, modelchecking
        3949, 50, modelchecking
        17810, 61, modelchecking
        35620, 61, modelchecking
        54322, 63, modelchecking
        14065, 58, modelchecking
        32402, 58, modelchecking
        42618, 58, modelchecking
        42193, 60, modelchecking
        34673, 84, modelchecking
        3458, 37, modelchecking
        1046, 32, modelchecking
        1621, 41, modelchecking
        433, 31, modelchecking
        326, 25, modelchecking
        1370, 39, modelchecking
        770, 32, modelchecking
        1012, 33, modelchecking
        506, 33, modelchecking
        937, 43, modelchecking
        1297, 33, modelchecking
        11362, 36, modelchecking
        3188, 41, modelchecking
        3228, 43, modelchecking
        356, 32, modelchecking
        2490, 32, modelchecking
        1066, 34, modelchecking
        2874, 33, modelchecking
        360, 33, modelchecking
        414, 40, modelchecking
        6563, 257, modelchecking
        19685, 513, modelchecking
        730, 36, modelchecking
        282, 43, modelchecking
        274, 34, modelchecking
        96, 25, modelchecking
        92, 32, modelchecking
        192, 33, modelchecking
        96, 33, modelchecking
        108, 40, modelchecking
        272, 41, modelchecking
        43, 16, modelchecking
        13, 10, modelchecking
        175, 23, modelchecking
        97, 19, modelchecking
        133, 21, modelchecking
        67, 17, modelchecking
        82434, 42, modelchecking
        2577, 36, modelchecking
        24593, 48, modelchecking
        17154, 36, modelchecking
        21770, 38, modelchecking
        23185, 46, modelchecking
        7729, 38, modelchecking
        2722, 38, modelchecking
        4243, 46, modelchecking
        43, 15, modelchecking
        2866, 36, modelchecking
        734, 33, modelchecking
        542, 32, modelchecking
        184, 33, modelchecking
        538, 34, modelchecking
        210, 40, modelchecking
        180, 32, modelchecking
        892, 41, modelchecking
        912, 43, modelchecking
        17773, 33, modelchecking
        2395, 30, modelchecking
        343, 27, modelchecking
        25, 12, modelchecking
        13, 10, modelchecking
        22873, 70, modelchecking
        1559, 133, mlsolver
        636, 41, mlsolver
        298, 25, mlsolver
        693, 88, mlsolver
        2858, 229, mlsolver
        1143, 65, mlsolver
        1126, 101, mlsolver
        467, 30, mlsolver
        12, 7, mlsolver
        22, 20, mlsolver
        974, 57, mlsolver
        2425, 197, mlsolver
        805, 49, mlsolver
        1992, 165, mlsolver
        1312, 73, mlsolver
        3291, 261, mlsolver
        6817, 38, mlsolver
        14053, 172, mlsolver
        1562, 25, mlsolver
        3029, 104, mlsolver
        3724, 293, mlsolver
        1481, 81, mlsolver
        336, 88, mlsolver
        15, 3, mlsolver
        191, 17, mlsolver
        365, 63, mlsolver
        76, 33, mlsolver
        9, 3, mlsolver
        912, 173, mlsolver
        21, 3, mlsolver
        45, 3, mlsolver
        8776, 813, mlsolver
        5736, 608, mlsolver
        39, 3, mlsolver
        27, 3, mlsolver
        1924, 288, mlsolver
        3492, 433, mlsolver
        33, 3, mlsolver
        9, 3, mlsolver
        319, 158, mlsolver
        160, 77, mlsolver
        9, 3, mlsolver
        1156, 581, mlsolver
        9, 3, mlsolver
        817, 410, mlsolver
        9, 3, mlsolver
        61, 26, mlsolver
        9, 3, mlsolver
        538, 269, mlsolver
        9, 3, mlsolver
        29335, 49, mlsolver
        62707, 243, mlsolver
        30, 18, mlsolver
        9, 3, mlsolver
        51, 3, mlsolver
        12732, 1048, mlsolver
        2014, 1013, mlsolver
        9, 3, mlsolver
        140, 32, mlsolver
        21, 3, mlsolver
        9, 3, mlsolver
        1555, 782, mlsolver
        15, 3, mlsolver
        75, 26, mlsolver
        33, 3, mlsolver
        330, 44, mlsolver
        200, 70, mlsolver
        47, 15, mlsolver
        455, 50, mlsolver
        39, 3, mlsolver
        45, 3, mlsolver
        600, 56, mlsolver
        765, 62, mlsolver
        51, 3, mlsolver
        105, 36, mlsolver
        34, 11, mlsolver
        63, 38, mlsolver
        19, 7, mlsolver
        1490, 526, mlsolver
        119, 39, mlsolver
        1190, 420, mlsolver
        107, 35, mlsolver
        924, 326, mlsolver
        95, 31, mlsolver
        83, 27, mlsolver
        692, 244, mlsolver
        33969, 141, mlsolver
        6182, 11, mlsolver
        456, 10, mlsolver
        2497, 74, mlsolver
        6431, 120, mlsolver
        730, 12, mlsolver
        8, 4, mlsolver
        16, 14, mlsolver
        543, 64, mlsolver
        69, 6, mlsolver
        348, 10, mlsolver
        3772, 100, mlsolver
        6346, 16, mlsolver
        60011, 182, mlsolver
        38075, 161, mlsolver
        3195, 14, mlsolver
        1264, 9, mlsolver
        36735, 174, mlsolver
        25971, 18, mlsolver
        49830, 20, mlsolver
        2861, 9, mlsolver
        94631, 218, mlsolver
        19, 6, mlsolver
        354, 98, mlsolver
        8, 4, mlsolver
        16, 14, mlsolver
        6831, 17, mlsolver
        33977, 130, mlsolver
        9585, 14, mlsolver
        6650, 9, mlsolver
        970, 156, mlsolver
        65, 11, mlsolver
        334, 96, mlsolver
        30, 9, mlsolver
        7, 4, mlsolver
        14, 10, mlsolver
        2443, 230, mlsolver
        148, 13, mlsolver
        332, 15, mlsolver
        5837, 319, mlsolver
        778, 17, mlsolver
        13687, 422, mlsolver
        89, 12, mlsolver
        2265, 212, mlsolver
        8, 4, mlsolver
        34, 15, mlsolver
        17, 8, mlsolver
        250, 71, mlsolver
        203, 64, mlsolver
        27, 8, mlsolver
        32233, 537, mlsolver
        1794, 19, mlsolver
        17, 13, mlsolver
        9, 4, mlsolver
        38, 10, mlsolver
        795, 128, mlsolver
        35037, 578, mlsolver
        1123, 18, mlsolver
        478, 16, mlsolver
        14399, 437, mlsolver
        5914, 314, mlsolver
        209, 14, mlsolver
        2611, 20, mlsolver
        82968, 739, mlsolver
        105, 16, mlsolver
        1546, 190, mlsolver
        15427, 9, mlsolver
        5596, 13, mlsolver
        44028, 1242, mlsolver
        2163, 64, mlsolver
        48211, 256, mlsolver
        10183, 128, mlsolver
        467, 32, mlsolver
        8691, 500, mlsolver
        100, 1, specialcases
        200, 1, specialcases
        200, 100, specialcases
        400, 200, specialcases
        1000, 500, specialcases
        2000, 1000, specialcases
        4000, 2000, specialcases
        10000, 5000, specialcases
        2000, 1, specialcases
        40000, 20000, specialcases
        20000, 10000, specialcases
        500, 1, specialcases
        301, 201, specialcases
        601, 401, specialcases
        1501, 1001, specialcases
        3001, 2001, specialcases
        6001, 4001, specialcases
        1000, 1, specialcases
        15001, 10001, specialcases
        30001, 20001, specialcases
        1000, 7, random
        5000, 8, random
        10000, 9, random
        1000, 7, random
        5000, 8, random
        500, 104, specialcases
        2500, 504, specialcases
        1000, 204, specialcases
        5000, 1004, specialcases
        10000, 9, random
        10000, 2004, specialcases
        1000, 7, random
        60001, 40001, specialcases
        5000, 8, random
        20000, 10, random
        25000, 5004, specialcases
        1000, 7, random
        10000, 10, random
        5000, 9, random
        50000, 10004, specialcases
        1000, 7, random
        5000, 8, random
        10000, 9, random
        20000, 9, random
        1000, 7, random
        10000, 8, random
        5000, 8, random
        20000, 10, random
        1000, 7, random
        10000, 9, random
        20000, 9, random
        5000, 8, random
        1000, 7, random
        10000, 8, random
        20000, 10, random
        5000, 8, random
        10000, 9, random
        1000, 6, random
        20000, 9, random
        5000, 9, random
        1000, 6, random
        5000, 8, random
        20000, 9, random
        10000, 8, random
        10000, 9, random
        1000, 6, random
        5000, 9, random
        20000, 10, random
        1000, 7, random
        10000, 9, random
        5000, 8, random
        20000, 10, random
        10000, 9, random
        1000, 7, random
        5000, 9, random
        20000, 9, random
        1000, 7, random
        10000, 9, random
        5000, 8, random
        20000, 10, random
        1000, 7, random
        10000, 8, random
        5000, 8, random
        20000, 9, random
        1000, 7, random
        5000, 10, random
        10000, 8, random
        10000, 9, random
        20000, 9, random
        1000, 7, random
        5000, 9, random
        20000, 9, random
        1000, 8, random
        10000, 10, random
        5000, 8, random
        20000, 9, random
        1000, 7, random
        5000, 9, random
        10000, 9, random
        20000, 10, random
        1000, 7, random
        10000, 9, random
        20000, 9, random
        5000, 9, random
        1000, 7, random
        5000, 8, random
        10000, 10, random
        1000, 6, random
        20000, 9, random
        10000, 9, random
        5000, 9, random
        1000, 6, random
        20000, 9, random
        10000, 9, random
        5000, 9, random
        20000, 9, random
        1000, 7, random
        10000, 9, random
        20000, 9, random
        5000, 8, random
        1000, 7, random
        10000, 9, random
        5000, 10, random
        20000, 9, random
        10000, 9, random
        1000, 6, random
        20000, 10, random
        1000, 7, random
        1000, 7, random
        1000, 7, random
        1000, 7, random
        1000, 8, random
        1000, 7, random
        1000, 7, random
        1000, 7, random
        1000, 7, random
        1000, 8, random
        1000, 8, random
        1000, 8, random
        1000, 7, random
        20000, 10, random
        1000, 6, random
        1000, 7, random
        1000, 6, random
        1000, 7, random
        1000, 7, random
        1000, 7, random
        1000, 6, random
        1000, 5, random
        1000, 6, random
        1000, 7, random
        1000, 7, random
        5000, 8, random
        5000, 6, random
        20000, 9, random
        5000, 7, random
        5000, 7, random
        5000, 7, random
        5000, 7, random
        5000, 7, random
        5000, 7, random
        5000, 7, random
        5000, 8, random
        5000, 6, random
        5000, 6, random
        5000, 7, random
        5000, 8, random
        5000, 7, random
        5000, 6, random
        5000, 7, random
        5000, 7, random
        5000, 7, random
        5000, 7, random
        5000, 6, random
        5000, 7, random
        5000, 6, random
        5000, 8, random
        5000, 7, random
        20000, 7, random
        20000, 7, random
        20000, 8, random
        20000, 8, random
        20000, 7, random
        20000, 7, random
        20000, 7, random
        20000, 7, random
        20000, 7, random
        20000, 6, random
        20000, 7, random
        20000, 8, random
        20000, 7, random
        20000, 7, random
        20000, 8, random
        20000, 7, random
        20000, 8, random
        20000, 7, random
        20000, 8, random
        20000, 7, random
        20000, 8, random
        20000, 7, random
        20000, 8, random
        20000, 8, random
        10000, 7, random
        10000, 7, random
        10000, 8, random
        10000, 8, random
        10000, 8, random
        10000, 7, random
        10000, 7, random
        10000, 6, random
        10000, 6, random
        10000, 7, random
        10000, 8, random
        10000, 8, random
        10000, 7, random
        10000, 6, random
        10000, 7, random
        10000, 7, random
        10000, 6, random
        10000, 8, random
        10000, 7, random
        10000, 8, random
        10000, 7, random
        10000, 7, random
        10000, 7, random
        10000, 7, random
        778, 39, modelchecking
        114, 43, modelchecking
        102, 36, modelchecking
        292, 37, modelchecking
        204, 36, modelchecking
        292, 43, modelchecking
        102, 27, modelchecking
        302, 45, modelchecking
        98, 35, modelchecking
        10000, 7, random
        1691, 28, modelchecking
        1691, 28, modelchecking
        882, 28, modelchecking
        882, 28, modelchecking
        846, 27, modelchecking
        763, 29, modelchecking
        486, 28, modelchecking
        486, 28, modelchecking
        27, 3, mlsolver
        225, 38, mlsolver
        59, 19, mlsolver
        330, 116, mlsolver
        71, 23, mlsolver
        494, 174, mlsolver
        86, 51, mlsolver
        15, 8, mlsolver
        7, 3, equivalence
        447, 51, equivalence
        447, 51, equivalence
        532, 46, equivalence
        310, 40, equivalence
        5925, 76, equivalence
        4921, 67, equivalence
        5925, 76, equivalence
        3, 2, equivalence
        6722, 49, equivalence
        6722, 49, equivalence
        7106, 41, equivalence
        816, 47, equivalence
        16119, 81, equivalence
        16119, 81, equivalence
        13425, 72, equivalence
        614, 40, equivalence
        12133, 76, equivalence
        12133, 76, equivalence
        10121, 67, equivalence
        3, 2, equivalence
        8961, 59, equivalence
        9553, 49, equivalence
        8961, 59, equivalence
        3, 2, equivalence
        412, 47, equivalence
        7855, 81, equivalence
        7855, 81, equivalence
        6511, 72, equivalence
        3, 2, equivalence
        3457, 59, equivalence
        3761, 49, equivalence
        3457, 59, equivalence
        3, 2, equivalence
        3, 2, equivalence
        3697, 48, equivalence
        3985, 46, equivalence
        3697, 48, equivalence
        11, 3, equivalence
        1095, 51, equivalence
        1095, 51, equivalence
        1264, 46, equivalence
        3, 2, equivalence
        13409, 48, equivalence
        14177, 46, equivalence
        13409, 48, equivalence
        11, 3, equivalence
        1127, 64, equivalence
        1127, 64, equivalence
        1306, 45, equivalence
        3, 2, equivalence
        34954, 83, equivalence
        34954, 83, equivalence
        30922, 74, equivalence
        3, 2, equivalence
        3, 2, equivalence
        1883, 100, equivalence
        6124, 71, equivalence
        7383, 78, equivalence
        7383, 78, equivalence
        3759, 100, equivalence
        15111, 78, equivalence
        12590, 71, equivalence
        15111, 78, equivalence
        7, 3, equivalence
        467, 64, equivalence
        467, 64, equivalence
        558, 45, equivalence
        3, 2, equivalence
        2266, 49, equivalence
        2410, 41, equivalence
        2266, 49, equivalence
        7, 3, equivalence
        353, 49, equivalence
        423, 40, equivalence
        353, 49, equivalence
        3, 2, equivalence
        3, 2, equivalence
        3977, 100, equivalence
        20317, 83, equivalence
        16910, 76, equivalence
        20317, 83, equivalence
        1993, 100, equivalence
        8206, 76, equivalence
        9907, 83, equivalence
        9907, 83, equivalence
        3, 2, equivalence
        3, 2, equivalence
        3, 2, equivalence
        816, 47, equivalence
        19045, 77, equivalence
        22915, 86, equivalence
        22915, 86, equivalence
        3, 2, equivalence
        3, 2, equivalence
        3, 2, equivalence
        44926, 85, equivalence
        44926, 85, equivalence
        11, 3, equivalence
        845, 49, equivalence
        983, 40, equivalence
        845, 49, equivalence
        3, 2, equivalence
        3, 2, equivalence
        4743, 66, equivalence
        16162, 76, equivalence
        19423, 81, equivalence
        19423, 81, equivalence
        3, 2, equivalence
        35234, 79, equivalence
        39822, 84, equivalence
        39822, 84, equivalence
        9205, 77, equivalence
        412, 47, equivalence
        11135, 86, equivalence
        11135, 86, equivalence
        2375, 66, equivalence
        9507, 81, equivalence
        7878, 76, equivalence
        9507, 81, equivalence
        28226, 66, equivalence
        38502, 73, equivalence
        3, 2, equivalence
        3, 2, equivalence
        3, 2, equivalence
        3, 2, equivalence
        43, 5, modelchecking
        61, 5, modelchecking
        43, 5, modelchecking
        61, 5, modelchecking
        3, 2, equivalence
        3, 2, equivalence
        35388, 9, mlsolver
        9101, 113, modelchecking
        48861, 113, modelchecking
        6, 3, modelchecking
        134162, 79, modelchecking
        134162, 79, modelchecking
        40200, 203, specialcases
        20100, 103, specialcases
        40100, 103, specialcases
        20200, 203, specialcases
        40400, 403, specialcases
        10100, 103, specialcases 
      };
      \legend{}
    \end{axis}
  \end{tikzpicture}

%% file: bfs_levels_vs_diameter_orig.tex
  \begin{tikzpicture}[mark size=2.5pt,remember picture,scale=0.7]
    \begin{axis}[axis x line=bottom,
                       axis y line=left,
                       xlabel={BFS Height},
                       ylabel={Diameter},
                       xmode={log},
                       ymode={log}
                       ,
                       xmin=0,
                       ymin=0,
                       scatter/classes={
                         modelchecking={mark=x},
                         equivalence={mark=o},
                         mlsolver={mark=+},
                         specialcases={mark=square},
                         random={mark=triangle}
                       },
                       legend pos=south east,
                       legend image post style={xshift=0.5cm},
                       legend cell align=left,
                       legend style={
                         nodes={right}
                       }]
      \addplot+[black,only marks,mark=x,scatter,scatter src=explicit symbolic] 
      table[col sep=comma,x=x, y=y, meta=cluster] { 
        x, y, cluster
        2, 1, modelchecking
        96, 95, modelchecking
        27, 59, modelchecking
        30, 80, modelchecking
        24, 29, modelchecking
        30, 29, modelchecking
        30, 29, modelchecking
        29, 28, modelchecking
        29, 28, modelchecking
        34, 33, modelchecking
        30, 29, modelchecking
        34, 33, modelchecking
        29, 28, modelchecking
        48, 49, modelchecking
        47, 48, modelchecking
        47, 48, modelchecking
        48, 48, modelchecking
        55, 54, modelchecking
        55, 54, modelchecking
        48, 49, modelchecking
        47, 48, modelchecking
        30, 38, modelchecking
        36, 47, modelchecking
        192, 191, modelchecking
        384, 383, modelchecking
        24, 38, modelchecking
        25, 34, modelchecking
        21, 30, modelchecking
        21, 30, modelchecking
        31, 40, modelchecking
        22, 30, modelchecking
        29, 38, modelchecking
        23, 32, modelchecking
        24, 31, modelchecking
        29, 38, modelchecking
        23, 31, modelchecking
        21, 30, modelchecking
        25, 34, modelchecking
        23, 31, modelchecking
        29, 38, modelchecking
        31, 40, modelchecking
        21, 30, modelchecking
        21, 30, modelchecking
        22, 30, modelchecking
        23, 32, modelchecking
        24, 31, modelchecking
        29, 38, modelchecking
        21, 30, modelchecking
        25, 34, modelchecking
        23, 31, modelchecking
        24, 31, modelchecking
        21, 30, modelchecking
        21, 30, modelchecking
        31, 40, modelchecking
        29, 38, modelchecking
        29, 38, modelchecking
        22, 22, modelchecking
        23, 32, modelchecking
        21, 30, modelchecking
        49, 48, modelchecking
        48, 47, modelchecking
        49, 48, modelchecking
        55, 54, modelchecking
        55, 54, modelchecking
        48, 47, modelchecking
        48, 47, modelchecking
        51, 50, modelchecking
        768, 767, modelchecking
        28, 36, modelchecking
        39, 48, modelchecking
        27, 36, modelchecking
        37, 46, modelchecking
        29, 38, modelchecking
        31, 38, modelchecking
        30, 38, modelchecking
        36, 46, modelchecking
        31, 42, modelchecking
        71, 77, modelchecking
        70, 76, modelchecking
        70, 76, modelchecking
        83, 84, modelchecking
        70, 76, modelchecking
        71, 76, modelchecking
        83, 84, modelchecking
        71, 77, modelchecking
        31, 38, modelchecking
        30, 38, modelchecking
        36, 46, modelchecking
        31, 42, modelchecking
        27, 36, modelchecking
        37, 46, modelchecking
        39, 48, modelchecking
        28, 27, modelchecking
        29, 38, modelchecking
        30, 39, modelchecking
        27, 35, modelchecking
        31, 45, modelchecking
        29, 43, modelchecking
        26, 35, modelchecking
        29, 36, modelchecking
        34, 43, modelchecking
        28, 37, modelchecking
        28, 36, modelchecking
        16, 15, modelchecking
        16, 15, modelchecking
        9, 8, modelchecking
        9, 8, modelchecking
        22, 32, modelchecking
        26, 33, modelchecking
        31, 39, modelchecking
        27, 33, modelchecking
        28, 32, modelchecking
        25, 33, modelchecking
        24, 31, modelchecking
        32, 43, modelchecking
        33, 41, modelchecking
        23, 31, modelchecking
        24, 32, modelchecking
        27, 37, modelchecking
        27, 33, modelchecking
        31, 39, modelchecking
        26, 33, modelchecking
        24, 31, modelchecking
        28, 32, modelchecking
        25, 33, modelchecking
        32, 43, modelchecking
        24, 32, modelchecking
        23, 31, modelchecking
        33, 41, modelchecking
        28, 36, modelchecking
        34, 43, modelchecking
        30, 39, modelchecking
        26, 35, modelchecking
        29, 36, modelchecking
        27, 35, modelchecking
        29, 43, modelchecking
        31, 45, modelchecking
        28, 37, modelchecking
        46, 50, modelchecking
        48, 61, modelchecking
        50, 61, modelchecking
        50, 63, modelchecking
        45, 58, modelchecking
        45, 58, modelchecking
        45, 58, modelchecking
        47, 60, modelchecking
        64, 84, modelchecking
        27, 37, modelchecking
        28, 32, modelchecking
        33, 41, modelchecking
        23, 31, modelchecking
        24, 25, modelchecking
        31, 39, modelchecking
        24, 32, modelchecking
        27, 33, modelchecking
        26, 33, modelchecking
        32, 43, modelchecking
        25, 33, modelchecking
        27, 36, modelchecking
        32, 41, modelchecking
        34, 43, modelchecking
        23, 32, modelchecking
        24, 32, modelchecking
        25, 34, modelchecking
        26, 33, modelchecking
        25, 33, modelchecking
        31, 40, modelchecking
        258, 257, modelchecking
        514, 513, modelchecking
        27, 36, modelchecking
        34, 43, modelchecking
        25, 34, modelchecking
        24, 25, modelchecking
        23, 32, modelchecking
        26, 33, modelchecking
        25, 33, modelchecking
        31, 40, modelchecking
        32, 41, modelchecking
        17, 16, modelchecking
        9, 10, modelchecking
        18, 23, modelchecking
        16, 19, modelchecking
        17, 21, modelchecking
        15, 17, modelchecking
        31, 42, modelchecking
        27, 36, modelchecking
        39, 48, modelchecking
        28, 36, modelchecking
        31, 38, modelchecking
        37, 46, modelchecking
        29, 38, modelchecking
        30, 38, modelchecking
        36, 46, modelchecking
        14, 15, modelchecking
        27, 36, modelchecking
        26, 33, modelchecking
        24, 32, modelchecking
        25, 33, modelchecking
        25, 34, modelchecking
        31, 40, modelchecking
        23, 32, modelchecking
        32, 41, modelchecking
        34, 43, modelchecking
        34, 33, modelchecking
        31, 30, modelchecking
        28, 27, modelchecking
        13, 12, modelchecking
        9, 10, modelchecking
        65, 70, modelchecking
        132, 133, mlsolver
        38, 41, mlsolver
        22, 25, mlsolver
        77, 88, mlsolver
        228, 229, mlsolver
        62, 65, mlsolver
        101, 101, mlsolver
        31, 30, mlsolver
        8, 7, mlsolver
        21, 20, mlsolver
        54, 57, mlsolver
        196, 197, mlsolver
        46, 49, mlsolver
        164, 165, mlsolver
        70, 73, mlsolver
        260, 261, mlsolver
        38, 38, mlsolver
        173, 172, mlsolver
        26, 25, mlsolver
        105, 104, mlsolver
        292, 293, mlsolver
        78, 81, mlsolver
        52, 88, mlsolver
        4, 3, mlsolver
        18, 17, mlsolver
        64, 63, mlsolver
        26, 33, mlsolver
        4, 3, mlsolver
        88, 173, mlsolver
        4, 3, mlsolver
        4, 3, mlsolver
        332, 813, mlsolver
        256, 608, mlsolver
        4, 3, mlsolver
        4, 3, mlsolver
        134, 288, mlsolver
        190, 433, mlsolver
        4, 3, mlsolver
        4, 3, mlsolver
        74, 158, mlsolver
        42, 77, mlsolver
        4, 3, mlsolver
        230, 581, mlsolver
        4, 3, mlsolver
        168, 410, mlsolver
        4, 3, mlsolver
        20, 26, mlsolver
        4, 3, mlsolver
        116, 269, mlsolver
        4, 3, mlsolver
        50, 49, mlsolver
        244, 243, mlsolver
        17, 18, mlsolver
        4, 3, mlsolver
        4, 3, mlsolver
        418, 1048, mlsolver
        384, 1013, mlsolver
        4, 3, mlsolver
        27, 32, mlsolver
        4, 3, mlsolver
        4, 3, mlsolver
        302, 782, mlsolver
        4, 3, mlsolver
        21, 26, mlsolver
        4, 3, mlsolver
        39, 44, mlsolver
        71, 70, mlsolver
        16, 15, mlsolver
        45, 50, mlsolver
        4, 3, mlsolver
        4, 3, mlsolver
        51, 56, mlsolver
        57, 62, mlsolver
        4, 3, mlsolver
        37, 36, mlsolver
        12, 11, mlsolver
        39, 38, mlsolver
        8, 7, mlsolver
        527, 526, mlsolver
        40, 39, mlsolver
        421, 420, mlsolver
        36, 35, mlsolver
        327, 326, mlsolver
        32, 31, mlsolver
        28, 27, mlsolver
        245, 244, mlsolver
        126, 141, mlsolver
        12, 11, mlsolver
        11, 10, mlsolver
        75, 74, mlsolver
        94, 120, mlsolver
        10, 12, mlsolver
        5, 4, mlsolver
        15, 14, mlsolver
        56, 64, mlsolver
        6, 6, mlsolver
        8, 10, mlsolver
        79, 100, mlsolver
        14, 16, mlsolver
        143, 182, mlsolver
        126, 161, mlsolver
        12, 14, mlsolver
        9, 9, mlsolver
        161, 174, mlsolver
        16, 18, mlsolver
        18, 20, mlsolver
        9, 9, mlsolver
        197, 218, mlsolver
        7, 6, mlsolver
        84, 98, mlsolver
        5, 4, mlsolver
        15, 14, mlsolver
        18, 17, mlsolver
        131, 130, mlsolver
        12, 14, mlsolver
        9, 9, mlsolver
        157, 156, mlsolver
        12, 11, mlsolver
        97, 96, mlsolver
        10, 9, mlsolver
        5, 4, mlsolver
        11, 10, mlsolver
        231, 230, mlsolver
        14, 13, mlsolver
        16, 15, mlsolver
        320, 319, mlsolver
        18, 17, mlsolver
        423, 422, mlsolver
        13, 12, mlsolver
        213, 212, mlsolver
        5, 4, mlsolver
        15, 15, mlsolver
        9, 8, mlsolver
        64, 71, mlsolver
        59, 64, mlsolver
        9, 8, mlsolver
        538, 537, mlsolver
        20, 19, mlsolver
        13, 13, mlsolver
        5, 4, mlsolver
        11, 10, mlsolver
        129, 128, mlsolver
        579, 578, mlsolver
        19, 18, mlsolver
        17, 16, mlsolver
        438, 437, mlsolver
        315, 314, mlsolver
        15, 14, mlsolver
        21, 20, mlsolver
        740, 739, mlsolver
        17, 16, mlsolver
        180, 190, mlsolver
        9, 9, mlsolver
        13, 13, mlsolver
        1222, 1242, mlsolver
        65, 64, mlsolver
        257, 256, mlsolver
        129, 128, mlsolver
        33, 32, mlsolver
        485, 500, mlsolver
        2, 1, specialcases
        2, 1, specialcases
        101, 100, specialcases
        201, 200, specialcases
        501, 500, specialcases
        1001, 1000, specialcases
        2001, 2000, specialcases
        5001, 5000, specialcases
        2, 1, specialcases
        20001, 20000, specialcases
        10001, 10000, specialcases
        2, 1, specialcases
        201, 201, specialcases
        401, 401, specialcases
        1001, 1001, specialcases
        2001, 2001, specialcases
        4001, 4001, specialcases
        2, 1, specialcases
        10001, 10001, specialcases
        20001, 20001, specialcases
        6, 7, random
        6, 8, random
        7, 9, random
        6, 7, random
        7, 8, random
        104, 104, specialcases
        504, 504, specialcases
        204, 204, specialcases
        1004, 1004, specialcases
        7, 9, random
        2004, 2004, specialcases
        5, 7, random
        40001, 40001, specialcases
        7, 8, random
        8, 10, random
        5004, 5004, specialcases
        5, 7, random
        6, 10, random
        7, 9, random
        10004, 10004, specialcases
        5, 7, random
        6, 8, random
        6, 9, random
        7, 9, random
        6, 7, random
        6, 8, random
        7, 8, random
        7, 10, random
        7, 7, random
        7, 9, random
        7, 9, random
        6, 8, random
        6, 7, random
        7, 8, random
        8, 10, random
        7, 8, random
        7, 9, random
        5, 6, random
        7, 9, random
        6, 9, random
        6, 6, random
        7, 8, random
        8, 9, random
        7, 8, random
        7, 9, random
        6, 6, random
        7, 9, random
        7, 10, random
        6, 7, random
        7, 9, random
        6, 8, random
        7, 10, random
        8, 9, random
        5, 7, random
        6, 9, random
        7, 9, random
        5, 7, random
        7, 9, random
        6, 8, random
        7, 10, random
        5, 7, random
        7, 8, random
        6, 8, random
        7, 9, random
        5, 7, random
        7, 10, random
        6, 8, random
        8, 9, random
        7, 9, random
        5, 7, random
        6, 9, random
        8, 9, random
        5, 8, random
        7, 10, random
        6, 8, random
        7, 9, random
        5, 7, random
        7, 9, random
        7, 9, random
        7, 10, random
        5, 7, random
        7, 9, random
        7, 9, random
        6, 9, random
        5, 7, random
        6, 8, random
        7, 10, random
        6, 6, random
        7, 9, random
        7, 9, random
        6, 9, random
        5, 6, random
        8, 9, random
        7, 9, random
        7, 9, random
        7, 9, random
        6, 7, random
        9, 9, random
        9, 9, random
        7, 8, random
        6, 7, random
        7, 9, random
        7, 10, random
        7, 9, random
        7, 9, random
        6, 6, random
        7, 10, random
        7, 7, random
        6, 7, random
        7, 7, random
        7, 7, random
        7, 8, random
        6, 7, random
        7, 7, random
        6, 7, random
        6, 7, random
        7, 8, random
        7, 8, random
        7, 8, random
        6, 7, random
        7, 10, random
        6, 6, random
        6, 7, random
        6, 6, random
        6, 7, random
        6, 7, random
        7, 7, random
        6, 6, random
        6, 5, random
        6, 6, random
        7, 7, random
        7, 7, random
        7, 8, random
        6, 6, random
        7, 9, random
        6, 7, random
        7, 7, random
        7, 7, random
        7, 7, random
        7, 7, random
        7, 7, random
        7, 7, random
        7, 8, random
        6, 6, random
        6, 6, random
        7, 7, random
        7, 8, random
        7, 7, random
        6, 6, random
        6, 7, random
        7, 7, random
        6, 7, random
        7, 7, random
        6, 6, random
        6, 7, random
        6, 6, random
        7, 8, random
        7, 7, random
        7, 7, random
        7, 7, random
        7, 8, random
        8, 8, random
        7, 7, random
        7, 7, random
        7, 7, random
        6, 7, random
        7, 7, random
        6, 6, random
        7, 7, random
        7, 8, random
        7, 7, random
        6, 7, random
        8, 8, random
        7, 7, random
        7, 8, random
        7, 7, random
        7, 8, random
        7, 7, random
        7, 8, random
        7, 7, random
        8, 8, random
        7, 8, random
        7, 7, random
        7, 7, random
        7, 8, random
        7, 8, random
        7, 8, random
        6, 7, random
        7, 7, random
        6, 6, random
        6, 6, random
        6, 7, random
        7, 8, random
        7, 8, random
        7, 7, random
        6, 6, random
        7, 7, random
        7, 7, random
        6, 6, random
        7, 8, random
        7, 7, random
        7, 8, random
        7, 7, random
        7, 7, random
        7, 7, random
        7, 7, random
        30, 39, modelchecking
        34, 43, modelchecking
        28, 36, modelchecking
        28, 37, modelchecking
        29, 36, modelchecking
        29, 43, modelchecking
        27, 27, modelchecking
        31, 45, modelchecking
        26, 35, modelchecking
        7, 7, random
        29, 28, modelchecking
        29, 28, modelchecking
        29, 28, modelchecking
        29, 28, modelchecking
        28, 27, modelchecking
        30, 29, modelchecking
        29, 28, modelchecking
        29, 28, modelchecking
        4, 3, mlsolver
        33, 38, mlsolver
        20, 19, mlsolver
        117, 116, mlsolver
        24, 23, mlsolver
        175, 174, mlsolver
        49, 51, mlsolver
        9, 8, mlsolver
        4, 3, equivalence
        40, 51, equivalence
        40, 51, equivalence
        28, 46, equivalence
        26, 40, equivalence
        51, 76, equivalence
        43, 67, equivalence
        51, 76, equivalence
        3, 2, equivalence
        38, 49, equivalence
        38, 49, equivalence
        29, 41, equivalence
        33, 47, equivalence
        56, 81, equivalence
        56, 81, equivalence
        49, 72, equivalence
        26, 40, equivalence
        51, 76, equivalence
        51, 76, equivalence
        43, 67, equivalence
        3, 2, equivalence
        46, 59, equivalence
        33, 49, equivalence
        46, 59, equivalence
        3, 2, equivalence
        33, 47, equivalence
        56, 81, equivalence
        56, 81, equivalence
        49, 72, equivalence
        3, 2, equivalence
        46, 59, equivalence
        33, 49, equivalence
        46, 59, equivalence
        3, 2, equivalence
        3, 2, equivalence
        34, 48, equivalence
        33, 46, equivalence
        34, 48, equivalence
        4, 3, equivalence
        40, 51, equivalence
        40, 51, equivalence
        28, 46, equivalence
        3, 2, equivalence
        34, 48, equivalence
        33, 46, equivalence
        34, 48, equivalence
        4, 3, equivalence
        53, 64, equivalence
        53, 64, equivalence
        31, 45, equivalence
        3, 2, equivalence
        60, 83, equivalence
        60, 83, equivalence
        49, 74, equivalence
        3, 2, equivalence
        3, 2, equivalence
        58, 100, equivalence
        47, 71, equivalence
        58, 78, equivalence
        58, 78, equivalence
        58, 100, equivalence
        58, 78, equivalence
        47, 71, equivalence
        58, 78, equivalence
        4, 3, equivalence
        53, 64, equivalence
        53, 64, equivalence
        31, 45, equivalence
        3, 2, equivalence
        38, 49, equivalence
        29, 41, equivalence
        38, 49, equivalence
        4, 3, equivalence
        38, 49, equivalence
        26, 40, equivalence
        38, 49, equivalence
        3, 2, equivalence
        3, 2, equivalence
        58, 100, equivalence
        60, 83, equivalence
        51, 76, equivalence
        60, 83, equivalence
        58, 100, equivalence
        51, 76, equivalence
        60, 83, equivalence
        60, 83, equivalence
        3, 2, equivalence
        3, 2, equivalence
        3, 2, equivalence
        33, 47, equivalence
        53, 77, equivalence
        61, 86, equivalence
        61, 86, equivalence
        3, 2, equivalence
        3, 2, equivalence
        3, 2, equivalence
        62, 85, equivalence
        62, 85, equivalence
        4, 3, equivalence
        38, 49, equivalence
        26, 40, equivalence
        38, 49, equivalence
        3, 2, equivalence
        3, 2, equivalence
        39, 66, equivalence
        49, 76, equivalence
        59, 81, equivalence
        59, 81, equivalence
        3, 2, equivalence
        51, 79, equivalence
        63, 84, equivalence
        63, 84, equivalence
        53, 77, equivalence
        33, 47, equivalence
        61, 86, equivalence
        61, 86, equivalence
        39, 66, equivalence
        59, 81, equivalence
        49, 76, equivalence
        59, 81, equivalence
        45, 66, equivalence
        51, 73, equivalence
        3, 2, equivalence
        3, 2, equivalence
        3, 2, equivalence
        3, 2, equivalence
        6, 5, modelchecking
        6, 5, modelchecking
        6, 5, modelchecking
        6, 5, modelchecking
        3, 2, equivalence
        3, 2, equivalence
        9, 9, mlsolver
        98, 113, modelchecking
        98, 113, modelchecking
        4, 3, modelchecking
        80, 79, modelchecking
        80, 79, modelchecking
        202, 203, specialcases
        102, 103, specialcases
        102, 103, specialcases
        202, 203, specialcases
        402, 403, specialcases
        102, 103, specialcases 
      };
      \legend{~modelchecking, ~equivalence, ~mlsolver, ~specialcases, ~random}
    \end{axis}
  \end{tikzpicture}

%% file: vertices_vs_avg_3_neighbourhood_orig.tex
  \begin{tikzpicture}[mark size=2.5pt,remember picture,scale=0.7]
    \begin{axis}[axis x line=bottom,
                       axis y line=left,
                       xlabel={Vertices},
                       ylabel={Avg 3-neighbourhood},
                       xmode={log},
                       ymode={log}
                       ,
                       xmin=0,
                       ymin=0,
                       scatter/classes={
                         modelchecking={mark=x},
                         equivalence={mark=o},
                         mlsolver={mark=+},
                         specialcases={mark=square},
                         random={mark=triangle}
                       },
                       legend pos=outer north east,
                       legend image post style={xshift=0.5cm},
                       legend cell align=left,
                       legend style={
                         nodes={right}
                       }]
      \addplot+[black,only marks,mark=x,scatter,scatter src=explicit symbolic] 
      table[col sep=comma,x=x, y=y, meta=cluster] { 
        x, y, cluster
        2, 0.5, modelchecking
        972, 6.47839506173, modelchecking
        2688, 7.40029761905, modelchecking
        15684, 7.32587350166, modelchecking
        588, 8.7193877551, modelchecking
        111456, 14.2523148148, modelchecking
        656, 7.88567073171, modelchecking
        308, 6.12012987013, modelchecking
        227, 5.72246696035, modelchecking
        328, 6.23475609756, modelchecking
        649, 6.38674884438, modelchecking
        656, 7.88567073171, modelchecking
        649, 6.38674884438, modelchecking
        227, 5.72246696035, modelchecking
        8626, 11.9291676327, modelchecking
        2126, 9.5305738476, modelchecking
        4313, 10.8678414097, modelchecking
        4021, 10.9510072121, modelchecking
        6359, 9.03632646643, modelchecking
        6359, 9.03632646643, modelchecking
        8626, 11.9291676327, modelchecking
        2126, 9.5305738476, modelchecking
        2832, 10.4774011299, modelchecking
        16356, 12.3464783566, modelchecking
        2916, 6.49279835391, modelchecking
        8748, 6.4975994513, modelchecking
        78732, 6.49973327237, modelchecking
        564, 7.34219858156, modelchecking
        876780, 16.1670099683, modelchecking
        2338, 6.20230966638, modelchecking
        147, 4.95918367347, modelchecking
        260, 4.89615384615, modelchecking
        747, 5.07764390897, modelchecking
        442, 4.8257918552, modelchecking
        731, 4.94801641587, modelchecking
        439, 6.62642369021, modelchecking
        598, 4.97157190635, modelchecking
        173, 4.68208092486, modelchecking
        150, 4.94, modelchecking
        236, 5.02118644068, modelchecking
        9282, 6.84776987718, modelchecking
        294, 5.66326530612, modelchecking
        341, 5.29618768328, modelchecking
        2643, 5.77563374953, modelchecking
        291, 5.66666666667, modelchecking
        516, 5.67635658915, modelchecking
        2034, 5.54228121927, modelchecking
        871, 7.25717566016, modelchecking
        2346, 5.68499573743, modelchecking
        2611, 5.63806970509, modelchecking
        468, 5.88247863248, modelchecking
        594, 5.82323232323, modelchecking
        78, 4.5, modelchecking
        156, 4.53205128205, modelchecking
        132, 4.46212121212, modelchecking
        75, 4.54666666667, modelchecking
        231, 4.67965367965, modelchecking
        89, 4.31460674157, modelchecking
        223, 4.56053811659, modelchecking
        78, 4.29487179487, modelchecking
        223, 6.25560538117, modelchecking
        120, 4.54166666667, modelchecking
        63907, 28.462234184, modelchecking
        31954, 28.7320523252, modelchecking
        63907, 28.462234184, modelchecking
        13222, 11.1575404629, modelchecking
        13222, 11.1575404629, modelchecking
        11977, 19.9194289054, modelchecking
        11977, 19.9194289054, modelchecking
        26996, 26.6053489406, modelchecking
        236196, 6.49991109079, modelchecking
        108336, 7.24597548368, modelchecking
        26244, 6.4991998171, modelchecking
        2946, 18.9467073999, modelchecking
        5649, 20.1933085502, modelchecking
        1041, 19.8386167147, modelchecking
        5201, 19.8438761777, modelchecking
        3121, 22.4206984941, modelchecking
        4486, 19.3680338832, modelchecking
        1122, 19.3573975045, modelchecking
        1619, 16.6856084002, modelchecking
        16642, 20.5355726475, modelchecking
        708588, 6.4999703636, modelchecking
        107276, 16.7234050487, modelchecking
        18625, 14.2500402685, modelchecking
        18625, 14.2500402685, modelchecking
        44773, 12.100708016, modelchecking
        53638, 17.1084119468, modelchecking
        51220, 17.2865872706, modelchecking
        44773, 12.100708016, modelchecking
        107276, 16.7234050487, modelchecking
        1028, 17.2947470817, modelchecking
        514, 17.2898832685, modelchecking
        691, 15.2561505065, modelchecking
        3714, 18.5075390415, modelchecking
        465, 17.8193548387, modelchecking
        1393, 17.8348887294, modelchecking
        1553, 18.1352221507, modelchecking
        450, 16.2955555556, modelchecking
        1393, 20.536252692, modelchecking
        12130, 7.03520197857, modelchecking
        2658, 5.69488337096, modelchecking
        3452, 5.99565469293, modelchecking
        3412, 5.83382180539, modelchecking
        380, 5.85526315789, modelchecking
        3066, 5.86073059361, modelchecking
        438, 5.49315068493, modelchecking
        1138, 7.44112478032, modelchecking
        384, 5.84375, modelchecking
        564914, 39.2768509897, modelchecking
        564914, 39.3851966848, modelchecking
        4860, 4.04897119342, modelchecking
        4860, 3.85, modelchecking
        2229, 16.4993270525, modelchecking
        2229, 16.5118887393, modelchecking
        5861, 17.8215321617, modelchecking
        2245634, 29.7177923918, modelchecking
        512018, 27.8643856271, modelchecking
        653574, 28.3818649457, modelchecking
        163394, 28.3817643243, modelchecking
        858114, 28.297432509, modelchecking
        388418, 21.0176742581, modelchecking
        140353, 30.4797118694, modelchecking
        211955, 29.3506451841, modelchecking
        869569, 29.7296281261, modelchecking
        353089, 18.87615587, modelchecking
        421057, 31.9761504974, modelchecking
        147458, 19.7688901246, modelchecking
        4898, 18.2290730911, modelchecking
        41762, 18.5934342225, modelchecking
        39178, 18.2322987391, modelchecking
        12818, 17.0837104072, modelchecking
        13825, 21.5822061483, modelchecking
        28964, 18.476315426, modelchecking
        10369, 13.9217860932, modelchecking
        45937, 18.5813396608, modelchecking
        4609, 18.6580603168, modelchecking
        9986, 14.087222111, modelchecking
        21506, 18.8533897517, modelchecking
        5958, 16.9776770728, modelchecking
        6866, 17.4067870667, modelchecking
        1490, 16.9704697987, modelchecking
        3344, 16.9844497608, modelchecking
        3530, 16.3045325779, modelchecking
        4033, 20.5440119018, modelchecking
        2977, 13.3459858918, modelchecking
        2690, 13.812267658, modelchecking
        1345, 17.575464684, modelchecking
        7897, 17.5827529442, modelchecking
        196, 5.02040816327, modelchecking
        222, 4.78378378378, modelchecking
        3058, 6.29332897319, modelchecking
        192, 5.046875, modelchecking
        782, 5.04475703325, modelchecking
        578, 4.87370242215, modelchecking
        956, 5.03870292887, modelchecking
        976, 5.19364754098, modelchecking
        574, 6.71777003484, modelchecking
        112514, 27.4663686297, modelchecking
        3949, 23.9549252975, modelchecking
        17810, 25.0052779337, modelchecking
        35620, 25.0054183043, modelchecking
        54322, 24.5213541475, modelchecking
        14065, 27.5894063278, modelchecking
        32402, 20.4201901117, modelchecking
        42618, 26.3935191703, modelchecking
        42193, 29.4123195791, modelchecking
        34673, 17.7291552505, modelchecking
        56241, 27.0272399139, modelchecking
        3458, 17.6283979179, modelchecking
        1046, 15.1003824092, modelchecking
        1621, 16.337446021, modelchecking
        433, 16.1224018476, modelchecking
        326, 14.4171779141, modelchecking
        1370, 15.7773722628, modelchecking
        770, 13.3545454545, modelchecking
        1012, 15.4318181818, modelchecking
        506, 15.4268774704, modelchecking
        937, 12.545357524, modelchecking
        1297, 19.1565150347, modelchecking
        11362, 6.92140468227, modelchecking
        3188, 5.71675031368, modelchecking
        3228, 5.81691449814, modelchecking
        356, 5.77808988764, modelchecking
        2490, 5.67429718876, modelchecking
        1066, 7.39587242026, modelchecking
        2874, 5.78462073765, modelchecking
        360, 5.76666666667, modelchecking
        414, 5.40579710145, modelchecking
        6563, 12.3239372238, modelchecking
        19685, 12.3302006604, modelchecking
        730, 6.0397260274, modelchecking
        282, 4.90425531915, modelchecking
        274, 6.52554744526, modelchecking
        96, 4.55208333333, modelchecking
        92, 4.8152173913, modelchecking
        192, 4.77604166667, modelchecking
        96, 4.75, modelchecking
        108, 4.55555555556, modelchecking
        272, 4.84191176471, modelchecking
        43, 10.0930232558, modelchecking
        13, 5.53846153846, modelchecking
        175, 13.1085714286, modelchecking
        97, 12.1855670103, modelchecking
        133, 12.7218045113, modelchecking
        67, 11.4029850746, modelchecking
        82434, 23.2162578548, modelchecking
        2577, 22.454016298, modelchecking
        24593, 22.7381775302, modelchecking
        17154, 21.9238078582, modelchecking
        21770, 21.9991272393, modelchecking
        23185, 22.4531378046, modelchecking
        7729, 24.8506921982, modelchecking
        2722, 21.9922850845, modelchecking
        4243, 18.384633514, modelchecking
        43, 10.1860465116, modelchecking
        2866, 6.37857641312, modelchecking
        734, 5.17847411444, modelchecking
        542, 5.06457564576, modelchecking
        184, 5.15217391304, modelchecking
        538, 6.8624535316, modelchecking
        210, 4.88571428571, modelchecking
        180, 5.18333333333, modelchecking
        892, 5.15807174888, modelchecking
        912, 5.2423245614, modelchecking
        17773, 76.2781747595, modelchecking
        59051, 12.3322890383, modelchecking
        2395, 41.3962421712, modelchecking
        343, 21.0524781341, modelchecking
        177149, 12.3329852271, modelchecking
        926213, 44.056811986, modelchecking
        3471554, 39.8455930687, modelchecking
        607753, 29.486036268, modelchecking
        867889, 63.6911989897, modelchecking
        308738, 44.0567568618, modelchecking
        638065, 44.2555397961, modelchecking
        579745, 44.6813944062, modelchecking
        289297, 45.5939086821, modelchecking
        1350433, 43.8426156648, modelchecking
        1191962, 45.0623652432, modelchecking
        1278433, 50.0891114356, modelchecking
        474914, 44.0644579861, modelchecking
        579745, 44.6813944062, modelchecking
        25, 8.24, modelchecking
        13, 5.53846153846, modelchecking
        655362, 38.0068374425, modelchecking
        170753, 28.0609476847, modelchecking
        297089, 40.9017365167, modelchecking
        177668, 41.1593815431, modelchecking
        81921, 42.8794326241, modelchecking
        88834, 41.1593534007, modelchecking
        245761, 60.5456439386, modelchecking
        153985, 42.0700717602, modelchecking
        258050, 42.1077465607, modelchecking
        164353, 41.8609213096, modelchecking
        153985, 42.0700717602, modelchecking
        65506, 40.2853326413, modelchecking
        185089, 46.0806098688, modelchecking
        540737, 9.72600173467, modelchecking
        1081474, 9.72682745956, modelchecking
        1093761, 9.67847180508, modelchecking
        1787138, 34.6111285195, modelchecking
        22873, 31.3456914266, modelchecking
        742970, 35.8406328654, modelchecking
        223393, 36.3982622553, modelchecking
        294434, 31.7835032639, modelchecking
        588868, 31.7835117548, modelchecking
        944090, 31.2177959728, modelchecking
        586658, 25.2021296906, modelchecking
        576385, 21.161860562, modelchecking
        670177, 36.9027077921, modelchecking
        917713, 34.6946627105, modelchecking
        1559, 4.11161000641, mlsolver
        636, 9.24842767296, mlsolver
        298, 9.72818791946, mlsolver
        693, 4.17604617605, mlsolver
        2858, 4.08817354794, mlsolver
        1143, 8.98075240595, mlsolver
        1126, 4.13143872114, mlsolver
        467, 9.46680942184, mlsolver
        12, 3.83333333333, mlsolver
        22, 3.40909090909, mlsolver
        974, 9.03901437372, mlsolver
        2425, 4.09319587629, mlsolver
        805, 9.12173913043, mlsolver
        1992, 4.10040160643, mlsolver
        1312, 8.9375, mlsolver
        3291, 4.08447280462, mlsolver
        6817, 15.7570778935, mlsolver
        14053, 4.45442254323, mlsolver
        1562, 14.5243277849, mlsolver
        3029, 4.46384945527, mlsolver
        3724, 4.08163265306, mlsolver
        1481, 8.90411883862, mlsolver
        336, 3.64880952381, mlsolver
        15, 7.66666666667, mlsolver
        191, 11.6020942408, mlsolver
        365, 4.35616438356, mlsolver
        76, 3.80263157895, mlsolver
        9, 3.88888888889, mlsolver
        912, 3.5076754386, mlsolver
        21, 11.5714285714, mlsolver
        45, 27.4444444444, mlsolver
        8776, 3.25922971741, mlsolver
        5736, 3.29567642957, mlsolver
        39, 23.4615384615, mlsolver
        27, 15.5185185185, mlsolver
        1924, 3.4106029106, mlsolver
        3492, 3.34392898053, mlsolver
        33, 19.4848484848, mlsolver
        9, 3.88888888889, mlsolver
        319, 3.26018808777, mlsolver
        160, 3.39375, mlsolver
        9, 3.88888888889, mlsolver
        1156, 3.11851211073, mlsolver
        9, 3.88888888889, mlsolver
        817, 3.14565483476, mlsolver
        9, 3.88888888889, mlsolver
        61, 3.77049180328, mlsolver
        9, 3.88888888889, mlsolver
        538, 3.18773234201, mlsolver
        9, 3.88888888889, mlsolver
        29335, 16.3516959264, mlsolver
        62707, 4.44424067488, mlsolver
        30, 4.3, mlsolver
        9, 3.88888888889, mlsolver
        51, 31.431372549, mlsolver
        12732, 3.23075714735, mlsolver
        2014, 3.08589870904, mlsolver
        9, 3.88888888889, mlsolver
        140, 5.20714285714, mlsolver
        21, 11.5714285714, mlsolver
        9, 3.88888888889, mlsolver
        1555, 3.09967845659, mlsolver
        15, 7.66666666667, mlsolver
        75, 4.92, mlsolver
        33, 19.4848484848, mlsolver
        330, 5.48181818182, mlsolver
        200, 3.275, mlsolver
        47, 4.25531914894, mlsolver
        455, 5.55824175824, mlsolver
        39, 23.4615384615, mlsolver
        45, 27.4444444444, mlsolver
        600, 5.615, mlsolver
        765, 5.65882352941, mlsolver
        51, 31.431372549, mlsolver
        105, 3.34285714286, mlsolver
        34, 4.32352941176, mlsolver
        63, 3.63492063492, mlsolver
        19, 4, mlsolver
        1490, 3.16979865772, mlsolver
        119, 4.40336134454, mlsolver
        1190, 3.17647058824, mlsolver
        107, 4.39252336449, mlsolver
        924, 3.18506493506, mlsolver
        95, 4.37894736842, mlsolver
        83, 4.36144578313, mlsolver
        692, 3.19653179191, mlsolver
        33969, 3.94721657982, mlsolver
        6182, 37.6239081203, mlsolver
        456, 13.4407894737, mlsolver
        2497, 3.96876251502, mlsolver
        6431, 4.68667392318, mlsolver
        730, 24.3164383562, mlsolver
        8, 2.875, mlsolver
        16, 3.3125, mlsolver
        543, 4.49907918969, mlsolver
        69, 14.0144927536, mlsolver
        348, 36.0890804598, mlsolver
        3772, 4.96924708378, mlsolver
        6346, 37.0520012606, mlsolver
        60011, 4.74289713553, mlsolver
        38075, 5.01087327643, mlsolver
        3195, 59.2078247261, mlsolver
        110328, 16.7313827859, mlsolver
        259455, 4.41634194754, mlsolver
        1264, 86.7816455696, mlsolver
        36735, 5.45313733497, mlsolver
        25971, 91.5036001694, mlsolver
        325637, 5.02396533563, mlsolver
        49830, 55.3956050572, mlsolver
        492107, 4.76515066845, mlsolver
        2861, 156.06571129, mlsolver
        94631, 5.38083714639, mlsolver
        19, 7.89473684211, mlsolver
        354, 4.00564971751, mlsolver
        8, 2.875, mlsolver
        16, 3.3125, mlsolver
        6831, 35.2241253111, mlsolver
        33977, 4.960502693, mlsolver
        9585, 248.828273344, mlsolver
        174667, 5.41644386175, mlsolver
        6650, 322.094135338, mlsolver
        234577, 5.36419597829, mlsolver
        970, 4.25670103093, mlsolver
        65, 4.10769230769, mlsolver
        334, 4.02395209581, mlsolver
        30, 4.1, mlsolver
        7, 3.71428571429, mlsolver
        14, 3.21428571429, mlsolver
        2443, 4.33483422022, mlsolver
        148, 4.0472972973, mlsolver
        332, 4.09939759036, mlsolver
        5837, 4.36457084119, mlsolver
        778, 4.07583547558, mlsolver
        13687, 4.35625045664, mlsolver
        89, 5.84269662921, mlsolver
        2265, 3.93465783664, mlsolver
        8, 3.5, mlsolver
        34, 3.5, mlsolver
        17, 5, mlsolver
        250, 3.8, mlsolver
        203, 4.26108374384, mlsolver
        27, 4.37037037037, mlsolver
        32233, 4.37182390718, mlsolver
        1794, 4.08528428094, mlsolver
        17, 3.41176470588, mlsolver
        9, 4, mlsolver
        38, 5.55263157895, mlsolver
        795, 3.88301886792, mlsolver
        35037, 3.97905071781, mlsolver
        1123, 6.15939447907, mlsolver
        478, 6.11087866109, mlsolver
        14399, 3.97263698868, mlsolver
        5914, 3.96533648969, mlsolver
        209, 5.97607655502, mlsolver
        2611, 6.17426273458, mlsolver
        82968, 3.97220615177, mlsolver
        105, 4.26666666667, mlsolver
        1546, 4.65006468305, mlsolver
        15427, 709.426200817, mlsolver
        567753, 5.33321884693, mlsolver
        5596, 362.117762688, mlsolver
        140309, 3.96219772074, mlsolver
        44028, 4.88359680204, mlsolver
        2163, 4.06287563569, mlsolver
        48211, 4.01178154363, mlsolver
        1031612, 4.92754058696, mlsolver
        10183, 4.02749680841, mlsolver
        214775, 4.9141054592, mlsolver
        228787, 4.00500028411, mlsolver
        4924413, 4.93349258074, mlsolver
        467, 4.13704496788, mlsolver
        8691, 4.81475089173, mlsolver
        100, 99, specialcases
        200, 199, specialcases
        200, 6, specialcases
        400, 6, specialcases
        1000, 6, specialcases
        2000, 6, specialcases
        4000, 6, specialcases
        10000, 6, specialcases
        40000, 6, specialcases
        20000, 6, specialcases
        500, 499, specialcases
        301, 4.32558139535, specialcases
        601, 4.32945091514, specialcases
        1501, 4.33177881412, specialcases
        100000, 6, specialcases
        3001, 4.33255581473, specialcases
        6001, 4.33294450925, specialcases
        1000, 999, specialcases
        15001, 4.33317778815, specialcases
        30001, 4.33325555815, specialcases
        1000, 616.288, random
        5000, 1040.6814, random
        10000, 1157.7516, random
        1000, 638.588, random
        5000, 1059.416, random
        500, 12.078, specialcases
        2500, 12.1756, specialcases
        1000, 12.139, specialcases
        5000, 12.1878, specialcases
        10000, 1170.8897, random
        10000, 12.1939, specialcases
        1000, 632.18, random
        60001, 4.33329444509, specialcases
        5000, 1068.2502, random
        20000, 1191.25425, random
        25000, 12.19756, specialcases
        1000, 657.205, random
        10000, 1150.3214, random
        5000, 1047.2814, random
        50000, 12.19878, specialcases
        1000, 656.416, random
        5000, 1087.2992, random
        10000, 1146.2152, random
        20000, 1208.66295, random
        1000, 646.1, random
        10000, 1175.9571, random
        50000, 1262.95928, random
        5000, 1054.5598, random
        20000, 1215.062, random
        1000, 615.443, random
        10000, 1201.5702, random
        50000, 1269.84456, random
        20000, 1222.04695, random
        5000, 1095.5076, random
        1000, 606.67, random
        10000, 1154.5732, random
        20000, 1221.98175, random
        5000, 1071.0144, random
        50000, 1254.08354, random
        10000, 1150.2535, random
        1000, 643.625, random
        20000, 1184.16135, random
        5000, 1088.6994, random
        50000, 1254.06, random
        1000, 611.475, random
        50000, 1257.59272, random
        5000, 1047.3436, random
        20000, 1223.64275, random
        10000, 1181.1945, random
        10000, 1171.2802, random
        1000, 623.747, random
        5000, 1055.905, random
        50000, 1255.08738, random
        20000, 1212.4384, random
        1000, 618.88, random
        10000, 1170.0699, random
        5000, 1052.037, random
        20000, 1196.14795, random
        50000, 1260.90418, random
        10000, 1194.2413, random
        1000, 607.004, random
        5000, 1091.2376, random
        50000, 1262.2042, random
        20000, 1214.15805, random
        1000, 621.773, random
        10000, 1162.0385, random
        5000, 1051.442, random
        50000, 1257.0844, random
        20000, 1238.8184, random
        1000, 625.822, random
        10000, 1210.2641, random
        5000, 1075.6788, random
        50000, 1258.32664, random
        20000, 1237.85315, random
        1000, 641.976, random
        5000, 1069.8648, random
        10000, 1177.3784, random
        50000, 1265.94872, random
        10000, 1160.7261, random
        20000, 1231.79865, random
        1000, 641.063, random
        50000, 1237.42708, random
        5000, 1092.607, random
        20000, 1215.59835, random
        1000, 647.75, random
        50000, 1272.81766, random
        10000, 1142.2885, random
        5000, 1083.4614, random
        20000, 1251.23875, random
        1000, 630.39, random
        5000, 1083.5702, random
        10000, 1135.527, random
        50000, 1253.64846, random
        20000, 1230.13185, random
        1000, 630.043, random
        50000, 1267.77052, random
        10000, 1173.2941, random
        20000, 1218.03415, random
        5000, 1089.717, random
        1000, 649.112, random
        5000, 1075.1758, random
        10000, 1178.0433, random
        1000, 632.733, random
        50000, 1257.08418, random
        20000, 1225.9263, random
        10000, 1175.9961, random
        5000, 1074.7128, random
        50000, 1248.49752, random
        1000, 653.063, random
        20000, 1210.214, random
        10000, 1181.9517, random
        5000, 1053.6582, random
        50000, 1272.92706, random
        20000, 1198.554, random
        1000, 626.818, random
        10000, 1159.3715, random
        20000, 1229.86735, random
        50000, 1268.42054, random
        5000, 1064.5738, random
        1000, 619.043, random
        10000, 1195.4883, random
        5000, 1081.1154, random
        50000, 1261.9623, random
        20000, 1211.62925, random
        10000, 1205.3828, random
        50000, 1245.35526, random
        1000, 566.009, random
        20000, 1225.0181, random
        1000, 304.603, random
        1000, 376.247, random
        1000, 329.17, random
        1000, 223.042, random
        1000, 313.358, random
        1000, 307.605, random
        1000, 361.332, random
        1000, 404.363, random
        1000, 374.23, random
        50000, 1270.2446, random
        1000, 267.721, random
        1000, 236.239, random
        1000, 265.988, random
        1000, 403.649, random
        20000, 1235.5673, random
        1000, 400.209, random
        1000, 404.73, random
        1000, 486.008, random
        1000, 532.464, random
        1000, 292.808, random
        1000, 392.717, random
        1000, 493.723, random
        1000, 744.094, random
        1000, 500.556, random
        1000, 352.919, random
        1000, 362.32, random
        5000, 600.0284, random
        5000, 1404.2752, random
        50000, 1267.75322, random
        20000, 1214.8601, random
        5000, 1002.4698, random
        5000, 786.3128, random
        5000, 676.9138, random
        5000, 921.8886, random
        5000, 729.4516, random
        5000, 728.2072, random
        5000, 858.512, random
        5000, 747.7272, random
        5000, 1122.1762, random
        5000, 1302.236, random
        5000, 970.5276, random
        5000, 730.5052, random
        5000, 812.3532, random
        5000, 1399.5054, random
        5000, 1015.3394, random
        50000, 1259.76932, random
        5000, 767.9544, random
        5000, 769.971, random
        5000, 1074.9048, random
        5000, 2096.9958, random
        5000, 1076.2358, random
        5000, 1228.7176, random
        5000, 592.887, random
        5000, 954.6564, random
        50000, 1254.9401, random
        20000, 1590.1139, random
        20000, 1681.93835, random
        20000, 1280.50885, random
        20000, 977.14765, random
        20000, 1338.8286, random
        20000, 1496.3231, random
        20000, 1589.02805, random
        20000, 2610.8728, random
        20000, 1489.7307, random
        20000, 2686.5831, random
        20000, 1429.8264, random
        20000, 1024.2735, random
        20000, 2358.74605, random
        20000, 1686.8695, random
        20000, 825.50015, random
        20000, 1712.5403, random
        20000, 1254.32445, random
        20000, 1269.83585, random
        20000, 1092.62625, random
        20000, 2327.19915, random
        20000, 2532.88875, random
        20000, 950.31885, random
        20000, 1208.75445, random
        20000, 1090.76565, random
        20000, 1245.6824, random
        50000, 1990.41152, random
        50000, 1399.9095, random
        50000, 3089.54924, random
        50000, 2141.78608, random
        50000, 1591.56294, random
        50000, 2308.72914, random
        50000, 1457.7007, random
        50000, 1773.44494, random
        50000, 1570.4663, random
        50000, 2331.07794, random
        50000, 1844.68678, random
        50000, 1626.2939, random
        50000, 2054.7263, random
        50000, 1774.65866, random
        50000, 2433.8576, random
        50000, 3043.41594, random
        50000, 1607.13628, random
        50000, 2309.31932, random
        10000, 1279.3833, random
        10000, 1090.221, random
        50000, 2433.582, random
        50000, 3192.55928, random
        50000, 1786.68624, random
        10000, 961.6495, random
        10000, 932.6267, random
        10000, 867.2195, random
        50000, 1384.01092, random
        10000, 1198.1051, random
        10000, 1387.4966, random
        10000, 1504.2111, random
        10000, 1793.572, random
        10000, 1813.0635, random
        10000, 1095.3237, random
        10000, 811.6552, random
        10000, 1172.6608, random
        50000, 1350.32382, random
        10000, 1933.9295, random
        10000, 1327.7505, random
        10000, 1387.1753, random
        10000, 2171.3671, random
        10000, 778.9984, random
        10000, 1077.4132, random
        10000, 761.4016, random
        50000, 4939.22878, random
        10000, 959.7388, random
        10000, 1082.1432, random
        10000, 1255.438, random
        50000, 1629.2048, random
        10000, 1016.4927, random
        778, 5.85989717224, modelchecking
        114, 4.36842105263, modelchecking
        102, 4.52941176471, modelchecking
        292, 6.29452054795, modelchecking
        204, 4.55392156863, modelchecking
        292, 4.59246575342, modelchecking
        102, 4.28431372549, modelchecking
        302, 4.73841059603, modelchecking
        98, 4.58163265306, modelchecking
        10000, 991.3674, random
        1691, 15.3311649911, modelchecking
        1691, 15.3311649911, modelchecking
        882, 9.30839002268, modelchecking
        882, 9.30839002268, modelchecking
        846, 13.1619385343, modelchecking
        763, 12.4154652687, modelchecking
        486, 10.3086419753, modelchecking
        486, 10.3086419753, modelchecking
        27, 15.5185185185, mlsolver
        225, 5.37333333333, mlsolver
        59, 4.30508474576, mlsolver
        330, 3.23636363636, mlsolver
        71, 4.33802816901, mlsolver
        494, 3.21255060729, mlsolver
        86, 3.6511627907, mlsolver
        15, 4.26666666667, mlsolver
        7, 2.42857142857, equivalence
        447, 7.98210290828, equivalence
        447, 7.98210290828, equivalence
        532, 8.26127819549, equivalence
        310, 7.33870967742, equivalence
        5925, 13.426835443, equivalence
        4921, 15.9051005893, equivalence
        5925, 13.426835443, equivalence
        3, 1, equivalence
        6722, 17.6019041952, equivalence
        6722, 17.6019041952, equivalence
        7106, 18.3595553054, equivalence
        816, 7.67647058824, equivalence
        16119, 13.7216948942, equivalence
        16119, 13.7216948942, equivalence
        13425, 16.0960148976, equivalence
        614, 7.61400651466, equivalence
        12133, 13.6525179263, equivalence
        12133, 13.6525179263, equivalence
        10121, 16.0380397194, equivalence
        3, 1, equivalence
        8961, 21.7029349403, equivalence
        9553, 23.3894064692, equivalence
        8961, 21.7029349403, equivalence
        3, 1, equivalence
        966897, 38.4443379181, equivalence
        912641, 29.3604100627, equivalence
        912641, 29.3604100627, equivalence
        412, 7.37378640777, equivalence
        7855, 13.4744748568, equivalence
        7855, 13.4744748568, equivalence
        6511, 15.9494701275, equivalence
        3, 1, equivalence
        3457, 17.8831356668, equivalence
        3761, 20.2597713374, equivalence
        3457, 17.8831356668, equivalence
        3, 1, equivalence
        2076302, 25.6177661053, equivalence
        2076302, 25.6177661053, equivalence
        2162594, 31.2405916228, equivalence
        3, 1, equivalence
        3697, 19.5764133081, equivalence
        3985, 22.4725219573, equivalence
        3697, 19.5764133081, equivalence
        11, 2.81818181818, equivalence
        1095, 9.50502283105, equivalence
        1095, 9.50502283105, equivalence
        1264, 9.50712025316, equivalence
        3, 1, equivalence
        13409, 22.7471101499, equivalence
        14177, 24.9880087466, equivalence
        13409, 22.7471101499, equivalence
        11, 2.81818181818, equivalence
        1127, 9.12156166815, equivalence
        1127, 9.12156166815, equivalence
        1306, 9.06814701378, equivalence
        3, 1, equivalence
        3644173, 41.6013948844, equivalence
        3471553, 32.7419650514, equivalence
        3471553, 32.7419650514, equivalence
        34954, 21.7834296504, equivalence
        34954, 21.7834296504, equivalence
        30922, 25.9533988746, equivalence
        3, 1, equivalence
        3, 1, equivalence
        97762, 22.1681225834, equivalence
        86802, 26.2355590885, equivalence
        97762, 22.1681225834, equivalence
        1883, 7.76473712161, equivalence
        6124, 16.5782168517, equivalence
        7383, 13.8101042936, equivalence
        7383, 13.8101042936, equivalence
        3759, 7.81245011971, equivalence
        15111, 14.0300443386, equivalence
        12590, 16.6977760127, equivalence
        15111, 14.0300443386, equivalence
        7, 2.42857142857, equivalence
        467, 7.67451820128, equivalence
        467, 7.67451820128, equivalence
        558, 7.93010752688, equivalence
        3, 1, equivalence
        2266, 15.7925860547, equivalence
        2410, 17.2784232365, equivalence
        2266, 15.7925860547, equivalence
        7, 2.42857142857, equivalence
        353, 7.60906515581, equivalence
        423, 7.89361702128, equivalence
        353, 7.60906515581, equivalence
        3, 1, equivalence
        57905, 24.2116915638, equivalence
        57905, 24.2116915638, equivalence
        50865, 29.7823454242, equivalence
        134097, 24.8041641498, equivalence
        118225, 30.1803764009, equivalence
        3, 1, equivalence
        134097, 24.8041641498, equivalence
        3977, 7.86924817702, equivalence
        20317, 14.1159620023, equivalence
        16910, 16.6763453578, equivalence
        20317, 14.1159620023, equivalence
        1993, 7.80431510286, equivalence
        8206, 16.5480136485, equivalence
        9907, 13.8800847885, equivalence
        9907, 13.8800847885, equivalence
        3, 1, equivalence
        604354, 23.7931973645, equivalence
        604354, 23.7931973645, equivalence
        631474, 29.6099142641, equivalence
        3, 1, equivalence
        77529, 24.3619806782, equivalence
        77529, 24.3619806782, equivalence
        68025, 29.7947960309, equivalence
        3, 1, equivalence
        178233, 24.9904002065, equivalence
        178233, 24.9904002065, equivalence
        156729, 30.2986364999, equivalence
        816, 7.67647058824, equivalence
        19045, 16.0739826726, equivalence
        22915, 13.8166266638, equivalence
        22915, 13.8166266638, equivalence
        3, 1, equivalence
        167247, 28.0414656167, equivalence
        147323, 36.7153329758, equivalence
        167247, 28.0414656167, equivalence
        3, 1, equivalence
        74039, 27.3802320399, equivalence
        74039, 27.3802320399, equivalence
        65131, 36.4087300978, equivalence
        3, 1, equivalence
        39614, 26.0777755339, equivalence
        44926, 21.8029648756, equivalence
        44926, 21.8029648756, equivalence
        11, 2.81818181818, equivalence
        845, 9.00118343195, equivalence
        983, 8.98575788403, equivalence
        845, 9.00118343195, equivalence
        3, 1, equivalence
        109890, 26.4764309764, equivalence
        124458, 22.2290250526, equivalence
        124458, 22.2290250526, equivalence
        3, 1, equivalence
        107824, 22.5435617302, equivalence
        107824, 22.5435617302, equivalence
        95288, 26.886480984, equivalence
        4743, 9.04975753742, equivalence
        16162, 17.4961638411, equivalence
        19423, 14.7120424239, equivalence
        19423, 14.7120424239, equivalence
        3, 1, equivalence
        35234, 26.6521825509, equivalence
        39822, 22.2287931294, equivalence
        39822, 22.2287931294, equivalence
        9205, 15.9116784356, equivalence
        412, 7.37378640777, equivalence
        11135, 13.5409070498, equivalence
        11135, 13.5409070498, equivalence
        2375, 9.01010526316, equivalence
        9507, 14.5055222468, equivalence
        7878, 17.3941355674, equivalence
        9507, 14.5055222468, equivalence
        104074, 33.1878471088, equivalence
        700289, 60.5696690938, equivalence
        788225, 43.2952507216, equivalence
        788225, 43.2952507216, equivalence
        28226, 35.0091050804, equivalence
        214897, 54.4937760881, equivalence
        240097, 39.0422329309, equivalence
        240097, 39.0422329309, equivalence
        38502, 32.1149031219, equivalence
        245697, 59.8446256975, equivalence
        276225, 42.1550945787, equivalence
        276225, 42.1550945787, equivalence
        54758, 35.855491435, equivalence
        440569, 54.7921687636, equivalence
        488629, 39.5784102049, equivalence
        488629, 39.5784102049, equivalence
        12239050, 30.9107979786, equivalence
        12239050, 30.9107979786, equivalence
        3, 1, equivalence
        10685466, 40.0625557182, equivalence
        40556396, 31.4424375627, equivalence
        3, 1, equivalence
        35431922, 40.7015671066, equivalence
        9488018, 30.8960739746, equivalence
        9488018, 30.8960739746, equivalence
        3, 1, equivalence
        8326786, 39.7730145821, equivalence
        31611530, 31.4085914538, equivalence
        31611530, 31.4085914538, equivalence
        3, 1, equivalence
        27799634, 40.3009373073, equivalence
        7626354, 65.0113598189, equivalence
        33702306, 18.4560170452, modelchecking
        43, 6.04651162791, modelchecking
        29868274, 20.1499912248, modelchecking
        61, 6.22950819672, modelchecking
        33702306, 17.1265222623, modelchecking
        33702306, 17.1265222623, modelchecking
        43, 6.04651162791, modelchecking
        29868274, 19.5799061908, modelchecking
        61, 6.22950819672, modelchecking
        5322498, 62.7029787517, equivalence
        19026506, 64.721358509, equivalence
        10927074, 31.5553051073, equivalence
        10927074, 31.5553051073, equivalence
        3, 1, equivalence
        9558194, 41.1077938991, equivalence
        3, 1, equivalence
        37636481, 47.6426102377, equivalence
        37636481, 47.6426102377, equivalence
        3782172, 38.6322591886, equivalence
        14808231, 39.375416753, equivalence
        2544209, 5.0295345233, mlsolver
        196527, 139.063411134, mlsolver
        1167042, 4.97941547948, mlsolver
        62953, 456.845059012, mlsolver
        1348959, 5.30051395187, mlsolver
        35388, 1594.12693568, mlsolver
        3155971, 5.27066154917, mlsolver
        80133, 3583.28788389, mlsolver
        1021740, 4.39759429992, mlsolver
        426185, 16.9929021434, mlsolver
        1042752, 3.94452372184, mlsolver
        183781, 128.844194993, mlsolver
        2177202, 3.98552040647, modelchecking
        9101, 4.16536644325, modelchecking
        48861, 4.38486727656, modelchecking
        6, 2, modelchecking
        531443, 12.3332172971, modelchecking
        1594325, 12.3332946545, modelchecking
        13834801, 10.095862745, modelchecking
        27876961, 10.0613749469, modelchecking
        188570, 8.58341729862, modelchecking
        377113, 9.56654901846, modelchecking
        417016, 8.9628431523, modelchecking
        1285576, 8.67894313522, modelchecking
        571378, 8.56065336782, modelchecking
        1411274, 8.54914637413, modelchecking
        2341446, 33.0711393728, modelchecking
        1997579, 53.5015020683, modelchecking
        1997579, 53.5015020683, modelchecking
        134162, 13.6209507908, modelchecking
        134162, 13.6209507908, modelchecking
        998790, 61.5490333303, modelchecking
        788879, 57.2709388892, modelchecking
        267378, 41.4587999013, modelchecking
        267378, 41.4587999013, modelchecking
        20712450, 40.3490985856, modelchecking
        8964338, 44.5565215189, modelchecking
        6823297, 24.0419238676, modelchecking
        3487362, 37.2669679832, modelchecking
        6974724, 37.2669687001, modelchecking
        7767169, 42.9659498847, modelchecking
        11488274, 36.8488141038, modelchecking
        94710, 37.6336923239, modelchecking
        2589057, 43.679673719, modelchecking
        10782593, 40.8409864863, modelchecking
        7310722, 28.8677707893, modelchecking
        19550209, 22.3759418122, modelchecking
        8835074, 36.473352685, modelchecking
        22288897, 40.0285299896, modelchecking
        6690605, 38.7174932611, modelchecking
        7429633, 40.0626409138, modelchecking
        24565250, 25.577674927, modelchecking
        861780, 7.18047413493, modelchecking
        40200, 158.158905473, specialcases
        80400, 158.945124378, specialcases
        20100, 156.586467662, specialcases
        201000, 159.416855721, specialcases
        80200, 306.668179551, specialcases
        160400, 308.204413965, specialcases
        40100, 303.595710723, specialcases
        401000, 309.126154613, specialcases
        20200, 83.8904950495, specialcases
        40400, 84.3016831683, specialcases
        10100, 83.0681188119, specialcases
        101000, 84.5483960396, specialcases
        200200, 752.173766234, specialcases
        400400, 755.96000999, specialcases
        100100, 744.601278721, specialcases
        1001000, 758.231756244, specialcases
        100000, 12.19939, specialcases
        250000, 12.199756, specialcases 
      };
      \legend{}
    \end{axis}
  \end{tikzpicture}

%% file: size_vs_ad_orig.tex
  \begin{tikzpicture}[mark size=2.5pt,remember picture,scale=0.7]
    \begin{axis}[axis x line=bottom,
                       axis y line=left,
                       xlabel={Vertices},
                       ylabel={Alternation depth},
                       xmode={log},
                       ymode={log}
                       ,
                       xmin=0,
                       ymin=0,
                       scatter/classes={
                         modelchecking={mark=x},
                         equivalence={mark=o},
                         mlsolver={mark=+},
                         specialcases={mark=square},
                         random={mark=triangle}
                       },
                       legend image post style={xshift=0.5cm},
                       legend cell align=left,
                       legend style={
                         at={(1,0.77)},
                         nodes={right}
                       }]
      \addplot+[black,only marks,mark=x,scatter,scatter src=explicit symbolic] 
      table[col sep=comma,x=x, y=y, meta=cluster] { 
        x, y, cluster
        2, 1, modelchecking
        972, 2, modelchecking
        2688, 2, modelchecking
        15684, 2, modelchecking
        588, 2, modelchecking
        111456, 2, modelchecking
        656, 1, modelchecking
        308, 1, modelchecking
        227, 1, modelchecking
        328, 1, modelchecking
        649, 1, modelchecking
        656, 1, modelchecking
        649, 1, modelchecking
        227, 1, modelchecking
        8626, 1, modelchecking
        2126, 1, modelchecking
        4313, 1, modelchecking
        4021, 1, modelchecking
        6359, 1, modelchecking
        6359, 1, modelchecking
        8626, 1, modelchecking
        2126, 1, modelchecking
        2832, 2, modelchecking
        16356, 2, modelchecking
        2916, 2, modelchecking
        8748, 2, modelchecking
        78732, 2, modelchecking
        564, 2, modelchecking
        876780, 2, modelchecking
        2338, 2, modelchecking
        147, 1, modelchecking
        260, 2, modelchecking
        747, 2, modelchecking
        442, 1, modelchecking
        731, 1, modelchecking
        439, 1, modelchecking
        598, 2, modelchecking
        173, 2, modelchecking
        150, 2, modelchecking
        236, 2, modelchecking
        9282, 2, modelchecking
        294, 2, modelchecking
        341, 2, modelchecking
        2643, 2, modelchecking
        291, 1, modelchecking
        516, 2, modelchecking
        2034, 1, modelchecking
        871, 1, modelchecking
        2346, 2, modelchecking
        2611, 1, modelchecking
        468, 2, modelchecking
        594, 2, modelchecking
        78, 2, modelchecking
        156, 2, modelchecking
        132, 2, modelchecking
        75, 1, modelchecking
        231, 2, modelchecking
        89, 2, modelchecking
        223, 1, modelchecking
        78, 1, modelchecking
        223, 1, modelchecking
        120, 2, modelchecking
        63907, 1, modelchecking
        31954, 1, modelchecking
        63907, 1, modelchecking
        13222, 1, modelchecking
        13222, 1, modelchecking
        11977, 1, modelchecking
        11977, 1, modelchecking
        26996, 1, modelchecking
        236196, 2, modelchecking
        108336, 2, modelchecking
        26244, 2, modelchecking
        2946, 1, modelchecking
        5649, 2, modelchecking
        1041, 1, modelchecking
        5201, 1, modelchecking
        3121, 1, modelchecking
        4486, 2, modelchecking
        1122, 2, modelchecking
        1619, 2, modelchecking
        16642, 2, modelchecking
        708588, 2, modelchecking
        107276, 1, modelchecking
        18625, 1, modelchecking
        18625, 1, modelchecking
        44773, 1, modelchecking
        53638, 1, modelchecking
        51220, 1, modelchecking
        44773, 1, modelchecking
        107276, 1, modelchecking
        1028, 2, modelchecking
        514, 2, modelchecking
        691, 2, modelchecking
        3714, 2, modelchecking
        465, 1, modelchecking
        1393, 1, modelchecking
        1553, 2, modelchecking
        450, 1, modelchecking
        1393, 1, modelchecking
        12130, 2, modelchecking
        2658, 1, modelchecking
        3452, 2, modelchecking
        3412, 1, modelchecking
        380, 1, modelchecking
        3066, 2, modelchecking
        438, 2, modelchecking
        1138, 1, modelchecking
        384, 2, modelchecking
        564914, 1, modelchecking
        564914, 1, modelchecking
        4860, 1, modelchecking
        4860, 1, modelchecking
        2229, 1, modelchecking
        2229, 1, modelchecking
        5861, 1, modelchecking
        2245634, 2, modelchecking
        512018, 2, modelchecking
        653574, 2, modelchecking
        163394, 2, modelchecking
        858114, 1, modelchecking
        388418, 1, modelchecking
        140353, 1, modelchecking
        211955, 1, modelchecking
        869569, 2, modelchecking
        353089, 2, modelchecking
        421057, 1, modelchecking
        147458, 2, modelchecking
        4898, 2, modelchecking
        41762, 1, modelchecking
        39178, 2, modelchecking
        12818, 2, modelchecking
        13825, 1, modelchecking
        28964, 1, modelchecking
        10369, 2, modelchecking
        45937, 2, modelchecking
        4609, 1, modelchecking
        9986, 1, modelchecking
        21506, 2, modelchecking
        5958, 2, modelchecking
        6866, 1, modelchecking
        1490, 2, modelchecking
        3344, 1, modelchecking
        3530, 2, modelchecking
        4033, 1, modelchecking
        2977, 2, modelchecking
        2690, 1, modelchecking
        1345, 1, modelchecking
        7897, 2, modelchecking
        196, 2, modelchecking
        222, 2, modelchecking
        3058, 2, modelchecking
        192, 1, modelchecking
        782, 2, modelchecking
        578, 1, modelchecking
        956, 1, modelchecking
        976, 2, modelchecking
        574, 1, modelchecking
        112514, 2, modelchecking
        3949, 1, modelchecking
        17810, 2, modelchecking
        35620, 2, modelchecking
        54322, 1, modelchecking
        14065, 1, modelchecking
        32402, 1, modelchecking
        42618, 2, modelchecking
        42193, 1, modelchecking
        34673, 2, modelchecking
        56241, 2, modelchecking
        3458, 2, modelchecking
        1046, 2, modelchecking
        1621, 2, modelchecking
        433, 1, modelchecking
        326, 1, modelchecking
        1370, 1, modelchecking
        770, 1, modelchecking
        1012, 2, modelchecking
        506, 2, modelchecking
        937, 2, modelchecking
        1297, 1, modelchecking
        11362, 2, modelchecking
        3188, 1, modelchecking
        3228, 2, modelchecking
        356, 1, modelchecking
        2490, 1, modelchecking
        1066, 1, modelchecking
        2874, 2, modelchecking
        360, 2, modelchecking
        414, 2, modelchecking
        6563, 1, modelchecking
        19685, 1, modelchecking
        730, 2, modelchecking
        282, 2, modelchecking
        274, 1, modelchecking
        96, 1, modelchecking
        92, 1, modelchecking
        192, 2, modelchecking
        96, 2, modelchecking
        108, 2, modelchecking
        272, 1, modelchecking
        43, 2, modelchecking
        13, 1, modelchecking
        175, 2, modelchecking
        97, 2, modelchecking
        133, 2, modelchecking
        67, 2, modelchecking
        82434, 2, modelchecking
        2577, 1, modelchecking
        24593, 2, modelchecking
        17154, 1, modelchecking
        21770, 2, modelchecking
        23185, 1, modelchecking
        7729, 1, modelchecking
        2722, 2, modelchecking
        4243, 2, modelchecking
        43, 2, modelchecking
        2866, 2, modelchecking
        734, 2, modelchecking
        542, 1, modelchecking
        184, 2, modelchecking
        538, 1, modelchecking
        210, 2, modelchecking
        180, 1, modelchecking
        892, 1, modelchecking
        912, 2, modelchecking
        17773, 2, modelchecking
        59051, 1, modelchecking
        2395, 2, modelchecking
        343, 2, modelchecking
        177149, 1, modelchecking
        926213, 2, modelchecking
        3471554, 2, modelchecking
        607753, 2, modelchecking
        867889, 2, modelchecking
        308738, 2, modelchecking
        638065, 1, modelchecking
        579745, 1, modelchecking
        289297, 1, modelchecking
        1350433, 2, modelchecking
        1191962, 1, modelchecking
        1278433, 1, modelchecking
        474914, 1, modelchecking
        579745, 1, modelchecking
        25, 2, modelchecking
        13, 1, modelchecking
        655362, 2, modelchecking
        170753, 2, modelchecking
        297089, 2, modelchecking
        177668, 2, modelchecking
        81921, 1, modelchecking
        88834, 2, modelchecking
        245761, 2, modelchecking
        153985, 1, modelchecking
        258050, 1, modelchecking
        164353, 1, modelchecking
        153985, 1, modelchecking
        65506, 1, modelchecking
        185089, 1, modelchecking
        540737, 1, modelchecking
        1081474, 1, modelchecking
        1093761, 1, modelchecking
        1787138, 2, modelchecking
        22873, 1, modelchecking
        742970, 2, modelchecking
        223393, 1, modelchecking
        294434, 2, modelchecking
        588868, 2, modelchecking
        944090, 1, modelchecking
        586658, 1, modelchecking
        576385, 2, modelchecking
        670177, 1, modelchecking
        917713, 2, modelchecking
        1559, 3, mlsolver
        636, 2, mlsolver
        298, 2, mlsolver
        693, 3, mlsolver
        2858, 3, mlsolver
        1143, 2, mlsolver
        1126, 3, mlsolver
        467, 2, mlsolver
        12, 2, mlsolver
        22, 2, mlsolver
        974, 2, mlsolver
        2425, 3, mlsolver
        805, 2, mlsolver
        1992, 3, mlsolver
        1312, 2, mlsolver
        3291, 3, mlsolver
        6817, 4, mlsolver
        14053, 2, mlsolver
        1562, 3, mlsolver
        3029, 2, mlsolver
        3724, 3, mlsolver
        1481, 2, mlsolver
        336, 1, mlsolver
        15, 1, mlsolver
        191, 2, mlsolver
        365, 2, mlsolver
        76, 1, mlsolver
        9, 1, mlsolver
        912, 1, mlsolver
        21, 1, mlsolver
        45, 1, mlsolver
        8776, 1, mlsolver
        5736, 1, mlsolver
        39, 1, mlsolver
        27, 1, mlsolver
        1924, 1, mlsolver
        3492, 1, mlsolver
        33, 1, mlsolver
        9, 1, mlsolver
        319, 1, mlsolver
        160, 1, mlsolver
        9, 1, mlsolver
        1156, 1, mlsolver
        9, 1, mlsolver
        817, 1, mlsolver
        9, 1, mlsolver
        61, 1, mlsolver
        9, 1, mlsolver
        538, 1, mlsolver
        9, 1, mlsolver
        29335, 4, mlsolver
        62707, 2, mlsolver
        30, 1, mlsolver
        9, 1, mlsolver
        51, 1, mlsolver
        12732, 1, mlsolver
        2014, 1, mlsolver
        9, 1, mlsolver
        140, 1, mlsolver
        21, 1, mlsolver
        9, 1, mlsolver
        1555, 1, mlsolver
        15, 1, mlsolver
        75, 1, mlsolver
        33, 1, mlsolver
        330, 1, mlsolver
        200, 2, mlsolver
        47, 2, mlsolver
        455, 1, mlsolver
        39, 1, mlsolver
        45, 1, mlsolver
        600, 1, mlsolver
        765, 1, mlsolver
        51, 1, mlsolver
        105, 2, mlsolver
        34, 2, mlsolver
        63, 2, mlsolver
        19, 1, mlsolver
        1490, 2, mlsolver
        119, 2, mlsolver
        1190, 2, mlsolver
        107, 2, mlsolver
        924, 2, mlsolver
        95, 2, mlsolver
        83, 2, mlsolver
        692, 2, mlsolver
        33969, 3, mlsolver
        6182, 4, mlsolver
        456, 4, mlsolver
        2497, 2, mlsolver
        6431, 1, mlsolver
        730, 1, mlsolver
        8, 1, mlsolver
        16, 2, mlsolver
        543, 1, mlsolver
        69, 1, mlsolver
        348, 1, mlsolver
        3772, 1, mlsolver
        6346, 1, mlsolver
        60011, 1, mlsolver
        38075, 1, mlsolver
        3195, 1, mlsolver
        110328, 4, mlsolver
        259455, 2, mlsolver
        1264, 3, mlsolver
        36735, 3, mlsolver
        25971, 1, mlsolver
        325637, 1, mlsolver
        49830, 1, mlsolver
        492107, 1, mlsolver
        2861, 3, mlsolver
        94631, 3, mlsolver
        19, 1, mlsolver
        354, 3, mlsolver
        8, 1, mlsolver
        16, 2, mlsolver
        6831, 3, mlsolver
        33977, 2, mlsolver
        9585, 4, mlsolver
        174667, 4, mlsolver
        6650, 3, mlsolver
        234577, 3, mlsolver
        970, 1, mlsolver
        65, 1, mlsolver
        334, 1, mlsolver
        30, 1, mlsolver
        7, 1, mlsolver
        14, 1, mlsolver
        2443, 1, mlsolver
        148, 1, mlsolver
        332, 1, mlsolver
        5837, 1, mlsolver
        778, 1, mlsolver
        13687, 1, mlsolver
        89, 1, mlsolver
        2265, 1, mlsolver
        8, 1, mlsolver
        34, 1, mlsolver
        17, 1, mlsolver
        250, 1, mlsolver
        203, 1, mlsolver
        27, 1, mlsolver
        32233, 1, mlsolver
        1794, 1, mlsolver
        17, 1, mlsolver
        9, 1, mlsolver
        38, 1, mlsolver
        795, 1, mlsolver
        35037, 1, mlsolver
        1123, 1, mlsolver
        478, 1, mlsolver
        14399, 1, mlsolver
        5914, 1, mlsolver
        209, 1, mlsolver
        2611, 1, mlsolver
        82968, 1, mlsolver
        105, 1, mlsolver
        1546, 1, mlsolver
        15427, 3, mlsolver
        567753, 3, mlsolver
        5596, 3, mlsolver
        140309, 3, mlsolver
        44028, 1, mlsolver
        2163, 1, mlsolver
        48211, 1, mlsolver
        1031612, 1, mlsolver
        10183, 1, mlsolver
        214775, 1, mlsolver
        228787, 1, mlsolver
        4924413, 1, mlsolver
        467, 1, mlsolver
        8691, 1, mlsolver
        100, 100, specialcases
        200, 200, specialcases
        200, 2, specialcases
        400, 2, specialcases
        1000, 2, specialcases
        2000, 2, specialcases
        4000, 2, specialcases
        10000, 2, specialcases
        2000, 2000, specialcases
        40000, 2, specialcases
        20000, 2, specialcases
        500, 500, specialcases
        301, 2, specialcases
        601, 2, specialcases
        1501, 2, specialcases
        100000, 2, specialcases
        3001, 2, specialcases
        6001, 2, specialcases
        1000, 1000, specialcases
        15001, 2, specialcases
        5000, 5000, specialcases
        30001, 2, specialcases
        10000, 10000, specialcases
        1000, 11, random
        5000, 11, random
        10000, 11, random
        1000, 11, random
        5000, 11, random
        500, 101, specialcases
        2500, 501, specialcases
        1000, 201, specialcases
        5000, 1001, specialcases
        10000, 11, random
        10000, 2001, specialcases
        1000, 11, random
        60001, 2, specialcases
        5000, 11, random
        20000, 11, random
        25000, 5001, specialcases
        1000, 11, random
        10000, 11, random
        5000, 11, random
        50000, 10001, specialcases
        1000, 11, random
        5000, 11, random
        10000, 11, random
        20000, 11, random
        1000, 11, random
        10000, 11, random
        50000, 11, random
        5000, 11, random
        20000, 11, random
        1000, 11, random
        10000, 11, random
        50000, 11, random
        20000, 11, random
        5000, 11, random
        1000, 11, random
        10000, 11, random
        20000, 11, random
        5000, 11, random
        50000, 11, random
        10000, 11, random
        1000, 11, random
        20000, 11, random
        5000, 11, random
        50000, 11, random
        1000, 11, random
        50000, 11, random
        5000, 11, random
        20000, 11, random
        10000, 11, random
        10000, 11, random
        1000, 11, random
        5000, 11, random
        50000, 11, random
        20000, 11, random
        1000, 11, random
        10000, 11, random
        5000, 11, random
        20000, 11, random
        50000, 11, random
        10000, 11, random
        1000, 11, random
        5000, 11, random
        50000, 11, random
        20000, 11, random
        1000, 11, random
        10000, 11, random
        5000, 11, random
        50000, 11, random
        20000, 11, random
        1000, 11, random
        10000, 11, random
        5000, 11, random
        50000, 11, random
        20000, 11, random
        1000, 11, random
        5000, 11, random
        10000, 11, random
        50000, 11, random
        10000, 11, random
        20000, 11, random
        1000, 11, random
        50000, 11, random
        5000, 11, random
        20000, 11, random
        1000, 11, random
        50000, 11, random
        10000, 11, random
        5000, 11, random
        20000, 11, random
        1000, 11, random
        5000, 11, random
        10000, 11, random
        50000, 11, random
        20000, 11, random
        1000, 11, random
        50000, 11, random
        10000, 11, random
        20000, 11, random
        5000, 11, random
        1000, 10, random
        5000, 11, random
        10000, 11, random
        1000, 11, random
        50000, 11, random
        20000, 11, random
        10000, 11, random
        5000, 11, random
        50000, 11, random
        1000, 11, random
        20000, 11, random
        10000, 11, random
        5000, 11, random
        50000, 11, random
        20000, 11, random
        1000, 11, random
        10000, 11, random
        20000, 11, random
        50000, 11, random
        5000, 11, random
        1000, 11, random
        10000, 11, random
        5000, 11, random
        50000, 11, random
        20000, 11, random
        10000, 11, random
        50000, 11, random
        1000, 16, random
        20000, 11, random
        1000, 14, random
        1000, 11, random
        1000, 13, random
        1000, 12, random
        1000, 14, random
        1000, 14, random
        1000, 12, random
        1000, 14, random
        1000, 13, random
        50000, 11, random
        1000, 12, random
        1000, 12, random
        1000, 13, random
        1000, 13, random
        20000, 11, random
        1000, 11, random
        1000, 14, random
        1000, 17, random
        1000, 14, random
        1000, 14, random
        1000, 13, random
        1000, 16, random
        1000, 18, random
        1000, 14, random
        1000, 13, random
        1000, 12, random
        5000, 14, random
        5000, 22, random
        50000, 11, random
        20000, 11, random
        5000, 19, random
        5000, 16, random
        5000, 15, random
        5000, 17, random
        5000, 16, random
        5000, 17, random
        5000, 19, random
        5000, 17, random
        5000, 19, random
        5000, 20, random
        5000, 20, random
        5000, 16, random
        5000, 19, random
        5000, 19, random
        5000, 18, random
        50000, 11, random
        5000, 17, random
        5000, 18, random
        5000, 20, random
        5000, 24, random
        5000, 19, random
        5000, 24, random
        5000, 17, random
        5000, 18, random
        50000, 11, random
        20000, 22, random
        20000, 22, random
        20000, 22, random
        20000, 19, random
        20000, 22, random
        20000, 23, random
        20000, 21, random
        20000, 24, random
        20000, 23, random
        20000, 27, random
        20000, 23, random
        20000, 19, random
        20000, 22, random
        20000, 22, random
        20000, 21, random
        20000, 23, random
        20000, 24, random
        20000, 20, random
        20000, 21, random
        20000, 25, random
        20000, 25, random
        20000, 22, random
        20000, 20, random
        20000, 20, random
        20000, 21, random
        50000, 25, random
        50000, 23, random
        50000, 27, random
        50000, 24, random
        50000, 22, random
        50000, 25, random
        50000, 23, random
        50000, 24, random
        50000, 24, random
        50000, 24, random
        50000, 24, random
        50000, 25, random
        50000, 24, random
        50000, 23, random
        50000, 25, random
        50000, 27, random
        50000, 22, random
        50000, 25, random
        10000, 20, random
        10000, 20, random
        50000, 26, random
        50000, 27, random
        50000, 25, random
        10000, 21, random
        10000, 17, random
        10000, 20, random
        50000, 22, random
        10000, 19, random
        10000, 20, random
        10000, 21, random
        10000, 22, random
        10000, 23, random
        10000, 21, random
        10000, 16, random
        10000, 19, random
        50000, 23, random
        10000, 24, random
        10000, 21, random
        10000, 19, random
        10000, 24, random
        10000, 17, random
        10000, 20, random
        10000, 18, random
        50000, 33, random
        10000, 19, random
        10000, 19, random
        10000, 19, random
        50000, 22, random
        10000, 18, random
        778, 2, modelchecking
        114, 2, modelchecking
        102, 2, modelchecking
        292, 1, modelchecking
        204, 2, modelchecking
        292, 1, modelchecking
        102, 1, modelchecking
        302, 2, modelchecking
        98, 1, modelchecking
        10000, 18, random
        1691, 1, modelchecking
        1691, 1, modelchecking
        882, 1, modelchecking
        882, 1, modelchecking
        846, 1, modelchecking
        763, 1, modelchecking
        486, 1, modelchecking
        486, 1, modelchecking
        27, 1, mlsolver
        225, 1, mlsolver
        59, 2, mlsolver
        330, 2, mlsolver
        71, 2, mlsolver
        494, 2, mlsolver
        86, 1, mlsolver
        15, 1, mlsolver
        7, 1, equivalence
        447, 2, equivalence
        447, 2, equivalence
        532, 2, equivalence
        310, 1, equivalence
        5925, 2, equivalence
        4921, 2, equivalence
        5925, 2, equivalence
        3, 1, equivalence
        6722, 2, equivalence
        6722, 2, equivalence
        7106, 2, equivalence
        816, 1, equivalence
        16119, 2, equivalence
        16119, 2, equivalence
        13425, 2, equivalence
        614, 1, equivalence
        12133, 2, equivalence
        12133, 2, equivalence
        10121, 2, equivalence
        3, 1, equivalence
        8961, 2, equivalence
        9553, 2, equivalence
        8961, 2, equivalence
        3, 1, equivalence
        966897, 2, equivalence
        912641, 2, equivalence
        912641, 2, equivalence
        412, 1, equivalence
        7855, 2, equivalence
        7855, 2, equivalence
        6511, 2, equivalence
        3, 1, equivalence
        3457, 2, equivalence
        3761, 2, equivalence
        3457, 2, equivalence
        3, 1, equivalence
        2076302, 2, equivalence
        2076302, 2, equivalence
        2162594, 2, equivalence
        3, 1, equivalence
        3697, 2, equivalence
        3985, 2, equivalence
        3697, 2, equivalence
        11, 1, equivalence
        1095, 2, equivalence
        1095, 2, equivalence
        1264, 2, equivalence
        3, 1, equivalence
        13409, 2, equivalence
        14177, 2, equivalence
        13409, 2, equivalence
        11, 1, equivalence
        1127, 2, equivalence
        1127, 2, equivalence
        1306, 2, equivalence
        3, 1, equivalence
        3644173, 2, equivalence
        3471553, 2, equivalence
        3471553, 2, equivalence
        34954, 2, equivalence
        34954, 2, equivalence
        30922, 2, equivalence
        3, 1, equivalence
        3, 1, equivalence
        97762, 2, equivalence
        86802, 2, equivalence
        97762, 2, equivalence
        1883, 1, equivalence
        6124, 2, equivalence
        7383, 2, equivalence
        7383, 2, equivalence
        3759, 1, equivalence
        15111, 2, equivalence
        12590, 2, equivalence
        15111, 2, equivalence
        7, 1, equivalence
        467, 2, equivalence
        467, 2, equivalence
        558, 2, equivalence
        3, 1, equivalence
        2266, 2, equivalence
        2410, 2, equivalence
        2266, 2, equivalence
        7, 1, equivalence
        353, 2, equivalence
        423, 2, equivalence
        353, 2, equivalence
        3, 1, equivalence
        57905, 2, equivalence
        57905, 2, equivalence
        50865, 2, equivalence
        134097, 2, equivalence
        118225, 2, equivalence
        3, 1, equivalence
        134097, 2, equivalence
        3977, 1, equivalence
        20317, 2, equivalence
        16910, 2, equivalence
        20317, 2, equivalence
        1993, 1, equivalence
        8206, 2, equivalence
        9907, 2, equivalence
        9907, 2, equivalence
        3, 1, equivalence
        604354, 2, equivalence
        604354, 2, equivalence
        631474, 2, equivalence
        3, 1, equivalence
        77529, 2, equivalence
        77529, 2, equivalence
        68025, 2, equivalence
        3, 1, equivalence
        178233, 2, equivalence
        178233, 2, equivalence
        156729, 2, equivalence
        816, 1, equivalence
        19045, 2, equivalence
        22915, 2, equivalence
        22915, 2, equivalence
        3, 1, equivalence
        167247, 2, equivalence
        147323, 2, equivalence
        167247, 2, equivalence
        3, 1, equivalence
        74039, 2, equivalence
        74039, 2, equivalence
        65131, 2, equivalence
        3, 1, equivalence
        39614, 2, equivalence
        44926, 2, equivalence
        44926, 2, equivalence
        11, 1, equivalence
        845, 2, equivalence
        983, 2, equivalence
        845, 2, equivalence
        3, 1, equivalence
        109890, 2, equivalence
        124458, 2, equivalence
        124458, 2, equivalence
        3, 1, equivalence
        107824, 2, equivalence
        107824, 2, equivalence
        95288, 2, equivalence
        4743, 1, equivalence
        16162, 2, equivalence
        19423, 2, equivalence
        19423, 2, equivalence
        3, 1, equivalence
        35234, 2, equivalence
        39822, 2, equivalence
        39822, 2, equivalence
        9205, 2, equivalence
        412, 1, equivalence
        11135, 2, equivalence
        11135, 2, equivalence
        2375, 1, equivalence
        9507, 2, equivalence
        7878, 2, equivalence
        9507, 2, equivalence
        104074, 1, equivalence
        700289, 2, equivalence
        788225, 2, equivalence
        788225, 2, equivalence
        28226, 1, equivalence
        214897, 2, equivalence
        240097, 2, equivalence
        240097, 2, equivalence
        38502, 1, equivalence
        245697, 2, equivalence
        276225, 2, equivalence
        276225, 2, equivalence
        54758, 1, equivalence
        440569, 2, equivalence
        488629, 2, equivalence
        488629, 2, equivalence
        12239050, 2, equivalence
        12239050, 2, equivalence
        3, 1, equivalence
        10685466, 2, equivalence
        40556396, 2, equivalence
        40556396, 2, equivalence
        3, 1, equivalence
        35431922, 2, equivalence
        9488018, 2, equivalence
        9488018, 2, equivalence
        3, 1, equivalence
        8326786, 2, equivalence
        31611530, 2, equivalence
        31611530, 2, equivalence
        3, 1, equivalence
        27799634, 2, equivalence
        7626354, 1, equivalence
        33702306, 2, modelchecking
        43, 1, modelchecking
        29868274, 2, modelchecking
        61, 1, modelchecking
        33702306, 2, modelchecking
        33702306, 2, modelchecking
        43, 1, modelchecking
        29868274, 2, modelchecking
        61, 1, modelchecking
        5322498, 1, equivalence
        19026506, 1, equivalence
        10927074, 2, equivalence
        10927074, 2, equivalence
        3, 1, equivalence
        9558194, 2, equivalence
        3, 1, equivalence
        37636481, 2, equivalence
        37636481, 2, equivalence
        3782172, 1, equivalence
        14808231, 1, equivalence
        2544209, 1, mlsolver
        196527, 1, mlsolver
        1167042, 2, mlsolver
        62953, 3, mlsolver
        1348959, 3, mlsolver
        35388, 3, mlsolver
        3155971, 3, mlsolver
        80133, 3, mlsolver
        1021740, 2, mlsolver
        426185, 4, mlsolver
        1042752, 3, mlsolver
        183781, 4, mlsolver
        2177202, 1, modelchecking
        9101, 1, modelchecking
        48861, 1, modelchecking
        6, 1, modelchecking
        531443, 1, modelchecking
        1594325, 1, modelchecking
        13834801, 1, modelchecking
        27876961, 1, modelchecking
        188570, 1, modelchecking
        377113, 1, modelchecking
        417016, 1, modelchecking
        1285576, 1, modelchecking
        571378, 1, modelchecking
        1411274, 1, modelchecking
        2341446, 1, modelchecking
        1997579, 1, modelchecking
        1997579, 1, modelchecking
        134162, 1, modelchecking
        134162, 1, modelchecking
        998790, 1, modelchecking
        788879, 1, modelchecking
        267378, 1, modelchecking
        267378, 1, modelchecking
        20712450, 2, modelchecking
        8964338, 2, modelchecking
        6823297, 2, modelchecking
        3487362, 2, modelchecking
        6974724, 2, modelchecking
        7767169, 1, modelchecking
        11488274, 1, modelchecking
        94710, 1, modelchecking
        2589057, 1, modelchecking
        10782593, 2, modelchecking
        7310722, 1, modelchecking
        19550209, 2, modelchecking
        8835074, 2, modelchecking
        22288897, 1, modelchecking
        6690605, 1, modelchecking
        7429633, 1, modelchecking
        24565250, 1, modelchecking
        861780, 2, modelchecking
        40200, 3, specialcases
        80400, 3, specialcases
        20100, 3, specialcases
        201000, 3, specialcases
        80200, 3, specialcases
        160400, 3, specialcases
        40100, 3, specialcases
        401000, 3, specialcases
        20200, 3, specialcases
        40400, 3, specialcases
        10100, 3, specialcases
        101000, 3, specialcases
        200200, 3, specialcases
        400400, 3, specialcases
        100100, 3, specialcases
        1001000, 3, specialcases
        100000, 20001, specialcases
        250000, 50001, specialcases 
      };
      \legend{~modelchecking, ~equivalence, ~mlsolver, ~specialcases, ~random}
    \end{axis}
  \end{tikzpicture}